\documentclass[a4paper,12pt]{article}
\pdfoutput=1

\usepackage{jheppub}
\usepackage[sort&compress]{natbib}
\usepackage[utf8]{inputenc}

\usepackage{amsmath,amssymb,amsthm}
\usepackage{physics,mathtools}
\usepackage{float}
\usepackage{subfig}
\usepackage[svgnames]{xcolor}

\usepackage{comment}
\usepackage[normalem]{ulem}
\usepackage[english]{babel}
\usepackage{longtable}
\usepackage{array}
\usepackage{tabularx}
\usepackage{booktabs}
\usepackage{rotating}

\usepackage{graphicx}

\hypersetup{linkcolor=DarkBlue,citecolor=DarkBlue, urlcolor=black}
\makeatletter
\def\@fpheader{\vspace{1mm}}
\makeatother

\title{\LARGE{Bulk phase shift and singularity}}

\author{\hspace{0.02cm} Yueke Jia$^{\color{black}a}$}
\author{and Manuela Kulaxizi$^{\color{black}a,b}$\linebreak\vspace{-0.2cm}}
\emailAdd{yujia@tcd.ie, manuela@maths.tcd.ie}

\affiliation[{\color{black}a}]{School of Mathematics and Hamilton Mathematics Institute,
\setlength{\parskip}{0pt}\newline\indent Trinity College, Dublin 2, Ireland}
\affiliation[{\color{black}b}] {Division of Nuclear and Particle Physics, Department of Physics, \setlength{\parskip}{0pt}\newline\indent National and Kapodistrian University of Athens, Greece}

\preprint{XXX}

\abstract{High-energy scattering of a light particle off a black hole at fixed impact parameter is described by an eikonal phase, which encodes the resulting time delay and angular deflection. This bulk phase shift admits a holographic interpretation as of the thermal momentum-space two-point function of a scalar operator in the CFT in the Regge limit. At small impact parameter, the phase shift acquires an imaginary part signaling inelastic scattering, obscuring the interpretation of time delay and deflection which become complex-valued. However, in a holographic CFT these quantities can also be extracted from the so called bulk-cone singularities of the position space thermal correlator. Extending this analysis to small impact parameters, we find that the position-space correlator develops a singularity precisely when a null geodesic reflects off the black hole singularity and reaches the opposite boundary. This provides a more robust identification of bulk time delay and angular deflection through singularities of position-space correlators.}

\begin{document}

\maketitle
\flushbottom
\newpage

\section{Introduction}
\label{sec.introduction}

The Shapiro time delay and gravitational light deflection remain among the most fundamental tests of General Relativity~\cite{Will:2014kxa}. Closely related to these observables is the notion of the \emph{eikonal phase shift} governing high--energy gravitational scattering at fixed impact parameter. The phase can be computed in the Regge regime of high energy, fixed momentum/angle scattering, from the resummation of multi--graviton exchanges in a $2\to2$ scattering amplitude~\cite{Amati:1987wq,Muzinich:1987in,Sundborg:1988tb}. Equivalently, in the high--energy, fixed--impact--parameter limit, one may describe the process in terms of a massless probe propagating through the shockwave geometry produced by a highly energetic lightlike particle; in this regime, the time delay experienced by the probe’s null geodesic, multiplied by its light--cone momentum, reproduces the eikonal phase shift \cite{Dray:1984ha,tHooft:1987vrq}.

At leading order, the gravitational eikonal phase is governed by single tree--level graviton exchange, which exponentiates in impact--parameter space within the elastic eikonal approximation \cite{Kabat:1992tb} (for a recent study of the exponentiation, see e.g. \cite{Du:2024rkf}). Subleading corrections arise from more intricate diagrams and have been explored in a variety of kinematic limits.
For a recent review see \cite{DiVecchia:2023frv} and references therein.

A particularly interesting case is that of a light probe scattering off a much heavier object, where the ratio of the Schwarzschild radius of the heavy object to the impact parameter provides a natural expansion parameter. In this context, moving from the elastic to the inelastic regime, arising for sufficiently small impact parameters (smaller than a certain critical value) received attention early on \cite{Matzner:1968ufm}.

The AdS/CFT correspondence~\cite{Maldacena:1997re,Witten:1998qj,Gubser:1998bc} offers a complementary perspective on this problem. In an asymptotically AdS$_{d+1}$ spacetime, the analogue of high--energy scattering is encoded in a dual CFT four--point function evaluated in the Regge limit~\cite{Cornalba:2006xk,Cornalba:2006xm,Cornalba:2007zb,Cornalba:2007fs,Costa:2012cb}. The dominant contribution to the correlator comes from the stress--tensor conformal block, followed by towers of double--trace (and higher) operators.
Focusing on the light-heavy particle scattering in gravity in the impact--parameter representation, the eikonal phase emerges from a Fourier transform of the heavy--heavy--light--light (HHLL) correlator in the dual CFT description. In holographic theories at large $N$ and in the Regge limit, only stress--tensor exchanges contribute to the leading eikonal phase. Contributions from double--trace operators appear to be parametrically suppressed at this order. Crossing symmetry then implies that, under these assumptions, the eikonal phase is related to the anomalous dimensions of large--spin heavy--light double--trace operators in the dual CFT. A partial list of references can be found in \cite{Kulaxizi:2018dxo, Karlsson:2019qfi,Kulaxizi:2019tkd, Karlsson:2019txu,Meltzer:2019pyl, Karlsson:2020ghx,Giusto:2020mup,Bianchi:2020yzr,Parnachev:2020zbr, Chandorkar:2021viw, Ceplak:2021wak, Kim:2021hqy, Geytota:2021ycx, Bianchi:2022wku, deRham:2022hpx, Hartman:2022njz, Giusto:2023awo, Dodelson:2023nnr, Fardelli:2024heb, Salgarkar:2023sya, Chen:2024iuv}.

Turning to the AdS realization of heavy--light high--energy scattering, one finds a close parallel with flat space. The amplitude appears to exponentiate in the impact--parameter representation, with the leading eikonal phase arising from single graviton exchange, and the first correction corresponding to the AdS analogue of a triangle Witten diagram, often interpreted as a correction induced by the heavy operator’s worldline. In this setting, the expansion may be parameterized by a dimensionless quantity $\mu \sim$ (Schwarzschild radius of the AdS--Schwarzschild black hole)/(AdS curvature radius), with the amplitude also depending on the AdS impact parameter $\mathbf{b}$. As shown for example in~\cite{Kulaxizi:2018dxo}, the AdS eikonal phase can be computed to all orders in $\mu$ using null--geodesic techniques, and reduces smoothly to the flat--space result in the limit where both the AdS radius and the Schwarzschild radius are large compared to the impact parameter.

It is natural to ask what occurs when the impact parameter drops below its critical value, such that the probe particle no longer returns to the AdS boundary. This regime was recently investigated in~\cite{Parnachev:2020zbr}, where the eikonal phase in heavy--light AdS scattering was studied via analytic continuation. It was found that, for sufficiently small impact parameter, the eikonal phase develops a large imaginary part, which may be interpreted as signaling the onset of inelastic effects such as absorption or capture by the black hole. While this interpretation is natural from the scattering perspective, its meaning in terms of time delay and angular deflection of null geodesics is less transparent, since both quantities generically become complex in this regime.

In the AdS--Schwarzschild geometry, a null geodesic that crosses the horizon is associated with a complex time delay, whose imaginary part equals $\beta/2$ and reflects the presence of the second asymptotic boundary in the Penrose diagram. By contrast, the angular deflection defined along such null geodesics remains real, even when the corresponding time delay becomes complex.

In the regime of elastic scattering, the time delay and angular deflection can alternatively be inferred from the locations in time and angle at which a thermal CFT two--point function on the boundary becomes singular, giving rise to the so--called bulk--cone singularities. These singularities were first advocated in~\cite{Hubeny:2006yu}, discussed in two--dimensional CFTs in~\cite{Kulaxizi:2018dxo}\footnote{The case of two-dimensional CFTs is special in this regard. In \cite{Kulaxizi:2018dxo} the singularity appears for the two-point correlator evaluated at a conical defect rather than a thermal state. The latter remains an interesting case to explore. For recent developments, see \cite{AliAhmad:2026wem}.}, and established for general holographic theories using bulk gravitational methods in~\cite{Dodelson:2023nnr} (see also \cite{Hashimoto:2019jmw, Hashimoto:2023buz, Riojas:2023pew, Riojas:2023pje}). In asymptotically flat black hole geometries, such singularities were previously investigated in \cite{Dolan:2011fh} (see also \cite{Buss:2017vud, Casals:2009zh, Casals:2016qyj, Zenginoglu:2012xe}).

In this note, we further explore this alternative viewpoint on bulk null--geodesic dynamics, focusing on the leading--order physics for impact parameters below the critical value. Concretely, we begin with the bulk phase--shift representation of the thermal two--point function in momentum space, analytically continue the phase shift to account for impact parameters beyond the critical one, and then Fourier transform back to position space to obtain the thermal correlator. The resulting behavior is intriguing: the thermal Wightman function develops a singularity that is naturally associated,  within a semiclassical geodesic approximation, with a null trajectory that enters the black hole interior and emerges on the opposite boundary, as originally discussed in~\cite{Fidkowski:2003nf,Festuccia:2005pi,Festuccia:2008zx}. The structure and scaling of the singularity coincide with those found in~\cite{Ceplak:2024bja} via an OPE resummation of the stress--tensor sector (see also \cite{Parisini:2023nbd})\footnote{For an earlier discussion of the phase shift in connection to singularities of the thermal correlator see \cite{Amado:2008hw}.}.

However, as was already pointed out in \cite{Fidkowski:2003nf} and confirmed explicitly in \cite{Buric:2025anb, Buric:2025fye}, the two-sided Wightman correlator does not exhibit any singularities. Contributions of the double-trace operators in the OPE cancel out the singularity of the stress-tensor sector.  The correlator studied herein is not the standard Wightman, but the one analytically continued to the Regge sheet,  as dictated by the Regge kinematics of the scattering problem and causality considerations.  However,  one would not expect it to develop a singularity unless a further analytic continuation, around the lightcone defined by the position of the two identical operators -- in the spirit of the ``bulk point'' \cite{Maldacena:2015iua} -- was performed.  By examining the region above the photon sphere and considering in particular a complexified time variable, we are effectively moving into the bulk-point realm. We view our results as complementary to the recent developments in~\cite{Afkhami-Jeddi:2025wra,Jia:2025jbi,Dodelson:2025jff, AliAhmad:2026wem}.

Returning to the bulk phase shift, these observations possibly suggest that in this regime, time delay and angular deflection are more appropriately extracted from the singularity structure of the boundary correlator, rather than directly from the bulk phase shift.

The paper is organized as follows. In Section~\ref{sec.two} we review the definition of the eikonal (bulk) phase shift in the AdS--Schwarzschild geometry. In Section~\ref{sec.three} we discuss its interpretation in the dual CFT as a finite--temperature momentum--space Wightman two--point function. In Section~\ref{sec.four} we summarize known results for the bulk phase shift beyond the critical impact parameter. Finally, in Section~\ref{sec.five} we compute the position--space two--point function near the singularity by explicitly integrating the bulk phase shift along the lines of \cite{Dodelson:2023nnr}. Section~\ref{sec.Conclusions} contains our conclusions together with open questions. Appendix~\ref{app:wkb} presents the WKB analysis of the radial equation, including the matching at the AdS boundary, the photon-sphere barrier and the continuation beyond the critical impact parameter. Appendix~\ref{appendixa} discusses the WKB evaluation of the phase shift,  exploring the possibility of obtaining the result via a near-boundary analysis. Appendix~\ref{app:GR} derives the position-space integral of Section~\ref{sec.five} from the retarded function, and Appendix~\ref{app:num} contains numerical tests in $d=4$. Appendices~\ref{appendixh}, \ref{appendixf}, \ref{appendixb}, \ref{appendixc} and~\ref{appendixLO} contain the CFT setup and further technical details.


\section{The bulk phase shift -- gravity}
\label{sec.two}

\noindent Consider an asymptotically AdS black hole in $(d+1)$ dimensions with AdS radius $R$:
\begin{equation}
\label{eq:adssch}
 ds^2 = -f dt^2 +f^{-1} dr^2 +r^2 d\Omega^2
\end{equation}
where
 \begin{equation}\label{eq:sphere metric}
 d\Omega^2  = d\theta^2 +\sin^2 \theta \ d\Omega_{d-2}^2\,,
 \end{equation}
and
\begin{equation}
\label{eq:metricfunctionbh}
f=1+\frac{r^2}{R^2} -\frac{\mu}{r^{d-2} }\,~~, \,~~
 \mu\equiv
 \left[\frac{d-1}{16 \pi }\Omega_{d-1}\right]^{-1} G_N M
\,.
\end{equation}
The Hawking temperature $T_H$ is~\cite{Witten:1998qj}
\begin{equation}
\label{eq:HawkingT}
     T_H =\frac{1}{\beta}= \frac{ {d r_H^2 +(d-2) R^2}}{ {4\pi R^2 r_H} }
\end{equation}
where $r_H$ denotes the position of the horizon:
 \begin{equation}
 \label{eq:rHdef}
f(r=r_H)=0\,.
\end{equation}
The two Killing vectors, $\partial_t$ and $\partial_\theta$, of \eqref{eq:adssch} allow one to define quantities conserved along the geodesics, {\it i.e.}, the energy and angular momentum:
\begin{equation}\label{eq:Killingv}
  p^t= \left( 1+ \frac{r^2}{ R^2} -\frac{\mu}{ r^{d-2} } \right) \frac{\partial t}{ \partial \lambda}, \qquad p^\theta = r^2 \frac{\partial \theta}{\partial \lambda}\,,
  \end{equation}
with $\lambda$ denoting an affine parameter.
The equation describing null geodesics becomes
\begin{equation}
\label{eq:nullgeodesicdef}
  \dot{r}^2 + (p^\theta)^2 V_{eik} (r)  =  (p^t)^2\,
\end{equation}
where
\begin{equation}
\label{eq:Veikdef}
   V_{eik} (r) = \frac{f(r)}{r^2}\,.
 \end{equation}

A light ray starting from the boundary, traversing the bulk and reemerging on the boundary, will experience both a time delay and a deflection which can be expressed as follows:
\begin{equation}
\label{eq:deltatandphi}
\begin{aligned}
    \Delta t &= 2\, b \int_{r_T}^\infty \frac{dr}{ f\sqrt{b^2-\frac{f(r)}{ r^2} } } \\
    \Delta \theta &=2\, \int_{r_T}^\infty \frac{dr}{r^2\sqrt{b^2-\frac{f(r)}{ r^2} }}\,.
\end{aligned}
\end{equation}
Here, $b=\frac{p^t}{p^\theta}$ and $r_T$ denotes the turning point of the geodesic, whose existence ensures that the light ray will reach the boundary. It is the minimum point of the trajectory, given by the loci of real and positive $r$ for which $\dot{r}=0$:
\begin{equation}
\label{eq:r0def}
b^2-\frac{f(r_T)}{r_T^2}=0\,.
\end{equation}
Note that $b$ is related to the impact parameter $\mathbf{b}$ as defined in pure AdS,
\begin{equation}\label{eq:bdef}
    \mathbf{b}=\left(b^2-\frac{1}{ R^2}\right)^{-\frac{1}{2}}\,,
\end{equation}
as can be seen from (\ref{eq:r0def}) setting $\mu=0$.
Clearly, $\mathbf{b}$ reduces to the familiar expression of flat space, $\mathbf{b} \approx \frac{p^\theta} {p^t}$, for large $R$ (or small $p^\theta/p^t$) whereas it diverges in the limit ${p^\theta/ p^t} \rightarrow R$.

In this note, we are interested in the bulk phase shift. For a particle described by a plane wave, the bulk phase shift is as follows:
\begin{equation}
\label{eq:deltadef}
\delta\equiv -p\cdot (\Delta x)=p^t (\Delta t)-p^\theta \,(\Delta\theta)\,,
\end{equation}
with $p^{t,\theta}$ denoting the momenta of the particle traversing the geometry. Combining (\ref{eq:deltadef}) with (\ref{eq:deltatandphi}) yields:
\begin{equation}\label{eq:deltadefint}
\delta(\sqrt{-p^2},b)=2\, p^\theta \int_{r_T}^\infty \, \frac{dr}{f(r)}\,
\sqrt{b^2-\frac{f(r)}{\,r^2}}
\end{equation}
In pure AdS, the bulk phase shift takes the form:
\begin{equation}\label{eq:deltaads}
\delta_{AdS}=\pi\,\big(R\,p^t-p^\theta\big)=\pi\,p^\theta\,\big(R\,b-1\big)\;\xrightarrow[R=1]{}\;\pi\,p^\theta\,(b-1)\,,
\end{equation}
since $\Delta t =R \,(\Delta \theta)= R\,\pi$: all null geodesics converge at the same point.\footnote{When comparing with the wave equation, $p^\theta$ is identified with $k=\ell+\frac{d-2}{2}$ rather than with $\ell$, see (\ref{eq:WKBnullgeodmatch}) and the end of Section~\ref{sec.three}.} It is not a scattering effect: in Fourier space it implements the shift $(t,\theta)\to(t+\pi,\theta+\pi)$ to the point antipodal to the source, where all null geodesics of pure AdS refocus. Its role in the comparison with the CFT definition of the phase shift is discussed at the end of Section~\ref{sec.three}.



\section{The bulk phase shift -- CFT}
\label{sec.three}

The bulk phase shift previously defined has a natural description in the dual CFT; it corresponds to the Fourier transform of the thermal two-point function on the sphere, of an operator of generic conformal dimension $\Delta\ll C_T$, weighted by the zero-temperature correlator. Here and in the rest of the paper we set the AdS radius $R=1$. Our starting point is the thermal Wightman function on the cylinder and its Fourier transform,
\begin{equation}
\label{eq:GWTdef}
G_+^T(t,\theta)=\langle \mathcal{O}_\Delta(x)\mathcal{O}_\Delta(0) \rangle_{T,S^{d-1}}\,,\qquad G_+^T(\omega,\ell)\equiv \int d^d x \, e^{ipx} \, G_+^T(t,\theta)\,.\footnotemark
\end{equation}
\footnotetext{This is schematic. On $S^{d-1}$ the exact transform is
\begin{equation*}
G_+^T(\omega,\ell)=\int_{-\infty}^{\infty}dt\,e^{i\omega t}\int_{S^{d-1}}d\Omega\;\frac{C_\ell^{\frac{d-2}{2}}(\cos\theta)}{C_\ell^{\frac{d-2}{2}}(1)}\,G_+^T(t,\theta)\,,
\end{equation*}
whose inverse is (\ref{eq:GWFourier}).}%
The superscript $T$ stands for finite temperature, whereas the variable $\ell$ corresponds to the angular momentum of a $d$-dimensional CFT placed on the sphere $S^{d-1}$.

Following \cite{Kulaxizi:2018dxo} (see Appendix~\ref{appendixh} for details), and assuming thermalization, the correlator can be written as a heavy-heavy-light-light four-point function,
\begin{equation}
\label{eq:GHHLLmain}
G_+^T(t,\theta)=\langle \mathcal{O}_H|\,\mathcal{O}_\Delta(t,\theta)\,\mathcal{O}_\Delta(0)\,|\mathcal{O}_H\rangle=u^{\Delta/2}\,\upsilon^{-\Delta}\,\mathcal{A}(u,\upsilon)\,,
\end{equation}
where $\mathcal{A}(u,\upsilon)$ depends only on the cross-ratios
\begin{equation}
\label{eq:zzbarmain}
u=z\bar z=e^{2it}\,,\qquad \upsilon=(1-z)(1-\bar z)=1+e^{2it}-2e^{it}\cos\theta\,,\qquad z=e^{ix^+}\,,\quad \bar z=e^{ix^-}\,,
\end{equation}
with $x^\pm=t\pm\theta$. The prefactor $u^{\Delta/2}\upsilon^{-\Delta}$ contains both the light-cone singularity and the conformal factor relating the cylinder to the plane, and $\mathcal{A}=1$ at zero temperature. The ordering of the operators in the Wightman function corresponds to $t\to t-i\epsilon$.

The phase shift is defined in a certain kinematical regime, the eikonal or Regge limit. In the CFT literature \cite{Kulaxizi:2018dxo,Kulaxizi:2017ixa} this is the limit where the second operator approaches the point shifted by $\pi$ in time and angle with respect to the first one (Appendix~\ref{appendixh}). It is described by the Regge variables
\begin{equation}
\label{eq:Reggevariables}
t'=t-\pi\,,\qquad \theta'=\theta-\pi\,,\qquad\text{i.e.}\qquad x'^{+}=x^+-2\pi\,,\quad x'^-=x^-\,,
\end{equation}
and corresponds to small $x'^{\pm}=t'\pm\theta'$, in the sense of (\ref{eq:Reggelimit}). To approach this configuration starting from the two operators placed close to each other on the spatial sphere and at the same time requires the analytic continuation $z\to z e^{2\pi i}$ at fixed $\bar z$, which corresponds to shifting $x^+\rightarrow x^{+}+2\pi$, and takes the correlator to the Regge sheet. We denote by $G^T_{\rm Regge}(t',\theta')$ the correlator continued in this way, as a function of the Regge variables, and by $G^T_{\rm Regge}(\omega,\ell)$ its Fourier transform, defined as in (\ref{eq:GWTdef}) with $(t,\theta)\to(t',\theta')$. The CFT phase shift is then defined in the Regge limit by
\begin{equation}
\label{eq:deltaCFTdef}
e^{i\delta_{CFT}(p)}\equiv \frac{G_{\rm Regge}^T(\omega, \ell) }{ G_{\rm Regge}^{T=0}(\omega,\ell) }\,,
\end{equation}
where numerator and denominator are understood as the leading behavior of the respective correlators in the Regge limit, rather than as exact Fourier transforms; this distinction is discussed in Section~\ref{sec.six}.

This appears different from the standard Wightman correlator; nevertheless, it is a physically natural definition for the kinematics of interest; we will see at the end of this section that it is in fact the standard Wightman correlator evaluated near the shifted point.
In Fourier space, the eikonal limit corresponds to large frequency and angular momentum, with their ratio being kept fixed:
\begin{equation}
\label{eq:eikonallimitedef}
\omega,k\gg 1,\quad b\equiv \frac{\omega}{k}=\text{finite}\,,
\end{equation}
where
\begin{equation}
k\equiv \ell+\frac{d-2}{2}\,.
\label{eq:ell_k_shift}
\end{equation}
At leading order $b\sim\omega/\ell$; the shift is motivated by holographic computations.\footnote{In essence, the shift of $\ell$ is natural for two reasons: the eigenvalue of the Laplacian on $S^{d-1}$ is $\ell(\ell+d-2)=k^2-\frac{(d-2)^2}{4}$, i.e.\ $k^2$ up to a constant, and $k$, rather than $\ell$, is the variable conjugate to $\theta$ in the large-spin form of the Gegenbauer polynomials, (\ref{eq:Ckapprox.}). It is immaterial at leading order, but it matters for the phases of order $k^0$ discussed below.}

In the Regge limit, and for holographic CFTs, {\it a.k.a.}, CFTs with an equivalent description of weakly coupled Einstein gravity,\footnote{In general, the Regge behavior of the phase shift is $\omega^{J-1}$ where $J$ is the spin of the Reggeon. In holography, $J=2$ leads to the linear growth of (\ref{eq:deltaCFTexpect}) with $\omega$ at fixed $b$.} $\delta_{CFT}(p)$ as defined in (\ref{eq:deltaCFTdef}) is given by the bulk phase shift (\ref{eq:deltadefint}) with the pure AdS contribution removed,
\begin{equation}
\label{eq:deltaCFTexpect}
\begin{gathered}
\delta_{CFT}=k\,\big(S(b)-S_{AdS}(b)\big)\,,\\
\delta_{AdS}=k\,S_{AdS}(b)=k\,\pi(b-1)=\pi(\omega-k)\,,
\end{gathered}
\end{equation}
with $b=\omega/k$ and $k$ given by (\ref{eq:ell_k_shift}). This relation holds to all orders in $\mu$ up to $\mathcal{O}(1/k)$ corrections in the phase; the pure AdS contribution $\delta_{AdS}$ is accounted for by the shift between Regge and cylinder variables. To leading order in $\mu$ it agrees with the Regge-theory computations of \cite{Kulaxizi:2018dxo, Parnachev:2020zbr, Karlsson:2019qfi, Karlsson:2019txu}.

To understand this, following \cite{Parnachev:2020zbr}, we consider the computation of the momentum space thermal two-point function holographically. The starting point is the equation of motion for a scalar field $\phi$ of mass $m$ in the black hole metric (\ref{eq:adssch}): $(\Box-m^2)\phi=0$. The mass of the field is related to the conformal dimension of the dual CFT operator via $m^2 R^2=\Delta (\Delta-d)$.
It is convenient to use the Fourier decomposition
\begin{equation}
\label{eq:phidecomp}
\phi(t,\Omega_{d-1},r)=e^{-i\omega t} Y_{\ell,m}(\Omega_{d-1}) r^{-\frac{d-1}{ 2}}\psi_{\omega\ell}(r)
\end{equation}
with $Y_{\ell,m}$ the spherical harmonics on $S^{d-1}$.
To study the eikonal limit, we set $\psi=e^{ikS}$ and retain the leading order terms in $\omega,k$, using $\ell(\ell+d-2)\simeq k^2$, which leads to the following equation for $S(r)$
\begin{equation}
\label{eq:WKBleading}
  \begin{aligned}
f(r)^2\,(\partial_r S)^2=b^2-\frac{f(r)}{ r^2}
\end{aligned}
\end{equation}
By comparing (\ref{eq:WKBleading}) with (\ref{eq:nullgeodesicdef}), we see that the equation which governs the null geodesics and the leading order equation which governs the dynamics of a scalar field in the same background for large frequency and momentum are the same with the simple identification:
\begin{equation}\label{eq:WKBnullgeodmatch}
  \begin{aligned}
  k\,f(r)(\partial_r S)=\dot{r},\quad p^\theta=k, \quad p^t=\omega
\end{aligned}
\end{equation}
Then, by combining equations (\ref{eq:WKBnullgeodmatch}) and (\ref{eq:deltadef}), we obtain the following relation:
\begin{equation}
  \begin{aligned}
k S(b)=\omega T(b)-k\,\Theta(b)=k\left(b T(b)-\Theta(b)\right)
\end{aligned}
\label{eq:Sdef}
\end{equation}
where
\begin{equation}
\begin{aligned}
T=&2\int_{r_T}^\infty \frac{b}{f(r)}\frac{dr}{\sqrt{b^2-\frac{f(r)}{r^2}}}\\
\Theta&=2\int_{r_T}^\infty \frac{1}{r^2}\frac{dr}{\sqrt{b^2-\frac{f(r)}{r^2}}}
\end{aligned}
\label{eq:TThetadefR}
\end{equation}
Here $T(b),\Theta(b)$ are equal to the time delay and angular deflection of a particle traversing a lightlike path in the AdS black hole geometry. We have named them differently here, to emphasize their nature as parts of the Fourier space correlator of the CFT. We retain this distinction henceforth and refer to $\Delta t,\Delta\theta$ when dealing with the bulk phase shift and to $T,\Theta$ when discussing the thermal CFT two-point function.

By virtue of their definition $(T(b),\Theta(b))$ satisfy the following identities:
\begin{equation}
\label{eq:propertiesTThetaA}
T(b)=\frac{\partial S(b)}{\partial b},\quad \Theta(b)= -S(b)+b\frac{\partial S}{\partial b}\,
\end{equation}
which combined with (\ref{eq:Sdef}) lead to
\begin{equation}
\label{eq:TThetafirstderid}
b\, T'(b)-\Theta'(b)=0\,.
\end{equation}
Similar identities relating their higher derivatives can be accordingly derived.
The identifications in (\ref{eq:Sdef}) are similar to those relating spacelike geodesics with the WKB phase describing the correlator for $\Delta\gg 1$ (see e.g. \cite{Festuccia:2005pi}).

The relationship between the null geodesics and the solutions to the scalar field equation is highly suggestive. In $d=2$, both a purely CFT and a fully fledged holographic computation have been performed and confirmed this relationship, albeit in the context of a conical defect state rather than a thermal one (see e.g. \cite{Kulaxizi:2018dxo})\footnote{$d=2$ is special: in the BTZ black hole, $V_{eik}=1-r_H^2/r^2$ has no maximum, so no null geodesic returns to the same boundary; this is not the case for a conical defect. Nor is there a bouncing singularity: the retarded correlator of the planar BTZ black hole, known in closed form, is singular only on the light cone and at its thermal images \cite{Grozdanov:2026cut}.}. In higher dimensions, a detailed holographic computation with the help of the WKB approximation was performed in \cite{Dodelson:2023nnr} (see eq.~(4.9) of said article). On the CFT side, the bulk phase shift was identified with the HHLL correlator via Regge theory to certain leading orders in $\mu$  \cite{Kulaxizi:2018dxo,  Karlsson:2019qfi, Karlsson:2019txu}. A partial list of references on the bulk phase shift in this context includes \cite{Meltzer:2019pyl, Giusto:2020mup, Bianchi:2020yzr, Chandorkar:2021viw, Ceplak:2021wak, Kim:2021hqy, Geytota:2021ycx, Bianchi:2022wku, Hartman:2022njz, Fardelli:2024heb, Salgarkar:2023sya, Giusto:2023awo, Chen:2024iuv}.

A derivation of this relationship from the wave equation, starting from the retarded function and treating the AdS boundary, the photon sphere and the horizon explicitly, is given in Appendices~\ref{app:wkb} and~\ref{app:GR} (see also \cite{Dodelson:2023nnr}); Appendix~\ref{appendixa} discusses to what extent it can be obtained from near-boundary data alone. The upshot is that, in the eikonal regime and in the cylinder variables $(t,\theta)$, the momentum-space thermal Wightman function of a holographic CFT takes the form
\begin{equation}
\label{eq:GTansatz}
G_{+}^T(\omega, \ell)=G_{+}^{T=0}(\omega, k)\, e^{-i\pi(\nu+1)}\, e^{i k\, S(b)}\,,\qquad \nu=\Delta-\frac{d}{2}\,,
\end{equation}
in the Regge limit. Throughout the rest of this paper we work with this Regge limit of the unshifted Wightman function.\footnote{The reader may be concerned that (\ref{eq:GTansatz}) is not real and positive, as the Fourier transform of a Wightman function should be. This is because the Regge limit retains only part of the full correlator; we defer this clarification to Section~\ref{sec.six}.} Here
\begin{equation}
\delta=k S(b)
\end{equation}
is the full bulk phase shift (\ref{eq:deltadefint}), including $\delta_{AdS}$, $S(b)$ satisfies eq.~(\ref{eq:WKBleading}), and $b=\omega/k$ with $k$ given by (\ref{eq:ell_k_shift}). The use of $k$ is natural within the WKB approximation, where $k^2$ is the coefficient of the eikonal potential (Appendices~\ref{app:wkb} and~\ref{appendixa}); with this choice (\ref{eq:GTansatz}) holds up to $\mathcal{O}(1/k)$ corrections in the phase (Appendix~\ref{app:wkb}).

The shift is also the origin of part of the constant phase in (\ref{eq:GTansatz}). Indeed, writing $e^{-i\pi(\nu+1)}=e^{-i\pi\Delta}\,e^{i\pi\frac{d-2}{2}}$ and $\delta_{AdS}=\pi(\omega-k)$, i.e.\ (\ref{eq:deltaads}) with $p^\theta\to k$, one finds
\begin{equation}
\label{eq:phaseorigin}
e^{-i\pi(\nu+1)}\,e^{i\delta_{AdS}}=e^{-i\pi\Delta}\,e^{i\pi(\omega-\ell)}\,.
\end{equation}
The factor $e^{i\pi(\omega-\ell)}$, with integer $\ell$, implements the shift (\ref{eq:Reggevariables}) to the antipodal point, and $e^{-i\pi\Delta}$ is the phase acquired by the correlator on the Regge sheet; both are discussed at the end of this section. The factor $e^{i\pi(d-2)/2}$ simply compensates for the use of $k$ rather than $\ell$ in the exponent of (\ref{eq:GTansatz}).

For the zero-temperature two-point function we use the standard Wightman function, whose Fourier transform is the positive spectral density\footnote{In \cite{Kulaxizi:2018dxo} this function was normalized with an additional factor $e^{i\pi\Delta}$. With the conventions used here and there ($t\to t-i\epsilon$, $z\to z e^{2\pi i}$, Fourier kernel $e^{i\omega t}$), the phase acquired by the cylinder correlator on the Regge sheet is $e^{-i\pi\Delta}$, see the end of this section. The factor $u^{\Delta/2}=(z\bar z)^{\Delta/2}$ relating the cylinder to the plane (Appendix~\ref{appendixh}) is close to one in modulus in the Regge limit, but contributes $e^{i\pi\Delta}$ to the monodromy and must be kept. Constant phases affect neither the location nor the exponent of the singularities, but they do affect the overall coefficients.}
\begin{equation}
\label{eq:GT0norm}
G_{+}^{T=0}(\omega, k)= \mathcal{N}_\Delta\, \left(\omega^2-k^2\right)^{\Delta-\frac{d}{2}}\,,\qquad \mathcal{N}_\Delta=\frac{ 2^{d+1-2\Delta} \pi^{1+\frac{d}{2}}}{ \Gamma\left[\Delta\right]\Gamma\left[\Delta-\frac{d}{ 2}+1\right] }\,,
\end{equation}
where we have suppressed the step function enforcing $\omega>k$.\footnote{At zero temperature the spectral density vanishes for $|\omega|<k$ and for $\omega<0$. At finite temperature this is no longer the case: the thermal Wightman function is nonzero for $|\omega|<k$, although exponentially small at large $k$, since the corresponding waves must tunnel through the region near the boundary where $V_{eik}>b^2$; and for $\omega<0$ it is related to its values at $\omega>0$ by the KMS condition, $G_+(-\omega)=e^{-\beta\omega}G_+(\omega)$. In (\ref{eq:GTansatz}), $G^{T=0}_+$ only provides the normalization for $b>1$, and these regions do not contribute to the singularity studied in Section~\ref{sec.five}; see Appendix~\ref{app:GR}.}

\subsection{Regge sheet versus cylinder variables} The CFT definition (\ref{eq:deltaCFTdef}) and the gravitational relation (\ref{eq:GTansatz}) describe the same correlator in the two sets of variables related by (\ref{eq:Reggevariables}). In terms of the variables (\ref{eq:zzbarmain}), the Wightman prescription $t\to t-i\epsilon$ gives $z=e^{\epsilon}e^{ix^+}$, $\bar z=e^{\epsilon}e^{ix^-}$, so that both lie just outside the unit circle. The continuation $z\to z e^{2\pi i}$ at fixed $\bar z$ is then a loop of $z$ around the origin at $|z|=e^{\epsilon}>1$.\footnote{The Regge loop therefore also encircles $z=1$. This is distinct from the further continuation around $\bar z=1$ discussed in Section~\ref{sec.six}.} In terms of $x^\pm$ the same loop is the real-time evolution $x^+\to x^++2\pi$ at fixed $x^-$ on the Lorentzian cylinder. Hence $G^T_{\rm Regge}(t',\theta')=G^T_+(t'+\pi,\theta'+\pi)$.

To express this in momentum space, recall that on the sphere the Fourier transform (\ref{eq:GWTdef}) amounts to the usual Fourier transform in time together with an expansion in hyperspherical harmonics; for a function of the angle $\theta$ between the two operators the latter is an expansion in Gegenbauer polynomials $C_\ell^{\frac{d-2}{2}}(\cos\theta)$. The shift then has the effect
\begin{equation}
\label{eq:shiftFourier}
\begin{aligned}
t\to t+\pi:&\qquad e^{i\omega t}\to e^{i\pi\omega}\,e^{i\omega t}\,,\\
\theta\to\theta+\pi:&\qquad C_\ell^{\frac{d-2}{2}}(\cos\theta)\to C_\ell^{\frac{d-2}{2}}(-\cos\theta)=(-1)^\ell\,C_\ell^{\frac{d-2}{2}}(\cos\theta)\,.
\end{aligned}
\end{equation}
In momentum space this reads
\begin{equation}
\label{eq:Reggecylinder}
G^T_+(\omega,\ell)=e^{i\pi(\omega-\ell)}\,G^T_{\rm Regge}(\omega,\ell)\,,\qquad e^{i\pi(\omega-\ell)}=e^{i\pi\frac{d-2}{2}}\,e^{i\delta_{AdS}}\,,
\end{equation}
as in (\ref{eq:phaseorigin}). The pure AdS phase shift is therefore, up to the constant $e^{i\pi(d-2)/2}$, the Fourier image of the shift to the antipodal point.\footnote{The constant phase $e^{i\pi(d-2)/2}$ is the one carried by the large-$k$ Gegenbauer polynomials across $\theta=\pi$, cf. the factor $e^{i\pi(d-2)/2\lfloor\theta/\pi\rfloor}$ in (\ref{eq:Ckapprox.}).}

The cylinder correlator at zero temperature reads
\begin{equation}
\label{eq:vacuumzzbar}
G^{T=0}_+=\frac{(z\bar z)^{\Delta/2}}{\big[(1-z)(1-\bar z)\big]^{\Delta}}=u^{\Delta/2}\,\upsilon^{-\Delta}\,.
\end{equation}
Under the loop described above the net phase is $e^{-i\pi\Delta}$ (since the Regge loop also encircles $z=1$). Therefore, in position and in momentum space,
\begin{equation}
\label{eq:GT0Regge}
G^{T=0}_{\rm Regge}(t',\theta')=e^{-i\pi\Delta}\,G^{T=0}_+(t',\theta')\quad\longrightarrow\quad G^{T=0}_{\rm Regge}(\omega,\ell)=e^{-i\pi\Delta}\,G^{T=0}_+(\omega,k)\,.
\end{equation}
Note that the standard Wightman function also acquires a phase $e^{-i\pi\Delta}$ inside the light-cone in position space; this is reproduced by the Fourier transform of the positive density (\ref{eq:GT0norm}). The phase in (\ref{eq:GT0Regge}) is an additional one, associated with the crossing of the antipodal light-cone.\footnote{At $T=0$, (\ref{eq:Reggecylinder}) and (\ref{eq:GT0Regge}) together require $e^{i\pi(\omega-\ell-\Delta)}=1$. This holds exactly, since the zero-temperature spectral density is supported at the normal-mode frequencies $\omega=\Delta+\ell+2n$, $n=0,1,2,\dots$; the smooth large-$k$ form (\ref{eq:GT0norm}) is an average over these frequencies.}

Combining (\ref{eq:GTansatz}), (\ref{eq:Reggecylinder}) and (\ref{eq:GT0Regge}), and using (\ref{eq:phaseorigin}), we find
\begin{equation}
\label{eq:deltaCFTvsbulk}
\frac{G^T_{\rm Regge}(\omega,\ell)}{G^{T=0}_{\rm Regge}(\omega,\ell)}=e^{i\delta_{CFT}}\,,\qquad \delta_{CFT}=k\big[S(b)-\pi(b-1)\big]=\delta-\delta_{AdS}\,.
\end{equation}
Thus $\delta_{CFT}$ vanishes for $\mu=0$, as expected from (\ref{eq:deltaCFTdef}), in agreement with (\ref{eq:deltaCFTexpect}), and is the quantity computed using Regge theory in \cite{Kulaxizi:2018dxo,Karlsson:2019qfi,Karlsson:2019txu} (its leading term is used in Appendix~\ref{appendixLO}).

Note that (\ref{eq:deltaCFTvsbulk}) is a function of $(\omega,\ell)$ through $b=\omega/k$. At leading order at large $\omega,k$ the distinction between $k$ and $\ell$ is immaterial. The constant phases matched above, however, are of order $k^0$, as is the difference
\begin{equation}
\label{eq:kvsell}
kS(\omega/k)-\ell S(\omega/\ell)=-\frac{d-2}{2}\,\Theta(b)+\mathcal{O}(1/k)\,,
\end{equation}
which follows from (\ref{eq:propertiesTThetaA}). The match is therefore meaningful at this order because (\ref{eq:GTansatz}) holds with $k$ rather than $\ell$ up to $\mathcal{O}(1/k)$ corrections in the phase, and because $k$ is the variable conjugate to $\theta$ in (\ref{eq:Ckapprox.}). For $\mu=0$, $\Theta=\pi$, and (\ref{eq:kvsell}) reduces to the constant $-\pi\frac{d-2}{2}$, which accounts for the phase $e^{i\pi(d-2)/2}$ in (\ref{eq:Reggecylinder}). For $\mu\neq0$, $\ell\big(S(\omega/\ell)-S_{AdS}(\omega/\ell)\big)=\delta_{CFT}+\frac{d-2}{2}\big(\Theta(b)-\pi\big)+\mathcal{O}(1/k)$, where the correction is suppressed by $1/k$ relative to $\delta_{CFT}$ but is a phase of order one. The two descriptions are equivalent; the cylinder variables are more convenient from the bulk point of view, since $T(b)$ and $\Theta(b)$ are then the actual time delay and angular deflection, and we use them in the rest of the paper.\footnote{In particular, the singularity found in Section~\ref{sec.five} at $t=t_c+i\beta/2$, $\theta=0$ (note that $t_c\to\pi$ as $\mu\to 0$) is located at $t'=t_c-\pi+i\beta/2$, $\theta'=-\pi$ in Regge variables.}

\section{Beyond the critical impact parameter}
\label{sec.four}

In this note, we are interested in the phase shift beyond the critical impact parameter.\footnote{Since $b=p^t/p^\theta$ is inversely related to the impact parameter $\mathbf{b}$, see (\ref{eq:bdef}), impact parameters smaller than the critical one correspond to $b>b_c$.} In any dimension $d\geq 3$ the time delay $T(b)=S'(b)$ diverges logarithmically when the impact parameter attains a certain critical value $b_c$, while the phase shift itself remains finite (Appendix~\ref{app:wkb:barrier}).  For $b=b_c$ the solution to the turning point equation is equal to the radius of the photon sphere and the divergence of the time delay signals the fact that the particle requires an infinite amount of time to return to the boundary.

The turning point equation in general $d$ can be expressed as
\begin{equation}
\label{eq:turningpointeq}
p[r]=(b^2-1)r^d-r^{d-2}+\mu=0\,.
\end{equation}
where, for simplicity, we have set the radius, $R$, of AdS spacetime to unity. The polynomial p[r] is minimized at $r_{min}=\sqrt{\frac{d-2}{d(b^2-1)}}$, leading to the existence of two real and distinct solutions to $p[r]=0$ when the value at $r_{min}$ lies below the axis. At the point where $p[r_{min}]=0$, the two distinct solutions merge into one, while for $p[r_{min}]>0$, cease to exist. The equation $p[r_{min}]=0$ gives rise to the value of the photon sphere $r_{photon}$ and the critical impact parameter below
\begin{equation}
\label{eq:photonsphere}
\begin{aligned}
r_T&=r_{photon}=\left(\frac{d\mu}{2}\right)^\frac{1}{d-2} \\
b_c^2&=1+\left(1-\frac{2}{d}\right) \left(\frac{2}{ d\mu}\right)^\frac{2}{d-2}
\end{aligned}
\end{equation}
All other solutions of the turning point equation are in general complex. It may be useful to note here that since for $b>b_c$ the turning point equation has no real and positive solutions, the expression under the square root in the definition of the bulk phase shift is nowhere negative.
\begin{equation}
\label{eq:positivityofsquareroot}
b^2-\frac{f(r)}{r^2}=
\frac{1}{r^d}\left((b^2-1) r^d-r^{d-2}+\mu \right)>0, \quad\text{for all $r\geq 0$}\quad.
\end{equation}

In an effort to explore what happens when $b>b_c$, \cite{Parnachev:2020zbr} analytically continued the bulk phase shift, observing that it attains a small imaginary part. Explicit computations in \cite{Parnachev:2020zbr} were performed for the asymptotically flat Schwarzschild black hole in four spacetime dimensions, but can be generalized both to asymptotically AdS black holes and any dimensionality $d$ (see Appendix~\ref{appendixb} for integration formulas that could be useful when treating generic $d$-dimensions).

For the case of $d=4$, the bulk phase shift, time delay and angular deflection for $b<b_c$ read as follows (see also Appendix B of \cite{Kulaxizi:2018dxo})\footnote{There is a typo in (B.11) of \cite{Kulaxizi:2018dxo} - the factors in front of the elliptic integrals should be divided by $\sqrt{1+4 \mathbf{b}^2 v_0^2(1-v_0^2)}$. Also note that $b$ there is $\mathbf{b}$ here.}
\begin{equation}
\label{eq:d4delta}
\delta=k\left(b\Delta t-\Delta\theta\right) \,,
\end{equation}
with
\begin{equation}
\label{eq:d4timedelayAngdeflection}
\begin{aligned}
\Delta t&=-\frac{ b }{r_T\, \sqrt{b^2-1}} \times\\
&\times\left\{\frac{1-\sqrt{1+4\mu}}{ \sqrt{1+4\mu}} \Pi\left[\frac{1-\sqrt{1-4\mu(b^2-1)}}{1+\sqrt{1+4\mu}},\frac{ 1-\sqrt{1-4\mu(b^2-1)}}{ 1+\sqrt{1-4\mu(b^2-1)}}\right]-\right.\\
&\left.-\frac{1+\sqrt{1+4\mu}}{\sqrt{1+4\mu}} \Pi\left[\frac{1-\sqrt{1-4\mu(b^2-1)}}{1-\sqrt{1+4\mu}},\frac{ 1-\sqrt{1-4\mu(b^2-1)}}{ 1+\sqrt{1-4\mu(b^2-1)}}\right]
\right\}\\
\Delta\theta&= \frac{2}{ r_T\sqrt{b^2-1}} K\left[ \frac{ 1-\sqrt{1-4\mu(b^2-1)}}{ 1+\sqrt{1-4\mu(b^2-1)}} \right]
\end{aligned}
\end{equation}
where $K[x],\Pi[y,x]$ denote the (generalized) elliptic integrals of the first and third kind, and the physical turning point is given by:
\begin{equation}
\label{eq:d4tpreal}
r_T^2=\frac{1+\sqrt{1-4\mu(b^2-1)}}{2(b^2-1)}
\end{equation}
When $b>b_c$, the analytic continuation of the time delay and angular deflection integrals generically produces a $b$-dependent imaginary part.
\begin{figure}
\includegraphics[width=0.8\textwidth]{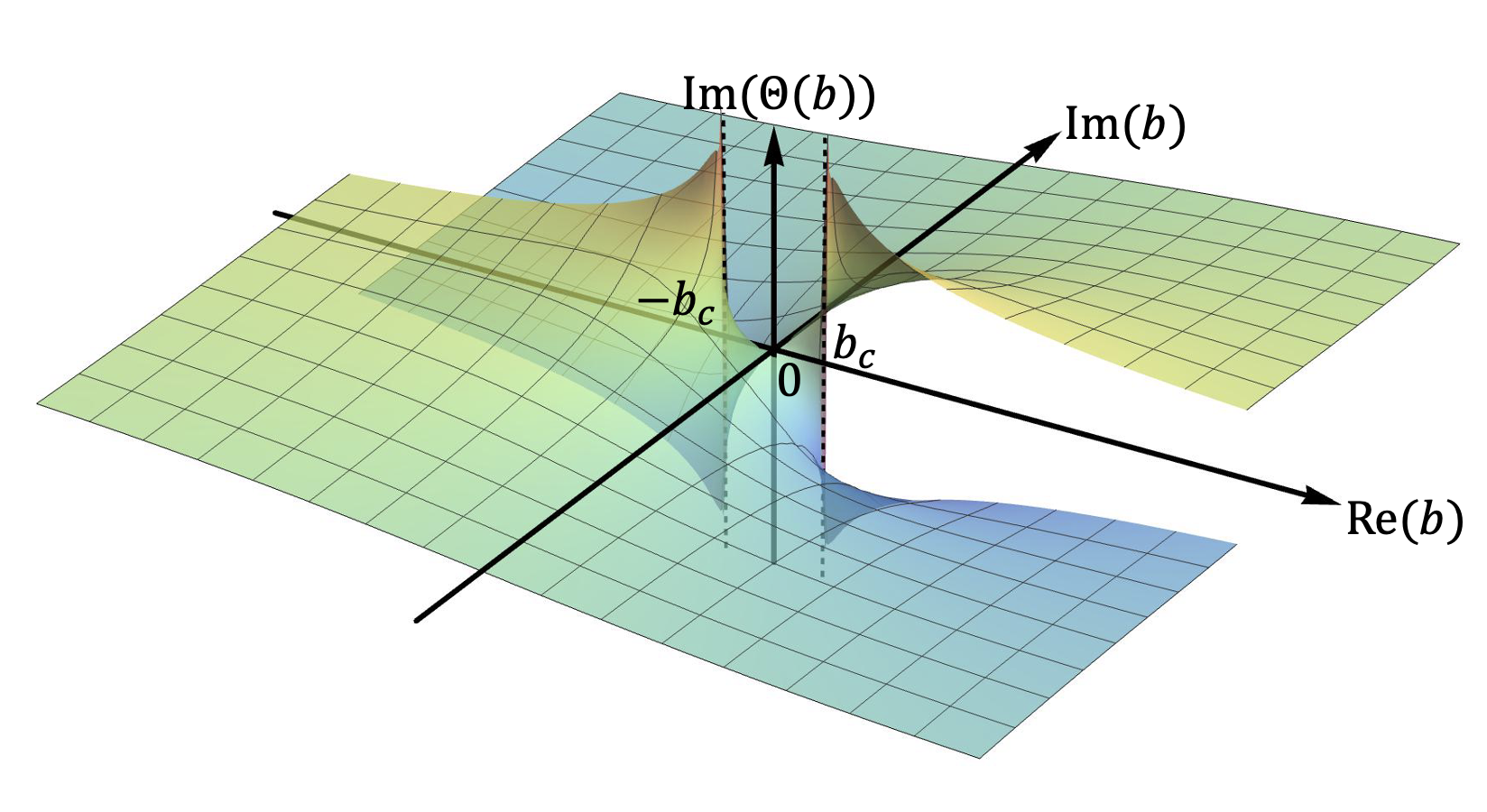}
\caption{This is a plot of the imaginary part of the angular deflection as a function of a general complex parameter $b={\omega\over k}$. For generic complex values of $b$, the angular deflection acquires an imaginary part.}
\label{fig:1}
\end{figure}
Figure \ref{fig:1} is a plot of the imaginary part of $\Delta\theta$ obtained via analytic continuation, for various (complex) values of $b$.
Curiously, for the limiting case $b\rightarrow\infty$, the angular deflection vanishes, while the imaginary part of the time delay approaches the value $\mathrm{Im}(\Delta t)\rightarrow_{b\rightarrow\infty}{\beta\over 2}$.

It is instructive to interpret the analytic continuation of the phase shift as resulting from the evaluation of the phase shift integral with a complex valued turning point. This makes sense in the context of a complex WKB approximation \cite{Kemble35, BerryMountWKB}. Which complex-valued turning point should be considered requires a detailed analysis. In this case, the physically consistent choice is the one which gives rise to a positive imaginary part; it is derived from the uniform WKB treatment of the photon-sphere barrier, with ingoing boundary conditions at the horizon, in Appendix~\ref{app:wkb:continuation} \footnote{See also \cite{Kulaxizi:2017ixa} for arguments on the positivity of the phase shift.}.

Specifically, for $b\in\mathbb{R}$, the integrand is an otherwise real, positive, well-defined function. Hence,
\begin{equation}
\begin{aligned}
S &\propto 2 \int_{r_{\pm}^c}^\infty \frac{dr}{f(r)} \sqrt{b^2-{f(r)\over r^2}} =\\
&=\lim_{r\rightarrow\infty}F(r)-F(r_\pm^c)=\lim_{r\rightarrow\infty}F(r)-\mathrm{Re}(F(r_\pm^c))-i \,\mathrm{Im}(F(r_\pm^c))
\end{aligned}
\label{bulkps}
\end{equation}
where $F(r)$ represents the anti-derivative of the integral which is finite at the boundary, and $r_{\pm}^c$ denote the two complex solutions to the turning point equation to which the original real ones reduce for $b>b_c$. Notice that $F(r_+^c)^\ast=F(r_{-}^c)$ (recall that the two roots are complex conjugates of each other). As a result, the two turning points yield opposite imaginary parts to the phase shift.  The physically consistent option corresponds to the one that gives rise to a decreasing exponential. The resulting bulk phase shift is then equal to the one produced by analytically continuing the expression for impact parameters below the photon-sphere to values above it.

It is curious that, as $b\rightarrow\infty$ the complex turning points approach the singularity, $r_T\rightarrow 0$. It is simple to understand where the imaginary part comes from then. Focusing on the bulk phase shift, in the large impact parameter region, we can express the relevant integral as follows:
\begin{equation}\label{eq:deltalimit}
\begin{aligned}
\delta&= 2 p^\theta \int_{r_T^c(b)}^\infty {dr\over f}\sqrt{b^2-{f(r)\over r^2} } \simeq 2 p^\theta b \int_0^\infty {dr\over f} =\omega \, t_c +i \omega {\beta\over 2}\,,
\end{aligned}
\end{equation}
where in the first line we denoted by $r_T^c$ the complex turning point and in the second line we approximated the integral for large $b$. Here $t_c$ denotes the time that a null geodesic needs to start from the boundary in order to reach the singularity \cite{Hubeny:2006yu}. In $d=4$ it is equal to:
\begin{equation}
\label{eq:trdef}
t_c={\beta\over 2} \,\sqrt{{1\over r_H^2}+1}\,.
\end{equation}
So, for large $b$, the imaginary part arises due to the branch cut of the logarithm at $r=r_H$. The same is true for the time-delay, while the angular deflection vanishes as $b\rightarrow \infty$, {\it i.e.},
\begin{equation}\label {eq:ThetaTinfty}
\begin{aligned}
\Delta\theta(b)&\approx 2\int_{b^{-{2\over d}}}^\infty \frac{dr}{r^2} \frac{1}{b}(1+\frac{1}{2}\frac{f(r)}{r^2b^2})\approx 0\,\\
\Delta t (b) &\approx 2\int_{b^{-{2\over d}}}^\infty \frac{dr}{f(r)}=t_c+i\frac{\beta}{2}
\end{aligned}
\end{equation}
The particle follows the null geodesic trajectory that bounces off the singularity and ``emerges on the other boundary''\cite{Fidkowski_2004, Festuccia:2008zx}.

We could expect the same to occur for generic values of $b>b_c$, since the particle no longer returns to the original boundary. However, for generic values $b>b_c$, the time delay and angular deflection as defined are complex valued, rendering their interpretation obscure.

In summary, the analytic continuation discussed above is the standard approach for dealing with the phase shift in the context of a scattering amplitude or, equivalently, the Fourier transform of the CFT two-point function. In the gravitational picture, however, this procedure raises the following question: what is the meaning of the time delay and angular deflection in this case? Analytically continued in the manner described above, these quantities acquire imaginary parts varying with the impact parameter, thus rendering their physical meaning obscure.

In an effort to reconcile the two pictures, one could take a different perspective on the matter and reconsider the computation of the phase shift as follows: For $b>b_c$,  the light particle does not return to the original boundary. One could then evaluate the relevant time and angular deflection integrals from the singularity all the way to the boundary, rather than from some complex turning point. Evaluated this way, the angular deflection remains real-valued, whilst the time delay acquires an imaginary part equal to $i{\beta\over 2}$, indicative of the presence of another boundary,
\begin{equation}
\label{eq:Deltatbouncing}
\begin{aligned}
&\Delta t=2 b\int_0^\infty {dr\over f\sqrt{b^2-{f(r)\over r^2} } }=2b\left[ \int_0^{r_H}  {dr\over f\sqrt{b^2-{f(r)\over r^2} } }+\int_{r_H}^\infty  {dr\over f\sqrt{b^2-{f(r)\over r^2} } }\right]=\\
&=2 \int_0^{r_H}  {dr\over f}\left[{b\over\sqrt{b^2-{f(r)\over r^2} }}-1\right]+2 \int_{r_H}^\infty  {dr\over f}\left[{b\over\sqrt{b^2-{f(r)\over r^2} }}-1\right] +2\int_0^\infty {dr\over f(r)}=\\
&= (\text{real part})+i {\beta\over 2}\,,
\end{aligned}
\end{equation}
where $\beta$ is defined in (\ref{eq:HawkingT}).
Notice that the first two integrals of the second line are well-defined,  without branch points or other singularities (recall \eqref{eq:positivityofsquareroot}), while the third one -- which corresponds to the time delay of the bouncing null geodesic for vanishing angular momentum -- contains the standard imaginary part ${\beta\over 2}$. Following this line of reasoning, the bulk phase shift would now also acquire the same imaginary part.
\begin{equation}\label{eq:deltanew}
\begin{aligned}
\delta&=2p^t\int_0^\infty {dr\over f}\sqrt{1-{f\over b^2\,r^2}} =\\
&=2 p^t \int_0^\infty {dr\over f} \left[ \sqrt{1-{f\over b^2\,r^2} }-1\right]+2p^t\int_0^\infty {dr\over f}=(\text{real part})+i p^t{\beta\over 2}
\end{aligned}
\end{equation}
What we have effectively advocated with this ad-hoc procedure is that the relevant boundary values of time delay and angular deflection should be evaluated directly from the singularity when $b>b_c$, thus disentangling the turning point from the impact parameter.

To investigate what actually happens we will study the thermal two-point function further. The position space thermal correlator provides an alternative perspective on the time delay and angular deflection when $b<b_c$; they correspond to the values of $(t,\theta)$, where the correlator is singular. Such singularities -- commonly referred to as bulk cone singularities \cite{Hubeny:2006yu, Kulaxizi:2018dxo, Dodelson:2020lal, Dodelson:2023nnr} -- when present in a generic thermal CFT two-point function, are often related to the existence of a dual gravitational description.



\section{Thermal two-point function and singularity}
\label{sec.five}

In this section, we investigate the behavior of the position space Wightman two-point function at finite temperature. Here we could directly use the expressions in \cite{Dodelson:2023nnr} by Wick rotating to Minkowski space. We will, however, simply consider the known expression of the Wightman function in momentum space, and compute the coordinate analog by the Fourier transform of $G_+(\omega, \ell)$ back to position space. This approach is simpler, but will not allow us to describe winding of the bulk null geodesics. These can be easily accounted for by comparing our expressions with \cite{Dodelson:2023nnr} and generalizing them appropriately.

Thus, we write:
\begin{equation}
  \begin{aligned}
G_+(t,\theta)=\sum_{\ell=0}^{\infty}\frac{k\,\Gamma(\frac{d}{2})}{(d-2)\pi^{d/2}}C_\ell^{\frac{d-2}{2}}(\cos{\theta})\int_{-\infty}^{\infty}\frac{d\omega}{2\pi}e^{-i\omega t}G_+(\omega,\ell)\,,
  \end{aligned}
\label{eq:GWFourier}
\end{equation}
where $k=\ell+\frac{d-2}{2}$ as in (\ref{eq:ell_k_shift}), and $C_\ell^{d-2\over 2} (\cos{\theta})$ represent the  Gegenbauer polynomials, related to hyperspherical harmonics in $d$--dimensions.

In the eikonal limit of interest here, the Wightman correlator for $b<b_c$ can be expressed as in (\ref{eq:GTansatz}). As shown in Appendices~\ref{app:wkb} and~\ref{app:GR}, the correlator takes the same form for $b>b_c$, with the eikonal phase evaluated using the appropriate complex turning point, or equivalently, analytically continued from the $b<b_c$ result. For $k\gg 1$ we use the following approximation of the Gegenbauer polynomials:
\begin{equation}
  \begin{aligned}
C_{\ell=k-{d-2\over 2}}^{\frac{d-2}{2}}(\cos \theta)\approx \frac{1}{2}\left(\frac{2}{k}\right)^{(4-d)/2}e^{-i\pi (d-2)/4}\frac{e^{i\pi (d-2)/2\lfloor \frac{\theta}{\pi}\rfloor} (e^{ik\theta}+e^{i\pi (d-2)/2}e^{-ik\theta})}{\Gamma(\frac{d-2}{2})(|\sin \theta|)^{(d-2)/2}}
  \end{aligned}
\label{eq:Ckapprox.}
\end{equation}
For $0<\theta<\pi$ this is (\ref{eq:gp4}) of Appendix~\ref{appendixc}, in any $d$.\footnote{For even $d$, the floor function and the absolute value of $\sin\theta$ extend (\ref{eq:Ckapprox.}) to other values of $\theta$. For odd $d$ the two terms acquire opposite phases: instead of the common factor $e^{i\pi(d-2)/2\lfloor\theta/\pi\rfloor}$, the term $e^{ik\theta}$ is multiplied by $e^{-i\pi(d-2)/2\lfloor\theta/\pi\rfloor}$ and the term $e^{-ik\theta}$ by $e^{+i\pi(d-2)/2\lfloor\theta/\pi\rfloor}$. The two prescriptions differ only for odd $\lfloor\theta/\pi\rfloor$. Only $0<\theta<\pi$ is used below.} Substituting into (\ref{eq:GWFourier}), we replace the summation over $\ell$ by an integration over $k$ to obtain:
\begin{equation}
\label{eq:GWFouriera}
\begin{aligned}
G_+(t,\theta)&\approx \frac{ e^{-i\pi (d-2)/4} e^{i\pi (d-2)/2\lfloor \frac{\theta}{\pi}\rfloor} }{(2\pi)^{d/2+1}(|\sin \theta|)^{d/2-1}} \times\\
&\times \int_0^{\infty} \int^{\infty}_{-\infty} dk\,d\omega \, k^{{d-2\over 2}}\, e^{-i\omega t} (e^{ik\theta}+e^{i \pi (d-2)/2} e^{-i k\theta}) G_+(\omega,k)
\end{aligned}
\end{equation}
Setting
\begin{equation}
\label{eq:gplusdef}
g_+(t,\theta)=\frac{ e^{-i\pi (d-2)/4} e^{i\pi (d-2)/2\lfloor \frac{\theta}{\pi}\rfloor} }{(2\pi)^{d/2+1}(|\sin \theta|)^{d/2-1}}\int_0^{\infty}dk \int^{\infty}_{-\infty} d\omega \, k^{d-2\over 2} e^{-i\omega t+ik\theta} G_+(\omega,k) \,,
\end{equation}
the position space Wightman function can be written as follows
\begin{equation}
\label{eq:GWgplusa}
G_+(t,\theta)=g_+(t,\theta)+ e^{i \pi (d-2)} g_+(t, 2\pi-\theta)\,.
\end{equation}
This is very similar to the $j=0$ term of eq.~(4.9) of \cite{Dodelson:2023nnr}, where the second term carries the factor $(-1)^{2\alpha}=e^{i\pi(d-2)}$, $\alpha=\frac{d-2}{2}$, with $g_+$ the analog of $g_E$ in eq.~(4.15) there, with the appropriate approximations for large momenta. We may equivalently express the correlator as:
\begin{equation}
\label{GWgplusb}
G_+(t,\theta)=g_+(t,\theta)+ e^{i \pi (d-2)/2} g_+(t, -\theta)\,,
\end{equation}
where in $g_+(t,-\theta)$ the floor function and $|\sin\theta|$ are evaluated at $|\theta|$, so that only the exponent $e^{ik\theta}$ in (\ref{eq:gplusdef}) changes sign.\footnote{If instead the floor function is evaluated at $-\theta$, $\lfloor-\theta/\pi\rfloor=-1$ for $0<\theta<\pi$, the coefficient of the second term in (\ref{GWgplusb}) becomes $e^{i\pi(d-2)}$.} The relative phase $e^{i\pi(d-2)/2}$ is the same as that between $X_j^+$ and $X_j^-$ in eq.~(4.27) of \cite{Dodelson:2023nnr}. It appears as if $g_+(t,\theta)$ is roughly equivalent to $g_E(\tau,\theta)$ after Wick rotation. However, note that $g_E(\tau,\theta)$ is manifestly KMS invariant, whereas $g_+(t,\theta)$ is not\footnote{It is the similarity of these expressions with those in \cite{Dodelson:2023nnr} which allows us to easily account for the winding number of the geodesics if desired.}.


To proceed, we change variables from $(\omega, k)$ to $(b={\omega\over k}, k)$ and focus on $g_+$ since the thermal two-point function can be completely determined in terms of $g_+$.
\begin{equation}
\label{eq:gplus0}
g_+(t,\theta)=\frac{ e^{-i\pi (d-2)/4} e^{i\pi (d-2)/2\lfloor \frac{\theta}{\pi}\rfloor} }{(2\pi)^{d/2+1}|\sin \theta|^{d/2-1}}\int_0^{\infty}dk \int^{\infty}_{-\infty} db \, k^{d\over 2} e^{-i k(b\, t-\theta)} G_+(b,k) \,,
\end{equation}

The next step is to substitute $G_{+}(b,k)$ from (\ref{eq:GTansatz}). The momentum space thermal Wightman function is given in terms of various branches, depending on the regime of the eikonal potential explored. Here, we focus on $b>b_c$ with $\omega>0$\footnote{The range $1<b<b_c$ is the region of the bulk-cone singularities \cite{Dodelson:2023nnr}. It is analytic at the singular point $t=T(\infty)$ found below and does not affect the analysis of this section (Appendix~\ref{app:GR}).} where $G_+(b,k)$ can be expressed as follows:
\begin{equation}
\label{eq:GWFourierbggbc}
G_+(b,k)= {e^{-i\pi (\Delta-{d\over 2}+1)} 2^{d+1-2\Delta} \pi^{1+{d\over 2}}\over \Gamma\left[\Delta\right]\Gamma\left[\Delta-{d\over 2}+1\right] } k^{2\Delta-d} \left(b^2-1\right)^{\Delta-{d\over 2}}\, e^{i k (b T(b)-\Theta(b))}
\end{equation}

with
\begin{equation}
\label{eq:TThetaCdef}
\begin{aligned}
T=2\int^\infty_{r_T^c}\frac{b}{f(r)}\frac{dr}{\sqrt{b^2-\frac{f(r)}{r^2}}}\,,\\
\Theta=2\int^\infty_{r_T^c}\frac{1}{r^2}\frac{dr}{\sqrt{b^2-\frac{f(r)}{r^2}}}\,, \\
\end{aligned}
\end{equation}
and $r_T^c$ the complex turning point leading to a positive imaginary part in the exponent of (\ref{eq:GWFourierbggbc}), or equivalently the analytic continuation of (\ref{eq:TThetadefR}) when $b>b_c$. Substituting into (\ref{eq:gplus0}) leads to
\begin{equation}
\label{gpluskbps}
\begin{aligned}
g_+(t,\theta)&=\frac{ e^{-i\pi \left(\Delta-{d\over 4}+{1\over 2}\right)} e^{i\pi (d-2)/2\lfloor \frac{\theta}{\pi}\rfloor} }{2^{2\Delta-d/2}|\sin \theta|^{d/2-1}\Gamma\left[\Delta\right]\Gamma\left[\Delta-{d\over 2}+1\right] } \times\\
\times\int_0^{\infty}dk &\int^{\infty}_{b_c} db \, k^{2\Delta-{d\over 2}} \,(b^2-1)^{\Delta-{d\over 2}}e^{-i k\left(b\,( t-T(b))-(\theta-\Theta(b))\right)} \,.
\end{aligned}
\end{equation}
Integrating over $k$ results in
\begin{equation}
\label{eq:gplusc}
\begin{aligned}
g_+(t,\theta)&=\frac{  e^{-i\pi\left(2\Delta-{d\over 2}+1\right)}\,e^{i\pi (d-2)/2\lfloor \frac{\theta}{\pi}\rfloor} \,\Gamma[2\Delta-{d\over 2}+1] } { 2^{2\Delta-d/2}|\sin \theta|^{d/2-1}\Gamma\left[\Delta\right]\Gamma[\Delta-{d\over 2}+1] } \times\\
 &\qquad\qquad\qquad\times \int_{b_c}^\infty db \, {(b^2-1)^{\Delta -{d\over 2}}\over \left(b\,( t-T(b))-(\theta-\Theta(b))-i\epsilon\right)^{2\Delta-{d\over 2}+1}}\,,
 \end{aligned}
\end{equation}
where we have included the $i\epsilon$ prescription to ensure convergence when/if the denominator in parenthesis is real valued.

Following \cite{Dodelson:2023nnr} we consider the conditions for a pinch singularity:
\begin{equation}
\begin{aligned}
b_\ast: -b_\ast t+\theta+b_\ast T(b_\ast)-\Theta(b_\ast)=0 \\
 \left.\frac{\partial}{\partial b}(-bt+\theta+bT(b)-\Theta(b))\right|_{b_\ast}=0\,.
\end{aligned}
\label{eq:singularityconditions}
\end{equation}
Combining these conditions with the property
\begin{equation}
\label{eq:derrelationsTTheta}
b T'(b)-\Theta'(b)=0
\end{equation}
which follows from the definition of $(T(b),\Theta(b))$, (see also (\ref{eq:propertiesTThetaA})), yields:
\begin{equation}\label{eq:conditionspinch}
\begin{aligned}
b_\ast: t&=T(b_\ast)\\
\theta&=\Theta(b_\ast)
\end{aligned}
\end{equation}
where $b_\ast$ represents the specific value(s) of $b$ within the integration contour for which  (\ref{eq:conditionspinch}) should be satisfied.

The conditions derived are formally the same as in \cite{Dodelson:2023nnr}. However,  $(T(b),\Theta(b))$ are evaluated with a complex turning point and for generic values of $b$ contain imaginary parts. For real $\theta$ and real $b$, the conditions (\ref{eq:conditionspinch}) can be satisfied only in the limit $b\to\infty$, where $\Theta\to0$ and $T\to t_c+i{\beta\over 2}$. We therefore find a candidate singularity at complex time, $\Im t=\beta/2$, outside the strip of analyticity. The singularity corresponds to the bouncing null geodesic for zero angular separation.

We are now ready to investigate the behavior of the correlator near the potential singularity. There are various ways of determining the form of the singularity in this case. We find it convenient to directly expand around $\theta=0$. To do so, we need to abandon
the approximative form of the Gegenbauer polynomials previously used, {\it i.e.}, (\ref{eq:Ckapprox.}), as this is not valid for $\theta=0$.   Notice instead that
\begin{equation}
\label{eq:Gegenbauerthetazero}
C_{\ell=k-{d-2\over 2}}^{d-2\over 2}(1)={\Gamma[ k+{d-2\over 2}]\over\Gamma[d-2]\Gamma[k-{d\over 2}+2]} \,.
\end{equation}
We proceed then by substituting (\ref{eq:Gegenbauerthetazero}) in (\ref{eq:GWFourier}), and making the standard approximations for $\ell \sim k\gg 1$. This yields
\begin{equation}
\label{eq:correlatorthetazero}
\begin{aligned}
G_+(t,0)&=\sum_{\ell=0}^{\infty}\frac{k \Gamma(\frac{d}{2})}{(d-2)\pi^{d/2}} {\Gamma[ k+{d-2\over 2}]\over\Gamma[d-2]\Gamma[k-{d\over 2}+2]}
\int_{-\infty}^{\infty}\frac{d\omega}{2\pi}e^{-i\omega t}G_+(\omega,\ell)=\\
&= {\Gamma[{d\over 2}]\over 2\,\Gamma[d-1]\,\pi^{{d\over 2}+1}}\int_0^\infty \int_{-\infty}^\infty d\omega \,dk \,k^{d-2}\, e^{-i\omega t}\,G_+(\omega,k)=\\
&={\Gamma[{d\over 2}]\over 2\,\Gamma[d-1]\,\pi^{{d\over 2}+1}}
\mathcal{N}_\Delta\, e^{-i\pi(\nu+1)}\int_0^\infty dk \int_{b_c}^\infty db \, k^{2\Delta-1}\, (b^2-1)^{\Delta-{d\over 2}} e^{-i k \left[b (t-T(b))+\Theta(b)\right]}\\
&={e^{-i\pi \left(2\Delta-{d\over 2}+1\right)}\,\mathcal{N}_\Delta\,\Gamma[2\Delta]\Gamma[{d\over 2}]\over 2\,\Gamma[d-1]\,\pi^{{d\over 2}+1}} \int_{b_c}^\infty db \, {(b^2-1)^{\Delta-{d\over 2}}\over \left(b(t-T(b))+\Theta(b)\right)^{2\Delta}}
\end{aligned}
\end{equation}

The integral in the last line of (\ref{eq:correlatorthetazero}) is expected to be singular for large $b$ when $t=\lim_{b\rightarrow\infty} T(b)=t_c+i{\beta\over 2}$. To determine the form of the singularity, we split the integral into two parts, denoting by $\delta I$ the putative singular term, for some finite $b_0\gg b_c$,
\begin{equation}
\label{eq:deltaIdef}
\begin{aligned}
\int_{b_c}^\infty db \, {(b^2-1)^{\Delta-{d\over 2}}\over \left(b(t-T(b))+\Theta(b)\right)^{2\Delta}} &=
\int_{b_c}^{b_0} db \, {(b^2-1)^{\Delta-{d\over 2}}\over \left(b(t-T(b))+\Theta(b)\right)^{2\Delta}} +\delta I\,, \\
\text{where}\qquad\delta I\equiv \int_{b_0}^\infty db \, {(b^2-1)^{\Delta-{d\over 2}}\over \left(b(t-T(b))+\Theta(b)\right)^{2\Delta}}&\simeq \int_{b_0}^\infty {db\over b^d\,(t-T(b) +b^{-1}\Theta(b))^{2\Delta} }\,,
\end{aligned}
\end{equation}
To approximate the integrand in $\delta I$ for $b\gg b_c>1$, we need to first investigate the behavior of the functions $T(b)$ and $\Theta(b)$ in this regime. For very large $b$, the complex turning points which arise from the merger of the two real, distinct roots that exist when $b<b_c$, behave as follows:
\begin{equation}
\label{eq:approximatrTcblarge}
b^2 r^d+\mu\simeq 0,\quad r_T^c\simeq \left({\mu\over b^2}\right)^{1\over d}e^{\pm {i\pi\over d}}
\end{equation}
where $\pm$ refers to the two complex-conjugate turning points that merge at $b=b_c$.\footnote{These are the two roots of $b^2r^d+\mu=0$ closest to the positive real axis; the one in the lower half plane gives $S_+$, with $\Im S_+>0$ (Appendix~\ref{app:wkb:largeb}). As $b\to\infty$ all $d$ roots approach $r=0$; see Fig.~\ref{fig:H:stokes} for $d=4$.}
Substituting into the integral expressions (\ref{eq:TThetaCdef}) yields:
\begin{equation}
\label{eq:seriesTThetaClargeb}
\begin{aligned}
\Theta(b)&\simeq_{b\gg b_c} a_1 b^{2/d-1}\\
T(b)&\simeq_{b\gg b_c} T(\infty)+d_1 b^{{2\over d}-2}
\end{aligned}
\end{equation}
where the numerical coefficients $(a_1,d_1)$ are related through the equation $b T'(b)-\Theta'(b)=0$. For simplicity, we have set $\lim_{b\rightarrow \infty} T(b)=T(\infty)$.

Using these expressions we can approximate $\delta I$ in (\ref{eq:deltaIdef}) for large $b$ by\footnote{To obtain (\ref{eq:deltaIdefa}) we have included the leading order behavior of $\Theta$. This changes the coefficient of the singular term, but not its exponent.}
\begin{equation}
\label{eq:deltaIdefa}
\delta I\simeq \int_{b_0}^\infty {db\over b^d\,(t-T(\infty)+\left({ d\over d-2}\right) d_1 b^{{2\over d}-2})^{2\Delta}} \,.
\end{equation}
Changing variables according to
\begin{equation}
b^{2(1- d)\over d}=u^2(t-T(\infty))
\label{eq:u_var_def}
\end{equation}
where we used $b^{2(1-d)/d}=b^{-q}$, with $q$ as in (\ref{eq:H:approach}), and $\frac{d}{d-2}\,d_1=a_1-d_1=-c$, by (\ref{eq:H:c1a1}), so that the bracket in (\ref{eq:deltaIdefa}) coincides with (\ref{eq:H:phase0}). This change of variables leads to
\begin{equation}
\label{eq:deltaIfinal}
\delta I\simeq {d\over (d-1)} {1\over (t-T(\infty))^{2\Delta-{d\over 2}}}
\int_0^{u_0} du {u^{d-1}\over \left(1+\left({d\over d-2}\right)d_1 u^2\right)^{2\Delta}}\,,\qquad u_0=b_0^{-{d-1\over d}}\,\big(t-T(\infty)\big)^{-{1\over 2}}\,.
\end{equation}
The coefficient of the singular term, the other contributions to the large-$k$ correlator and the absence of cancellations at $t=T(\infty)$ are discussed in Appendix~\ref{app:GR}, and the coefficient is checked numerically in Appendix~\ref{app:num}. As $t\to T(\infty)$, $u_0\to\infty$ and the integral in (\ref{eq:deltaIfinal}), along a ray in the complex $u$-plane that avoids the zero of the denominator (Appendix~\ref{app:GR:coef}), tends to a finite value for $4\Delta>d$\footnote{The special case of $d=2$ in the context of a conical defect geometry was analyzed in \cite{Kulaxizi:2018dxo}.  For recent developments,  see \cite{AliAhmad:2026wem}.}, hence a singularity at $t=T(\infty)$ is present, and its form -- given by (\ref{eq:deltaIfinal}) -- is consistent with \cite{Ceplak:2024bja}. We may therefore view this result as complementary to those of \cite{Afkhami-Jeddi:2025wra, Dodelson:2025jff, Jia:2025jbi}.

\section{Comments on the result}\label{sec.six}

We begin by clarifying in which sense (\ref{eq:GTansatz}) describes the Wightman function. The exact momentum-space Wightman function is real and positive,
\begin{equation}
\label{eq:Gplusrho}
G^T_+(\omega,\ell)=\tilde\rho(\omega,\ell)\,\big[1+n_B(\omega)\big]\geq0\,,
\end{equation}
where $\tilde\rho=-i(G_R-G_A)$ is the spectral function and $n_B(\omega)=(e^{\beta\omega}-1)^{-1}$; it therefore cannot be given by the single phase (\ref{eq:GTansatz}). In the large $\omega,k$ limit, the Wightman function contains, besides the eikonal term (\ref{eq:GTansatz}),
\begin{itemize}
\item its complex-conjugate partner $G^{T=0}_+\,e^{i\pi(\nu+1)}e^{-ik\bar S(b)}$;
\item a term without phase, associated with propagation near the boundary;
\item terms with phases $e^{iNkS}$, $N\geq2$, describing multiple reflections in the bulk;
\item for $\omega<0$, the Boltzmann-suppressed terms dictated by the KMS condition, $G_+(-\omega)=e^{-\beta\omega}G_+(\omega)$ (Appendix~\ref{app:GR}).
\end{itemize}
For $b<b_c$, where $S$ is real, the eikonal term and its partner combine into a standing wave, $\propto\cos\big(kS(b)-\pi(\nu+1)\big)$. For $b>b_c$ each is continued so that it remains bounded: $e^{ikS}$ through $\Im b>0$, which gives $S_+$, and $e^{-ikS}$ through $\Im b<0$, which gives $S_-=\bar S_+$. The two complex turning points discussed in Section~\ref{sec.four} thus both appear, each with its own exponential.

In position space these terms are localized in different regions. The eikonal term is singular near $(t,\theta)=(T(b),\Theta(b))$, i.e.\ close to the Regge point; its partner near $(-T(b),-\Theta(b))$; the term without phase near the light cone of the first operator; and the others elsewhere. The Regge-limit correlator in the CFT is different; it is obtained by retaining the leading terms as $x'^\pm\to0$ on the Regge sheet. This is why the Regge-limit correlator is described by a single phase $e^{i\delta}$ rather than by a standing wave, and why (\ref{eq:deltaCFTdef}) is a ratio of leading Regge-limit forms. Since the phase grows linearly with the energy at every order in $\mu$, the Regge limit at fixed $\mu$ retains the full phase $kS(b)$, to all orders in $\mu$.

In this limit the Wightman and retarded functions are effectively the same. The spectral function contains the eikonal term through $G_R$ and its partner through $G_A$, and for $\omega>0$ the eikonal term of $G_+$ equals $-i$ times the corresponding term of $G_R$ (Appendix~\ref{app:GR:norm}). Equivalently, near the singular point $t=t_c+i\beta/2$ the Wightman function and the commutator, $-iG_R$ for $t>0$, differ only by terms that are regular there. It is the analyticity of $G_R$ in the upper half $\omega$-plane that selects the continuation $S_+$, with $\Im S_+>0$, beyond the critical impact parameter (Appendix~\ref{app:wkb:continuation}). The position-space function obtained by Fourier transforming (\ref{eq:GTansatz}) is therefore the Regge-limit projection of the Wightman function, effectively the positive-frequency part of the eikonal term of the retarded function. Its singularities, and those of the remaining terms of $G_\pm$, $G_R$ and $G_A$, are listed in Table~\ref{tab:I:orderings} and in Appendix~\ref{app:GR:sing}.
\begin{table}[t]\centering
\begin{tabular}{lll}\toprule
Correlator & From $\omega>0$ & From $\omega<0$\\\midrule
$G_+$ & $t_c+i\beta/2$\,$^{(a)}$, \ $-t_c+i\beta/2$\,$^{(b)}$ & $-t_c-3i\beta/2$\,$^{*}$, \ $t_c-3i\beta/2$\,$^{*}$\\
$G_-$ & $t_c+3i\beta/2$\,$^{*}$, \ $-t_c+3i\beta/2$\,$^{*}$ & $-t_c-i\beta/2$, \ $t_c-i\beta/2$\\
$G_R$ & $t_c+i\beta/2$ & $t_c-i\beta/2$\\
$G_A$ & $-t_c+i\beta/2$ & $-t_c-i\beta/2$\\\bottomrule
\end{tabular}
\caption{Singular points of the eikonal term and of its complex-conjugate partner at $\theta=0$, for the two Wightman functions and for the retarded and advanced functions; $T(\infty)=t_c+i\beta/2$. (a): eikonal term (\ref{eq:GTansatz}); (b): its complex-conjugate partner; $^*$: Boltzmann suppressed in momentum space. Further points follow by shifts $\pm in\beta$ (thermal images) and from the terms with $N\geq2$.}
\label{tab:I:orderings}
\end{table}

The singularity found in Section~\ref{sec.five} is located at $t=t_c+i\beta/2$, i.e.\ at $\Im t=\beta/2$. This lies outside the strip $-\beta<\Im t<0$ in which the thermal Wightman function $G_+(t)$ is analytic, as a consequence of the KMS condition and of the positivity of the spectrum. Its presence is therefore not in conflict with the expected analyticity properties of the thermal correlator \cite{Hubeny:2006yu, Festuccia_2006}. In particular, the point $t=t_c-i\beta/2$, which lies inside the strip and corresponds to the real-time two-sided correlator, $G_{12}(t)=G_+(t-i\beta/2)$, is not a singular point (Appendix~\ref{app:GR:sing}), in agreement with \cite{Fidkowski:2003nf}. It is useful to state this also directly for the object we compute. The Fourier transform of (\ref{eq:GTansatz}), i.e.\ the Regge limit of the Wightman function, is supported at positive frequency, and therefore is analytic in the entire lower half of the complex $t$-plane, where $e^{-i\omega t}$ decays for $\omega>0$; in the integral (\ref{eq:correlatorthetazero}) this is the condition of convergence of the $k$-integral. All its singularities, at $t=t_c+i\beta/2$, at the points $N\,T(\infty)$ associated with further reflections, and at their thermal images, lie in the upper half plane, exactly where the positive-frequency structure allows them. In terms of the retarded function, of which this object is effectively the positive-frequency part, the relevant requirement is causality: $G_R(\omega)$ is analytic for $\Im\omega>0$, which is precisely what selects the continuation $S_+$ beyond the critical impact parameter, and the singular point has $\Re t=t_c>0$, consistent with the support of $G_R$ at $t>0$. Unlike the Wightman functions, $G_R=i\theta(t)(G_+-G_-)$ is not analytic in a strip around the real axis, and it is singular also at the conjugate point $t_c-i\beta/2$, through $G_-$ (Table~\ref{tab:I:orderings}). None of these points lies on the real axis, where real-time correlators are expected to be singular only on light cones and bulk cones. The singularity at $t_c+i\beta/2$ is reached from the strip by analytic continuation in complex time, across the real time axis, and therefore lies on a different sheet from the one on which the thermal correlator is analytic in the strip. As we now explain, this sheet is closely related to the second sheet obtained by the analytic continuation $(1-z)\rightarrow (1-z)e^{2\pi i}$, discussed extensively in \cite{Afkhami-Jeddi:2025wra}.

Given the presence of the singularity, the resulting correlator is dominated by the stress-tensor sector \cite{Ceplak:2021wak,Buric:2025fye, Buric:2025anb}. Indeed, the stress-tensor sector alone, which is not KMS invariant, is singular at the bouncing point inside the strip, $t=t_c-i\beta/2$ \cite{Ceplak:2024bja}, and the double-trace operators cancel this singularity there \cite{Buric:2025fye, Buric:2025anb}. To reach $t=t_c+i\beta/2$ from the strip one has to cross the real time axis. At $\theta=0$ all real $t\neq0$ correspond to timelike separations, where $G_+$ and $G_-$ differ by the commutator; the Euclidean correlator, periodic in imaginary time with period $\beta$, therefore has a cut along the timelike real axis. Continuing $G_+$ across it leads to a sheet that differs from the principal one by the analytically continued commutator, $G_+-G_-=-iG_R$ for $t>0$. On the principal sheet the point $t_c+i\beta/2$ is identified, by periodicity, with the two-sided point $t_c-i\beta/2$, where the correlator is regular. The singularity therefore arises from the continued commutator, i.e.\ from the retarded function, in line with the discussion of the Regge limit above. For timelike separations the commutator is the discontinuity around the light cone, $\bar z=1$. At leading order in $1/c$ the double-trace operators enter the light-light OPE with integer powers of $(1-z)(1-\bar z)$ relative to the overall prefactor, and do not contribute to this discontinuity, while the identity and the multi-stress tensors do. The singularity at $t_c+i\beta/2$ is therefore expected to be that of the stress-tensor sector, no longer canceled by the double traces, through the same mechanism that makes the second sheet of \cite{Afkhami-Jeddi:2025wra} singular. The analytic continuation beyond the critical impact parameter in Fourier space thus leads to a position space correlator obtained by two analytic continuations: one around $z=0$ to the Regge sheet and a further one around $\bar z=1$. It will be interesting to make this precise, by matching the monodromies explicitly and by comparing the coefficient of the singularity, (\ref{eq:I:Cinf}), with that of the stress-tensor sector \cite{Ceplak:2024bja}.

In this work, we analytically continued the eikonal phase beyond the critical impact parameter,  here $b_c$.  In the regime $1<b<b_c$ the eikonal phase,  when integrated back to position space,  gives rise to the standard bulk cone singularities \cite{Dodelson:2023nnr}.  Double-trace operators are expected to contribute to the Wightman CFT correlator in this case, contrary to what happens for the singularity at $t=t_c+i\beta/2$, which, as explained above, arises on the sheet where the double traces drop out.  This is verified by integrating the leading order in $\mu$ eikonal to position space,  and observing that it is devoid of singularities in the limit ${\theta'\over t'}\rightarrow 0$, with $(t',\theta')$ the Regge variables (\ref{eq:Reggevariables}), {\it i.e.},  the ``bulk-point'' limit (see Appendix~\ref{appendixLO}).  The Regge conformal blocks are singular there and regularity implies that double trace operators contribute in the OPE together with the stress-tensor to this order (see e.g. the discussion in section 3.2 of \cite{Kulaxizi:2017ixa}. The relevant limit is $z\rightarrow\bar{z}$ or $\rho\rightarrow 0$ in the notations therein and in \cite{Kulaxizi:2018dxo})\footnote{We thank Matthew Dodelson for correspondence on this point.}.

The remaining points among $\pm(t_c\pm i\beta/2)$, where the stress-tensor sector is singular \cite{Ceplak:2024bja}, are distributed among $G_+$, $G_-$, $G_R$ and $G_A$ as in Table~\ref{tab:I:orderings}. In particular, the integral over $b\in(-\infty,-b_c)$, i.e.\ over negative frequencies, gives the singularity at $t=-t_c-i\beta/2$ (Appendix~\ref{app:GR:orderings}); it belongs to the advanced function, equivalently to $G_-$, rather than to $G_+$.

From a technical point of view,  a singular behavior for any other pair of $(T(b),\Theta(b))$,  which would not correspond to an endpoint singularity from the point of view of the integral over $b$,  would be characterized by the bulk cone exponent,  {\it i.e.}, $\left(2\Delta-{d\over 2}+1\right)$ instead of $\left(2\Delta-\frac{d}{2}\right)$.

The singular behavior of the correlator provides a possible answer to the questions about time delay and deflection originally posed. These quantities can be read off from the location of the singularity that the thermal Wightman function in coordinate space exhibits. When $b>b_c$, a singularity exists only for vanishing angular deflection and time-delay associated to a bouncing null geodesic. This picture relies on extending the definition of the Wightman two-point function to include complex time, or equivalently both one-sided and two-sided correlators. In this context, the two-sided correlator provides meaning to the phase shift for impact parameters smaller than the critical one ($b>b_c$).

A careful treatment of the WKB approximation, including the matching at the AdS boundary (Appendix~\ref{app:wkb}), shows that the retarded function at large $k$ is a geometric series in the round-trip amplitude $Y$, (\ref{eq:I:bounce}). Its $N$-th term carries the phase $e^{iNkS(b)}$ with the explicit weight $(1-e^{2\pi i\nu})Y^N$, and describes $N-1$ further reflections of the null geodesic off the AdS boundary. The corresponding singularities of the position-space correlator are located at $t=N\,T(\infty)+in\beta$ beyond the critical impact parameter (Table~\ref{tab:I:sing}),\footnote{These points are consistent with the lattice of bouncing singularities obtained from the asymptotic quasinormal modes in \cite{Dodelson:2025jff}, see eq.~(2.14) there, which refers to the two-sided correlator. Shifting by $i\beta/2$ and identifying the lattice vectors with $T(\infty)$ and $-\overline{T(\infty)}$, the lattice reads $n\,T(\infty)-m\,\overline{T(\infty)}$, $n,m\geq0$. Since $T(\infty)-\overline{T(\infty)}=i\beta$, this is $N\,T(\infty)+i\min(n,m)\beta$ for $n\geq m$ and $-N\,\overline{T(\infty)}+i\min(n,m)\beta$ for $m>n$, with $N=|n-m|$: the number of bounces changes $\Re t$, while the thermal images shift $\Im t$ by multiples of $\beta$.} and within the allowed region at $t=N\,T(b)$, $\theta=N\,\Theta(b)$ up to windings around the sphere. These are the bulk-cone singularities with $(N-1)$ bounces from the boundary of \cite{Dodelson:2023nnr}, see eq.~(4.30) there, where $n$ corresponds to our $N$\footnote{Note that $S(b)$ here is equal to $S_m/2$ in \cite{Dodelson:2023nnr}.}, which are distinct from the windings around the sphere, labeled by $j$ there. In \cite{Dodelson:2023nnr}, where the analysis is performed at imaginary spin, these contributions appear as exponentially small corrections, of order $e^{-2nS(0,z_-)}$, to the locations and residues of the quasi-normal modes; this reproduces the location of the singularities, but their shape and size could not be matched to the numerics. At real frequency and spin the round-trip amplitude $Y$ is instead of order one, and each term of the series contributes at the same order; the full structure of the series, including the weights of all terms, is confirmed numerically against the exact retarded function in Appendix~\ref{app:num:GR}, and, at fixed $\ell$, at $b\to\infty$ through the residues of the quasi-normal modes in Appendix~\ref{app:num:qnm}. The explicit weights given here may therefore be useful for a quantitative comparison with the heights of these singularities.

We did not observe singularities for bouncing null geodesics with non-zero, but real valued, angular deflection. It is possible that certain analytic continuations that we missed could provide such points. It might also be worthwhile to relax the condition $b\in \mathbf{R}$. Focusing on $d=4$, it is easy to see -- considering for instance (\ref{eq:d4timedelayAngdeflection}) and (\ref{eq:d4tpreal}) -- that for $b\in \mathbf{I}$, the angular displacement remains real-valued, while the time displacement becomes generically complex with a fixed - independent from the value of $b$ - real value equal to ${\beta\over 2}$ (see also Fig.~\ref{fig:1}). For $b=i\tilde{b}$ with $\tilde{b}\in\mathbf{R}$ within the integration contour, the Regge-limit correlator exhibits a singularity at $(t,\theta)\sim(T(\tilde{b}),\Theta(\tilde{b}))$. In accordance with what was previously discussed, the power of the singularity in this case is that of the bulk-cone, {\it i.e.}, equal to $(2\Delta-{d\over 2}+1)$. A singularity of this type has not yet been directly observed from the stress-tensor sector OPE analysis.
Notice that, if $T(\Theta)$ denotes the value for which the singularity occurs, the behavior near the point $\Theta= 0$ is (the result below is valid in $d=4$),
\begin{equation}
\label{eq:imbcase}
T(\Theta)\simeq_{\Theta\ll 1} {\beta\over 2}+i \# \Theta^3+ \cdots\,.
\end{equation}
This behavior is different from the one naively coming from the individual conformal blocks and points to a scenario similar to the one discussed in section 3.5 of \cite{Ceplak:2025dds} \footnote{We thank A.~Parnachev for explaining this to us.}. It would be interesting to investigate this further.

Finally, let us remark that the result obtained here remains valid in the black brane case. We need not perform the computations anew, but rather scale the horizon over the AdS length to infinity. Equivalently, one may simply replace $f(r)$ in (\ref{eq:TThetaCdef}) with the black brane metric function, {\it i.e.}, $f(r)\rightarrow r^2-{\mu\over r^{d-2}}$, set $\Theta\sim X$, as well as, $\sin\theta\sim \theta\sim x$ in (\ref{gpluskbps}). The formulae for $\theta\sim x\sim 0$ are straightforwardly adapted to the black brane case, reducing to the same, expected, result\footnote{From a WKB point of view, the discussion here applies to the case $b>1$.}.

The phase-shift representation also provides a unified picture of the singularities of the thermal correlator. In (\ref{eq:gplusc}) the phase is $k\big[b\,(t-T(b))-(\theta-\Theta(b))\big]$, and singularities can only arise where $(t,\theta)=(T(b),\Theta(b))$, i.e.\ at the boundary endpoints of null geodesics with $b=\omega/k$, with $T(b)$ and $\Theta(b)$ continued analytically in $b$ where needed. For real $1<b<b_c$ these are the real null geodesics that return to the boundary, which give the bulk-cone singularities of \cite{Dodelson:2023nnr}, including those with winding and with multiple reflections off the AdS boundary. For $b>b_c$ the turning points are complex, and the singularity found here, at $t=t_c+i\beta/2$, arises in the limit $b\to\infty$ and is associated with the null geodesic bouncing off the black hole singularity. For $0<b<1$, i.e.\ $\omega<k$, the null geodesics are imaginary; in the planar limit they account for the singularities of the stress-tensor sector at spacelike separation, whose location also follows from the phase shift \cite{Araya:2026shz}. Imaginary $b$, finally, gives complex geodesics with real angular displacement and complex time, cf.\ (\ref{eq:imbcase}) and \cite{Araya:2026shz}. In all cases the singularities lie on null geodesics of the black hole geometry, real or complex, and the cases differ only in whether the singular point is a pinch of the $b$-integral at finite $b$ or an endpoint at $b\to\infty$, and in whether it is reached at real or at complex boundary positions. This is consistent with the general result that the retarded propagator is singular whenever the two points can be connected by a null geodesic, which follows from the Hadamard form of the bulk propagator, for any conformal dimension \cite{Grozdanov:2026cut}; see \cite{Grozdanov:2026ktq} for bouncing geodesics in asymptotically flat and de Sitter black holes and their relation to cavity quasinormal modes.

\section{Summary and conclusions}
\label{sec.Conclusions}

In this work, we examined the behavior of the bulk phase shift beyond the critical impact parameter by analytically continuing known expressions for elastic eikonal scattering of a light probe off a black hole (treated as a very heavy particle) in asymptotically anti–de Sitter spacetime. Interpreting the bulk phase shift in the dual CFT in terms of the Regge limit of a finite–temperature momentum–space Wightman two–point function, we then Fourier transformed back to position space. Using standard techniques, we found that the resulting correlator develops singularities at times corresponding to null geodesics that enter the black hole interior, effectively bounce off the singularity, and reappear on the opposite boundary. In this note, the analytic continuation of the correlator to the Regge sheet, dictated by the kinematics of high--energy scattering and standard causality considerations (see for example \cite{Penedones:2007ns,Hartman:2015lfa, Kulaxizi:2018dxo}), served to define the phase shift; the singularity, however, requires a further continuation. The bouncing singularities are characteristic of the stress–tensor sector of the full correlator. The singularity occurs at $t=t_c+i\beta/2$, outside the strip of analyticity of the Wightman function, on a sheet reached through the analytically continued commutator, where at leading order in $1/c$ the double-trace operators do not contribute;\footnote{This is an analytic continuation of the ``bulk-point'' type \cite{Maldacena:2015iua}, around the light cone $\bar z=1$, in addition to the one around $z=0$ which defines the Regge-limit correlator (see Section~\ref{sec.six}). It is analogous to the continuation leading to the second sheet of \cite{Afkhami-Jeddi:2025wra}.} its image on the principal sheet, $t_c-i\beta/2$, is regular, as required. For impact parameters larger than the critical one ($b<b_c$), which are responsible for the standard bulk-cone singularities, double-trace contributions are expected to contribute to the OPE.

The resulting picture appears opposite to ordinary scattering intuition. In conventional scattering,
\begin{equation}
{\rm Im}\,\delta >0
\end{equation}
typically signals access to additional intermediate states and therefore reveals more microscopic information about the target. Here, however, the analytically continued correlator seems to contain \emph{less} OPE data, as double traces disappear, and only the multi-stress-tensor sector survives.
This suggests that the imaginary part of the phase shift is not associated with opening ordinary OPE channels, but with the existence of a different saddle. This is possible because the analytic continuation, the Fourier transform and the $\mu$-expansion do not commute. In position space, the correlator obtained after the two monodromies, which exhibits the singularity, can be expanded in $\mu$, with the stress-tensor block as its leading term, since the double traces drop out. At this order it is singular at the bulk point, where the stress-tensor Regge block diverges (see the discussion in sections 3.2 and 3.3 of \cite{Kulaxizi:2017ixa}). The Fourier transform of the phase shift expanded to leading order in $\mu$ does not reproduce this expansion: it gives the correlator after the Regge monodromy only, which includes the double traces and is regular at the bulk point (Appendix~\ref{appendixLO}). In momentum space, the phase shift beyond $b_c$ is a nonperturbative reorganization of the same gravitational sector\footnote{Note that the analytically continued phase shift is obtained from a fully resummed eikonal answer $e^{i\delta_{\rm cont}(b)}$, and the continuation beyond $b_c$ generally produces a complex function with nontrivial branch structure. It is not obvious then that $\delta_{\rm cont}(b)=\sum_{n\geq0}\mu^n\,\delta^{(n)}(b)$ admits a uniformly valid expansion in the region relevant for the Fourier transform. This can be seen explicitly in $d=4$, where the phase shift is known in closed form, (\ref{eq:d4delta})--(\ref{eq:d4tpreal}). The turning point (\ref{eq:d4tpreal}) depends on $\mu$ through $\sqrt{1-4\mu(b^2-1)}$, whose branch point is at $b=b_c$, $b_c^2=1+\frac{1}{4\mu}$: the expansion in $\mu$ is an expansion in $\mu(b^2-1)$, which converges only for $b<b_c$, while for $b>b_c$ the continuation to $b+i\epsilon$, with $\epsilon>0$ infinitesimal, which gives $S_+$, is complex.}: the singularity, and with it the imaginary part of the phase shift, cannot be seen at any finite order in $\mu$, since its location $t=t_c+i\beta/2$ is not analytic at $\mu=0$, with $\beta/2\propto r_H\sim\mu^{1/(d-2)}$ for small black holes (Appendix~\ref{appendixLO}), and the region $b>b_c$ recedes to infinity as $\mu\to0$. Expanding in $\mu$ before Fourier transforming therefore leads to the correlator on the Regge sheet, whereas the resummed phase shift, continued beyond $b_c$, leads to the correlator after the two monodromies, so that the perturbative expansion in position space and the complex phase shift in momentum space are not in contradiction. It would be interesting to understand how geometric absorption in gravity is encoded in the CFT OPE after the second analytic continuation around $\bar z=1$.

The integration contour in $b$, and the way it is fixed by the retarded function, are discussed in Appendices~\ref{app:wkb:continuation} and~\ref{app:GR}; there it is also shown that, in even $d$, the singularity cannot be obtained by continuing the integral over $1<b<b_c$ alone onto another sheet. Our approach bears similarities in spirit with \cite{Jia:2025jbi}, which considered the spectral functions -- closely related to the Wightman correlator -- in the eikonal regime, and with the analysis of the correlator on the second sheet in \cite{Afkhami-Jeddi:2025wra}. It would be interesting to make the connection with these works more precise. Relatedly, certain discrepancies between the bulk phase shift computation and the position space integration were observed in \cite{Karlsson:2019txu}; it would be worthwhile to revisit them.

Returning to the bulk phase shift, we expect that the dominant contribution in the regime of interest is controlled by near–boundary physics. It would therefore be interesting to establish a derivation of the bulk phase–shift representation of the momentum–space correlator based solely on near–boundary considerations, in close analogy with the way stress–tensor OPE coefficients are obtained. This would complement existing approaches, such as the WKB analysis of~\cite{Dodelson:2023nnr}, which requires matching semiclassical phases from the boundary all the way to the horizon. An argument in this direction is presented in Appendix~\ref{appendixa}: for $b<b_c$ the phase is determined by the region outside the outer turning point, with the boundary condition at the horizon replaced by two conditions at zeroth and first order in $\mu$, while its continuation beyond $b_c$ follows from the analyticity of the retarded function.

Beyond the simplest null trajectories with vanishing angular deflection, there exist additional null bouncing geodesics for which the angular deflection $\Theta$ is nonzero. Correspondingly, the Regge-limit correlator exhibits singularities at these spacetime locations, provided that imaginary values of $b=\omega/k$ lie within the integration contour. A more systematic and rigorous analysis of these contributions would be valuable.

The singularity observed is linked to the bulk cone singularities, known to be present in holographic theories, or more generally, finite temperature QFTs in the large $N$ and large 't~Hooft coupling, $\lambda$, regime. Corrections to these parameters are expected to generically transform the bulk cone singularities to bumps \cite{Maldacena:2015iua, Dodelson:2022eiz}. For the case of stringy corrections (equivalently, 't~Hooft coupling corrections in the CFT language), their effective width was discussed in \cite{Dodelson:2023vrw} in the limits of early and late times.
It would be interesting to see how these formulas, if appropriate, change when the impact parameter is small. It is worthwhile exploring gravitational (large $c$ or $N$) corrections to the bulk phase shift in the same regime. We know for instance that no singularity is expected at zero coupling. Furthermore, it will be very interesting to explore the large $c$, weakly coupled regime explicitly using Regge theory, as was done for example in two-to-two Regge scattering in \cite{Costa:2012cb}.

Corrections to Regge scattering, or more generally to the eikonal kinematics, are computed in a variety of contexts (see for example, \cite{DiVecchia:2023frv} and references therein or \cite{Aoude:2024jxd} for a worldline approach). It is important to examine how our results are changed when they are taken into account.

More generally, while we focused on thermal correlators, the bulk phase shift of a holographic CFT can also be extracted from two–point functions evaluated in a generic heavy state, {\it i.e.}, created by an operator whose conformal dimension scales with the central charge \cite{Kulaxizi:2018dxo}. Such correlators are expected to differ from their thermal counterparts over certain ranges of impact parameter. In the bulk description, this question translates into replacing the black hole background with the geometry of a star or other massive object. Exploring this correspondence further, perhaps along the lines of \cite{Aprile:2025hlt}, would be an interesting direction for future work.

Natural extensions of this work involve the study of AdS-RN, AdS-Kerr and similar geometries. The former has been discussed in relation to null-bouncing geodesics fairly recently in \cite{Ceplak:2025dds}. It is important to check whether the techniques of this paper are applicable there.

Another class of gravitational backgrounds which have gained interest recently in the context of holography are the Robinson-Trautman geometries \cite{BernardideFreitas:2014eoi, Skenderis:2017dnh}. Understanding whether or not bulk-cone singularities, including those studied herein, are present, will provide additional information on the type of theories/states dual to such geometries \cite{Castro:2025itb}.

Asymptotically flat and dS black holes provide another direction of interest. Bulk-cone singularities were investigated some years ago for the asymptotically flat Schwarzschild geometry. See for instance, \cite{Dolan:2011fh, Zenginoglu:2012xe, Casals:2016qyj, Buss:2017vud}, where the relevant retarded two-point function was studied in detail. The main ideas and technical details involved are rather similar, while certain aspects of the flat space phase shift were already examined in \cite{Parnachev:2020zbr},  so it should not be very difficult to investigate these backgrounds.

Specifically in the context of flat space, the two-point function enters certain computations relevant in gravitational wave physics, see e.g.~\cite{Casals:2009zh, Wardell:2014kea, Casals:2013mpa}. Much like the case of bulk-cone singularities, it would be interesting to see if there is an imprint of the singularity on the waveform of the emitted waves.

Moving away from gravity, it is worth exploring field theoretic methods to obtain the bulk phase shift, and as a consequence these results.  The thermal bootstrap \cite{Iliesiu:2018fao}, conformal Regge theory \cite{Costa:2012cb}, or even partonic models \cite{Hatta:2007cs, Hatta:2008tx}, might be useful in this direction. We also expect that techniques involving light-ray operators \cite{Kravchuk:2018htv, Kologlu:2019mfz, Belin:2019mnx} can be particularly helpful in purely field theoretic discussions of the bulk phase shift.

Of special interest is the $d=2$ thermal CFT. As discussed earlier, bulk cone singularities have been observed in the CFT two-point function evaluated on the background of a conical defect geometry. However, the analogous result for the case of a thermal state is missing -- even though the correlator is known explicitly \cite{Cardy:1984epx, Son:2002sd}. From the gravitational point of view, the situation differs in that there exist no null geodesics returning to the boundary. A relevant discussion appeared recently in \cite{AliAhmad:2026wem}; see also \cite{Grozdanov:2026cut} for the singularities of the retarded BTZ correlator. It will be interesting to explore this further.

Finally, causality entered here only through the analyticity of the retarded function, which selects the continuation $S_+$ beyond $b_c$. In the standard impact-parameter regime, causality constraints are well understood and imply, for example, the positivity of the bulk phase shift, see for instance \cite{Kulaxizi:2017ixa}. From the perspective of finite-temperature position-space correlators, a basic requirement is that bulk-cone singularities do not lie inside the boundary light cone: a null geodesic through the bulk cannot arrive earlier than a null curve along the boundary \cite{Gao:2000ga}. In pure AdS all null geodesics from a boundary point reach the antipodal point at $t=\pi$, saturating this bound, and for real geodesics in the black hole the time delay $T(b)-\pi$ is positive, at leading order in $\mu$ the Shapiro time delay. Furthermore, the group velocity of the relevant momentum-space excitations is higher than unity, whilst causality is preserved \cite{Dodelson:2023nnr}. The bouncing singularity does not fit this pattern in a simple way. The theorem of \cite{Gao:2000ga} applies to real causal curves outside the horizon, whereas the bouncing geodesic crosses the horizon and the singularity lies at complex time. Its real part, $t_c$, is positive while the angular deflection vanishes, so that it lies outside the boundary light cone; however, $t_c$ is smaller than the pure AdS value $\pi$ -- in $d=4$, $t_c=\pi\sqrt{1+r_H^2}/(1+2r_H^2)$ -- and therefore corresponds to a time advance relative to pure AdS. It would be important to examine how these and possibly other causality conditions are modified once the impact parameter exceeds the critical value, and to understand the implications such constraints may have for the results presented here.

\section*{Acknowledgments}

We are grateful to Ilia Buri\'{c}, Nejc \v{C}eplak, Chantelle Esper, Robin Karlsson, Andrei Parnachev, Rodolfo Russo and Samuel Valach for discussions. The work of YJ is funded in part by the China Scholarship
Council (project ID 202309120013). MK acknowledges support in part from COST Action CA22113, for a short scientific mission to NKUA, Greece, where part of this work was completed.
MK is thankful to the Physics Department of NKUA, Greece, for hospitality, and to the Simons Center for Geometry and Physics for hospitality and support during the program ``Black Hole Physics from Strongly Coupled Thermal Dynamics,'' and the workshop ``Energy Operators in Particle Physics, theory and Gravity,'' where part of this work was completed. We acknowledge the
use of Anthropic’s Claude Code for editing, figures, and numerics.

\appendix{}

\section{WKB analysis of the radial equation beyond the critical impact parameter}
\label{app:wkb}

In this appendix we discuss the WKB treatment of the radial wave equation, also used in Appendix~\ref{appendixa}, with three aims:
\begin{enumerate}
\item to make the matching to the AdS boundary explicit, so that the full retarded correlator, and not only the phase $e^{ikS(b)}$, is obtained from near-boundary data plus a single ``inner'' reflection amplitude;
\item to describe the inner reflection uniformly in $b$, below, at and above the photon sphere $b=b_c$;
\item to show that this uniform description fixes the analytic continuation of $S(b)$ to $b>b_c$ used in Sections~\ref{sec.four} and~\ref{sec.five}, in particular the sign of $\Im S$.
\end{enumerate}
The material is standard in the semiclassical literature \cite{LandauLifshitz,BerryMountWKB,FromanFroman,Kemble35,MillerGood,FordHillWakanoWheeler}, and was applied to black-hole perturbations in \cite{SchutzWill,IyerWill} and to AdS black holes in \cite{Festuccia:2005pi,Festuccia:2008zx}. We present it in a form adapted to the eikonal limit used in the main text. Unless stated otherwise, formulae hold in general $d\ge3$; explicit expressions are given for $d=4$.

\subsection{Radial equation and WKB waves}
\label{app:wkb:waves}

We consider a scalar field of mass $m$, dual to a CFT operator of dimension $\Delta$ with $m^2=\Delta(\Delta-d)$, in the AdS-Schwarzschild geometry (\ref{eq:adssch}), decomposed as in (\ref{eq:phidecomp}). In terms of the tortoise coordinate $z$, defined via
\begin{equation}
\label{eq:tortoisedef}
dz=-{dr\over f(r)}\,,
\end{equation}
with $z=0$ at the AdS boundary and $z\to\infty$ at the horizon, the radial equation is a one-dimensional Schr\"odinger problem,
\begin{equation}\label{eq:eqSch}
\left[-\partial^2_z +V_\ell(z)\right]\psi(z)=\omega^2\psi(z)\,,
\end{equation}
with the potential
\begin{equation}
\label{Vldef}
V_\ell(z)=f(r)\left[ \frac{1}{r^2}\left( \Big(\ell+\frac{d}{2}-1\Big)^2-\frac{1}{4} \right)+\nu^2-\frac{1}{4}+\frac{\mu}{r^d}\frac{(d-1)^2}{4}  \right],
\end{equation}
and $\nu^2=m^2+{d^2\over 4}$. Writing $\omega=bk$, $k=\ell+\tfrac{d-2}{2}$, one has
\begin{equation}
V_\ell=k^2\,V_{\rm eik}(r)+O(k^0),\qquad V_{\rm eik}=\frac{f(r)}{r^2},
\end{equation}
so at large $k$ and fixed $b$ the equation is governed by $k^2\big(b^2-V_{\rm eik}\big)$, see \eqref{eq:eqEikonal}. The $O(k^0)$ terms in $V_\ell$ only matter where $b^2-V_{\rm eik}$ is small or near the boundary, and are treated separately below.

\begin{figure}[t]\centering
\includegraphics[width=\textwidth]{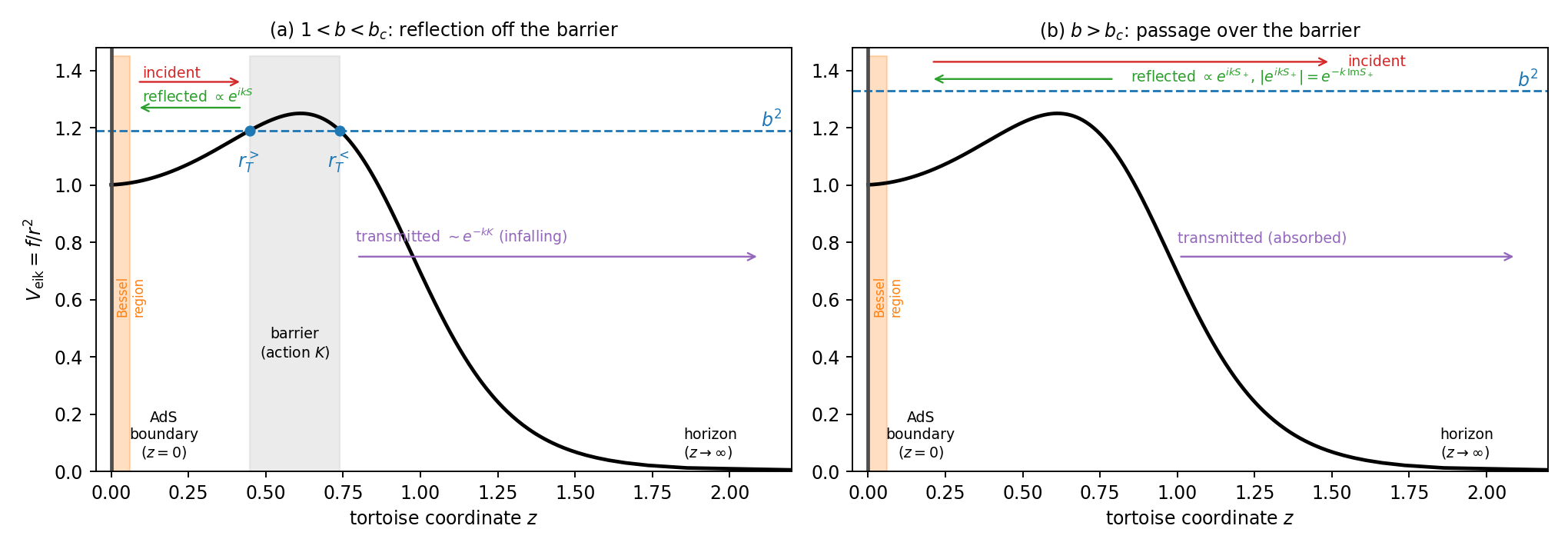}
\caption{The eikonal potential $V_{\rm eik}=f/r^2$ as a function of the tortoise coordinate $z$ for $d=4$, $\mu=1$. (a) For $1<b<b_c$ there are two turning points $r_T^>>r_T^<$. A wave sent in from the boundary is reflected at $r_T^>$, acquiring the phase $e^{ikS(b)}$, and tunnels through the barrier with amplitude $\sim e^{-kK}$. (b) For $b>b_c$ the wave passes over the barrier and is mostly absorbed by the horizon; the reflected wave is exponentially small and is described by the continued phase $S_+(b)$. The shaded strip at the boundary is the region where the Bessel description \eqref{eq:H:bessel} is needed.}
\label{fig:H:schematic}
\end{figure}

\paragraph{WKB waves.} Let $p(z)=\sqrt{\omega^2-V_\ell}\simeq k\sqrt{b^2-V_{\rm eik}}$ denote the local radial momentum. Substituting $\psi=e^{i\Phi}$ into \eqref{eq:eqSch} gives $\Phi'^2-i\Phi''=p^2$, which is solved order by order in $1/k$: $\Phi'=\pm p+\tfrac i2\,p'/p+O(1/k)$. Hence, in any region where $b^2>V_{\rm eik}$ and $|p'|\ll p^2$,
\begin{equation}\label{eq:H:WKBwaves}
\psi\simeq\frac{1}{\sqrt{p}}\Big(c_1\,e^{iW(z)}+c_2\,e^{-iW(z)}\Big),\qquad
W(z)=\int_z^{z_T}p\,dz'=k\int_{r_T}^{r}\frac{dr'}{f}\sqrt{b^2-V_{\rm eik}(r')} .
\end{equation}
The phase is referred to the turning point $r_T$. Since $W$ decreases as $z$ increases, with the time dependence $e^{-i\omega t}$ the term $c_1e^{iW}$ moves towards the boundary (outgoing) and $c_2e^{-iW}$ towards the horizon (ingoing). The approximation fails in three places:
\begin{itemize}
\item near turning points, $V_{\rm eik}(r_T)=b^2$, where $p\to0$;
\item near the AdS boundary, $z\lesssim1/k$, where $V_\ell\simeq(\nu^2-\tfrac14)/z^2$ is not small compared to $k^2$;
\item near the top of the potential, $b\simeq b_c$, where two turning points coalesce.
\end{itemize}
The three regions are shown in Fig.~\ref{fig:H:schematic}. Each is handled by a local ``comparison equation'' that can be solved exactly: the Airy, Bessel and Weber equations respectively.

\paragraph{An isolated turning point.} Near a simple turning point $b^2-V_{\rm eik}$ is linear in $z-z_T$, and \eqref{eq:eqSch} reduces to the Airy equation. The solution that decays into the classically forbidden region connects to
\begin{equation}\label{eq:H:airy}
\psi\propto\frac{2}{\sqrt p}\cos\Big(W-\frac\pi4\Big)\quad\Longrightarrow\quad \frac{c_1}{c_2}=e^{-i\pi/2}
\end{equation}
in the allowed region \cite{LandauLifshitz,BerryMountWKB}. Total reflection at a simple turning point thus comes with the phase $-\pi/2$ (the Maslov phase). In pure AdS this is the only inner reflection; in the black hole it is replaced by the barrier amplitude of Section~\ref{app:wkb:barrier}.

\subsection{The eikonal phase}
\label{app:wkb:phase}

Between the outer turning point $r_T\equiv r_T^>$ and the boundary, a wave picks up the phase $W(0)$ on the way out and the same on the way in. The round-trip phase is the bulk phase shift \eqref{eq:deltadefint},
\begin{equation}\label{eq:H:S}
\delta=kS(b)=2W(0)=2k\int_{r_T}^{\infty}\frac{dr}{f(r)}\sqrt{b^2-V_{\rm eik}(r)} ,
\end{equation}
where we have used $p\simeq k\sqrt{b^2-V_{\rm eik}}$. The integral converges at large $r$, where the integrand behaves as $\sqrt{b^2-1}/r^2$. Differentiating with respect to $b$ (the boundary term vanishes since the integrand vanishes at $r_T$) reproduces \eqref{eq:TThetadefR}--\eqref{eq:propertiesTThetaA}:
\begin{equation}\label{eq:H:legendre}
T(b)=S'(b),\qquad \Theta(b)=b\,S'(b)-S(b),\qquad S=b\,T-\Theta .
\end{equation}
So $S$ is the Legendre transform of the boundary-to-boundary time delay and angular deflection of the null geodesic with $p^t/p^\theta=b$. In position space, the stationary points in $b$ of $bt-\theta-S(b)$ are exactly the null geodesics, $t=T(b)$, $\theta=\Theta(b)$, as in \eqref{eq:singularityconditions}--\eqref{eq:conditionspinch}.

For $\mu=0$ the integral is elementary and gives $S_{\rm AdS}=\pi(b-1)$, i.e.\ $T=\Theta=\pi$, in agreement with \eqref{eq:deltaads}.

\subsection{Matching at the AdS boundary}
\label{app:wkb:boundary}

The WKB waves \eqref{eq:H:WKBwaves} are not valid for $z\lesssim1/k$. There $V_{\rm eik}=1+O(z^2)$, while the mass and curvature terms of \eqref{Vldef} combine into an inverse-square potential, giving \eqref{eq:eomnearboundary}:
\begin{equation}\label{eq:H:bessel}
\partial_z^2\psi+\Big(k^2(b^2-1)-\frac{\nu^2-\frac14}{z^2}\Big)\psi\simeq0.
\end{equation}
For the retarded correlator it is more convenient to use the basis
\begin{equation}
\psi=\alpha\,\sqrt z\,J_\nu\big(kz\sqrt{b^2-1}\big)+\gamma\,\sqrt z\,J_{-\nu}\big(kz\sqrt{b^2-1}\big),
\end{equation}
because each term has a definite falloff: $J_\nu$ corresponds to the normalizable and $J_{-\nu}$ to the non-normalizable mode. Using $J_{\pm\nu}(x)\simeq(x/2)^{\pm\nu}/\Gamma(1\pm\nu)$ at small $x$, the retarded function, defined as the ratio of the normalisable to the non-normalisable coefficient for the solution that is ingoing at the horizon \cite{Son:2002sd}, is
\begin{equation}\label{eq:H:GRalpha}
G_R(\omega,k)=\Big(\frac{k\sqrt{b^2-1}}{2}\Big)^{2\nu}\frac{\Gamma(1-\nu)}{\Gamma(1+\nu)}\,\frac{\alpha}{\gamma}
=-\frac{\Gamma(-\nu)}{\Gamma(\nu)}\Big(\frac{k\sqrt{b^2-1}}{2}\Big)^{2\nu}\frac{\alpha}{\gamma}.
\end{equation}
(Overall normalizations of $G_R$ drop out of everything below, since we normalize the position-space correlator by its short-distance limit.)

Because $k\sqrt{b^2-1}=O(k)$, there is an intermediate region $1/k\ll z\ll1$ in which both descriptions are valid:
\begin{itemize}
\item the Bessel functions take their asymptotic form $J_{\pm\nu}(x)\simeq\sqrt{2/(\pi x)}\cos(x-\varphi_\pm)$, with $\varphi_\pm=\pm\frac{\nu\pi}{2}+\frac\pi4$;
\item the WKB waves \eqref{eq:H:WKBwaves} apply, with $p\simeq k\sqrt{b^2-1}$, so that $W(z)\simeq\frac{\delta}{2}-kz\sqrt{b^2-1}$.
\end{itemize}
In this region the WKB solution reads
\begin{equation}
\psi\propto c_1\,e^{i\delta/2}\,e^{-ikz\sqrt{b^2-1}}+c_2\,e^{-i\delta/2}\,e^{ikz\sqrt{b^2-1}}\,.
\end{equation}
We define the inner reflection amplitude, referred to the turning point,
\begin{equation}\label{eq:H:Rdef}
\mathcal R\equiv\frac{c_1}{c_2}.
\end{equation}
It is fixed entirely by the physics at $r\lesssim r_T$, \emph{including the boundary condition in the interior}: regularity at $r=0$ in pure AdS, which gives $\mathcal R=e^{-i\pi/2}$ (an isolated turning point), and the ingoing condition at the horizon in the black hole, which gives $\mathcal R=e^{-i\pi/2}\times$(barrier factor) (Section~\ref{app:wkb:barrier}). Note that the horizon condition does not set $c_1$ or $c_2$ to zero; it fixes their ratio. Both waves are present in $r>r_T$: the wave $c_2$ sent in from the boundary, and the wave $c_1$ reflected back towards it.

Equating the coefficients of $e^{\pm ikz\sqrt{b^2-1}}$ gives
\begin{equation}
\alpha\,e^{-i\varphi_+}+\gamma\,e^{-i\varphi_-}\ \propto\ c_2\,e^{-i\delta/2},\qquad
\alpha\,e^{i\varphi_+}+\gamma\,e^{i\varphi_-}\ \propto\ c_1\,e^{i\delta/2},
\end{equation}
with the same constant of proportionality. Taking the ratio and using $e^{2i\varphi_\pm}=ie^{\pm i\pi\nu}$, one finds
\begin{equation}\label{eq:H:alphagamma}
\frac{\alpha}{\gamma}=-e^{-i\pi\nu}\,\frac{1-e^{2\pi i\nu}\,Y}{1-Y},\qquad
\boxed{Y(\omega,k)\equiv e^{-i\pi(\nu+\frac12)}\,\mathcal R\;e^{ikS(b)}} .
\end{equation}
$Y$ is the amplitude for one complete round trip, boundary $\to$ inner reflection $\to$ boundary. It is the product of three factors:
\begin{itemize}
\item the eikonal phase $e^{ikS}$ accumulated in the bulk;
\item the inner reflection amplitude $\mathcal R$;
\item the phase $e^{-i\pi(\nu+1/2)}$ of reflection off the AdS boundary, which is twice the phase in the asymptotic form of $J_\nu$.
\end{itemize}
Inserting \eqref{eq:H:alphagamma} into \eqref{eq:H:GRalpha} gives
\begin{equation}\label{eq:H:GRY}
G_R(\omega,k)\simeq\mathcal G_0\;e^{-i\pi\nu}\,\frac{1-e^{2\pi i\nu}Y}{1-Y},\qquad
\mathcal G_0=\frac{\Gamma(-\nu)}{\Gamma(\nu)}\Big(\frac k2\Big)^{2\nu}\big(b^2-1\big)^{\nu}.
\end{equation}
This is the main result of the boundary analysis. It expresses the retarded function at large $k$ in terms of the eikonal phase and a single inner reflection amplitude. Its expansion in powers of $Y$ is the expansion in the number of bounces, discussed in Appendix~\ref{app:GR}. The next step is to determine the inner reflection amplitude $\mathcal R$, or equivalently $Y$. Before doing so, we check its value for pure AdS.

In pure AdS, with $\mathcal R=e^{-i\pi/2}$ and $kS_{\rm AdS}=k\pi(b-1)=\pi(\omega-k)$, \eqref{eq:H:alphagamma} gives
\begin{equation}\label{eq:H:YAdS}
Y_{\rm AdS}=e^{-i\pi(\nu+1)}e^{i\pi(\omega-k)}=e^{i\pi(\omega-\ell-\Delta)} .
\end{equation}
The poles of \eqref{eq:H:GRY}, at $Y=1$, are then at the normal modes $\omega=\Delta+\ell+2m$. In Appendix~\ref{app:GR:AdS} we check \eqref{eq:H:GRY} against the exact AdS correlator.

\subsection{The photon-sphere barrier}
\label{app:wkb:barrier}

\begin{figure}[t]\centering
\includegraphics[width=.62\textwidth]{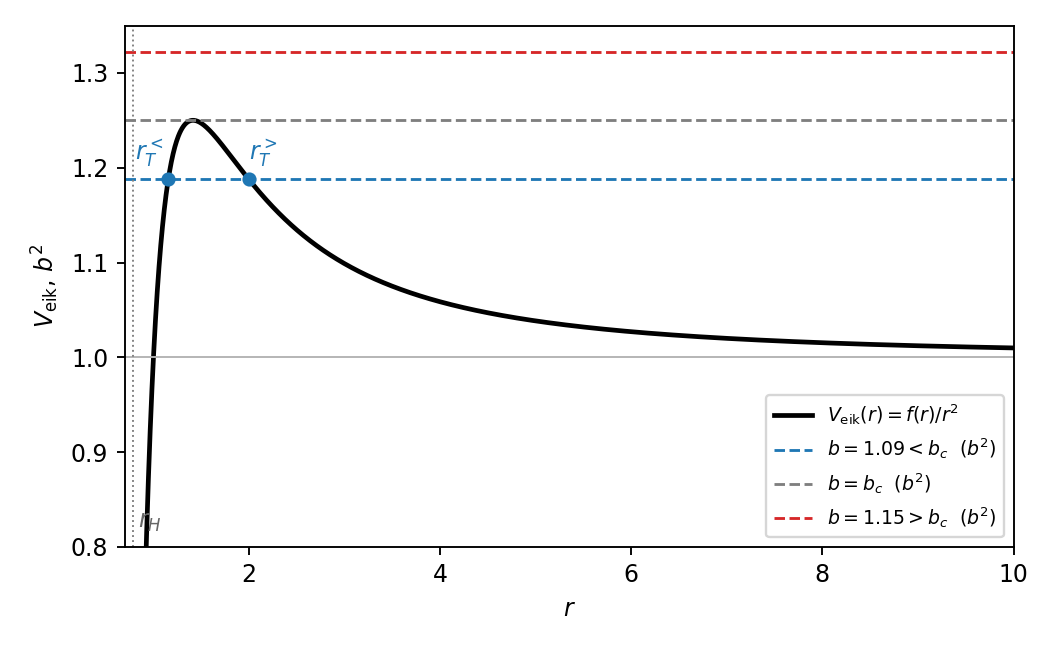}
\caption{Eikonal potential $V_{\rm eik}=f/r^2$ for $d=4$, $\mu=1$, with three values of $b^2$ and the two turning points for $b=1.09$.}
\label{fig:J:potential}
\end{figure}

To compute $Y$ in the black hole we need to consider carefully the photon-sphere barrier. $V_{\rm eik}$ rises from $0$ at the horizon to a maximum $b_c^2$ at the photon sphere, \eqref{eq:photonsphere}, and decreases to $1$ at the boundary (Fig.~\ref{fig:J:potential}). In $d=4$ the maximum is at $r=\sqrt{2\mu}$ with $b_c^2=1+1/(4\mu)$.
\begin{itemize}
\item For $1<b<b_c$ there are two real turning points, $r_T^<<r_T^>$. The barrier between them has action
\begin{equation}\label{eq:H:K}
K(b)=\int_{r_T^<}^{r_T^>}\frac{dr}{f}\sqrt{V_{\rm eik}-b^2}>0,
\end{equation}
and the tunneling probability is $\simeq e^{-2kK}$.
\item For $b=b_c$ the two turning points coalesce. Continuing $b$ once around $b_c$ in the complex plane exchanges $r_T^<$ and $r_T^>$, so the lower limit of \eqref{eq:H:S} moves from $r_T^>$ to $r_T^<$ and
\begin{equation}\label{eq:H:monodromy}
S\ \longrightarrow\ S+2\int_{r_T^<}^{r_T^>}\frac{dr}{f}\sqrt{b^2-V_{\rm eik}}=S\pm2iK .
\end{equation}
A function with this monodromy and no other singularity near $b_c$ has the local form
\begin{equation}\label{eq:H:Slog}
S(b)=P(b)+\frac{K(b)}{\pi}\,\ln(b_c-b),
\end{equation}
with $P$ analytic at $b=b_c$. Since $K\propto b_c-b$ near the top, $S$ is finite at $b_c$, while $T=S'$ diverges logarithmically: this is the familiar logarithmic time delay of geodesics that orbit close to the photon sphere. The branch point of $S$ at $b_c$ is the origin of the two possible continuations $S_\pm$ discussed in Section~\ref{sec.four}. Which one is realized cannot be decided from \eqref{eq:H:S} alone; it requires the wave equation near the top of the barrier.
\item For $b>b_c$ there is no real turning point; the corresponding geodesics fall into the horizon.
\end{itemize}
To determine the reflection amplitude near and above the top of the barrier, we first approximate the barrier by a parabola, and then use the comparison-equation method, which maps the actual barrier onto a parabolic one \cite{Kemble35,MillerGood,FordHillWakanoWheeler,BerryMountWKB,FromanFroman}.

\paragraph{Parabolic barrier.} Near the barrier top the eikonal potential takes the form
\begin{equation}
V_{\rm eik}\simeq b_c^2-\tfrac12V_2\,(z-z_0)^2,\qquad V_2\equiv\big|\partial_z^2V_{\rm eik}\big|_{z_0}=2(d-2)\,b_c^4\,,
\end{equation}
with $z_0$ the position of the maximum. Expressed in terms of a new variable
\begin{equation}
x=(2k^2V_2)^{1/4}\,(z-z_0)\,,
\end{equation}
\eqref{eq:eqSch} reduces to Weber's equation,
\begin{equation}
\partial_x^2\psi+\Big(\frac{x^2}{4}-\frac{\eta}{\pi}\Big)\psi=0,\qquad
\eta=\frac{\pi k\,(b_c^2-b^2)}{\sqrt{2V_2}}=\frac{\pi k\,(b_c^2-b^2)}{2\sqrt{d-2}\,b_c^2}=k\,K(b) ,
\end{equation}
so $\eta>0$ below the top and $\eta<0$ above. Its solutions are parabolic-cylinder functions. The reflection amplitude is known exactly \cite{LandauLifshitz,FordHillWakanoWheeler} up to constant phases fixed by the normalization of the parabolic-cylinder functions,
\begin{equation}\label{eq:H:parabolicR}
\mathcal R\propto\frac{\Gamma\big(\frac12+\frac{i\eta}{\pi}\big)}{\sqrt{2\pi}}\,e^{\eta/2}\,,\qquad |\mathcal R|^2=\frac{1}{1+e^{-2\eta}}\,.\footnotemark
\end{equation}
\footnotetext{For the modulus we used $\big|\Gamma\big(\frac12+\frac{i\eta}{\pi}\big)\big|^2=\pi/\cosh\eta$.}%
This is the origin of the $\Gamma$-function in \eqref{eq:H:kemble} below.

\paragraph{Uniform (Kemble--Miller--Good) amplitude.} The comparison-equation method gives a reflection amplitude valid uniformly in $b$: below, near and above the top. Define the \emph{barrier parameter}
\begin{equation}\label{eq:H:eta}
\eta(b)=k\,K(b)\quad(b<b_c),\qquad \eta(b)=-k\,\Im S_+(b)\quad(b>b_c).
\end{equation}
It is analytic across $b_c$ and reduces to the parameter $\eta$ of Weber's equation for a parabolic barrier. Write
\begin{equation}
\mathcal R=e^{-i\pi/2}\,e^{-ikS}\,X ,
\end{equation}
i.e.\ $X$ is the reflection amplitude with the Maslov phase removed and the eikonal phase to the boundary included. Then $Y=e^{-i\pi(\nu+1)}X$, and
\begin{equation}\label{eq:H:kemble}
\boxed{X=e^{ik\hat S(b)}\,\frac{\Gamma\big(\frac12+\frac{i\eta}{\pi}\big)}{\sqrt{2\pi}}\,e^{\eta/2}\,e^{-i\frac{\eta}{\pi}\left(\ln|\eta/\pi|-1\right)}
=e^{ik\hat S(b)}\,\big(1+e^{-2\eta}\big)^{-1/2}\,e^{i\varphi(\eta)}}
\end{equation}
with
\begin{equation}
\hat S=\begin{cases}S,&b<b_c,\\ \Re S_+,&b>b_c,\end{cases}\qquad
\varphi(\eta)=\arg\Gamma\Big(\frac12+\frac{i\eta}{\pi}\Big)-\frac{\eta}{\pi}\ln\Big|\frac\eta\pi\Big|+\frac\eta\pi .
\end{equation}
where in deriving (\ref{eq:H:kemble}) we used $|\Gamma(\tfrac12+\tfrac{i\eta}{\pi})|^2=\pi/\cosh\eta$. Unlike the parabolic amplitude (\ref{eq:H:parabolicR}), (\ref{eq:H:kemble}) contains no undetermined constant: its phases are fixed by matching to the WKB waves (\ref{eq:H:WKBwaves}) on both sides of the barrier, so that deep below the top $X\to e^{ikS}$, i.e.\ $\mathcal R\to e^{-i\pi/2}$, the isolated-turning-point result (\ref{eq:H:airy}). In the Regge limit this is the complete answer; beyond it, a further phase $\delta_\infty$ arises as $b\to\infty$, see Appendix~\ref{app:wkb:delta}. For a parabolic barrier \eqref{eq:H:kemble} reduces to the exact Weber result. The same construction underlies the WKB computation of black-hole quasinormal modes in \cite{SchutzWill,IyerWill}. Its accuracy is $O(1/k)$ uniformly in $b$;\footnote{More precisely, the error is $O(1/k)$ uniformly for $b$ in any bounded range around $b_c$. The uniformity breaks down as $b\to\infty$, where the turning points approach the singularity $r=0$ (Fig.~\ref{fig:H:stokes}(c)) and the connection there is no longer that of an isolated barrier, for $b\sim k^{d/(d-2)}$; see Appendix~\ref{app:wkb:delta}.} this is tested against the exact radial equation in Appendix~\ref{app:num:GR}.

The amplitude \eqref{eq:H:kemble} is the reflection coefficient of a scattering problem with \emph{no wave incident from the horizon side}. Between the barrier and the horizon, the exact potential vanishes exponentially fast in $z$ and there is no further turning point for real $b>0$. The WKB waves therefore remain valid all the way to the horizon, where $p(z)\to\omega$. There the transmitted wave is $e^{+i\omega z}\propto(r-r_H)^{-i\omega/f'(r_H)}$, which with $e^{-i\omega t}$ is the ingoing solution; the solution $e^{-i\omega z}$ (outgoing from the horizon) is absent. This is precisely the condition that defines the retarded function \cite{Son:2002sd}, so \eqref{eq:H:GRY} with \eqref{eq:H:kemble} is the large-$k$ form of $G_R$ itself,
\begin{equation}\label{eq:H:GRfinal}
\boxed{\begin{aligned}
&G_R(\omega,k)\simeq\mathcal G_0\;e^{-i\pi\nu}\,\frac{1-e^{2\pi i\nu}Y}{1-Y},\qquad
\mathcal G_0=\frac{\Gamma(-\nu)}{\Gamma(\nu)}\Big(\frac k2\Big)^{2\nu}\big(b^2-1\big)^{\nu},\\
&Y=e^{-i\pi(\nu+1)}X,\qquad X=e^{ik\hat S(b)}\,\big(1+e^{-2\eta}\big)^{-1/2}\,e^{i\varphi(\eta)},
\end{aligned}}
\end{equation}
and its poles at $Y=1$ are quasinormal modes. Flux conservation for real $\omega$ is the statement $|X|^2=(1+e^{-2\eta})^{-1}\le1$, the transmission probability being $1-|X|^2=(1+e^{2\eta})^{-1}$. Imposing instead the outgoing condition gives the advanced function. For real $\omega$ the radial equation is real, so that the outgoing solution is the complex conjugate of the ingoing one. This exchanges $c_1\leftrightarrow\bar c_2$, i.e.\ $\mathcal R\to1/\bar{\mathcal R}$ and $X\to1/\bar X$, and \eqref{eq:H:GRfinal} then gives $G_A=\overline{G_R}$; expanded in $\bar Y$, it yields the advanced terms $\bar Y^N$ of \eqref{eq:I:rho}.

The same condition determines the continuation of Section~\ref{app:wkb:continuation}. The retarded function is analytic for $\Im\omega>0$, i.e.\ $\Im b>0$ at fixed real $k$, and the continuation of $S$ around $b_c$ through $\Im b>0$ is exactly the one selected there by the poles of the $\Gamma$-function; it is also what fixes, as $b\to\infty$, the side on which the contour in \eqref{eq:H:Tinf} passes the horizon pole of $1/f$, and hence the sign of $\Im T(\infty)=+\beta/2$. The advanced function would give the conjugate continuation $S_-$ and $\overline{T(\infty)}$. In the numerical tests of Appendix~\ref{app:num}, the ingoing condition is imposed explicitly at $r=r_H$.

\subsection{Continuation beyond the critical impact parameter}
\label{app:wkb:continuation}

Here we consider the continuation beyond the critical impact parameter in more detail. Let us first look at two special limits of the uniform amplitude. With the help of Stirling's formula, i.e.
\begin{equation}\label{eq:H:stirling}
\ln\Gamma\Big(\frac12+\frac{i\eta}{\pi}\Big)\simeq\frac{i\eta}{\pi}\ln\Big(\frac{i\eta}{\pi}\Big)-\frac{i\eta}{\pi}+\frac12\ln2\pi\,,\qquad \Big|\arg\Big(\frac{i\eta}{\pi}\Big)\Big|<\pi\,,
\end{equation}
we find:
\begin{itemize}
\item \emph{deep below the top}, $\eta\to+\infty$: $\Gamma\big(\frac12+\frac{i\eta}{\pi}\big)/\sqrt{2\pi}$ has modulus $e^{-\eta/2}$, which cancels the explicit $e^{\eta/2}$, and its phase cancels the last factor in \eqref{eq:H:kemble}, so $X\to e^{ikS}$. This is total reflection, i.e.\ ordinary WKB;
\item \emph{far above the top}, $\eta\to-\infty$: $\Gamma\big(\frac12+\frac{i\eta}{\pi}\big)/\sqrt{2\pi}$ has modulus $e^{\eta/2}$, which combines with the explicit $e^{\eta/2}$, and the phases again cancel. Hence
\begin{equation}\label{eq:H:Splus}
X\ \to\ e^{ik\Re S_+}\,e^{\eta}=e^{ik\Re S_+}\,e^{-k\Im S_+}=e^{ikS_+(b)} .
\end{equation}
\end{itemize}
The leading term above $b_c$ is therefore precisely $e^{ikS_+}$, the analytic continuation of the phase below the barrier. This justifies the prescription of Sections~\ref{sec.four}--\ref{sec.five}, \eqref{bulkps} and \eqref{eq:GWFourierbggbc}.

Let us see this in detail. Below $b_c$, $S(b)$ is real; above $b_c$ the two continuations $S_\pm(b)$, obtained by passing above or below $b_c$, are complex conjugates of each other. The exact amplitude selects one of them:
\begin{itemize}
\item the Stirling form (\ref{eq:H:stirling}) is analytic except on the half-line $\arg(i\eta/\pi)=\pi$, i.e.\ the positive imaginary $\eta$-axis. This is where the poles of $\Gamma(\tfrac12+\tfrac{i\eta}{\pi})$ lie, at $\eta=i\pi(n+\tfrac12)$, $n=0,1,\dots$;
\item to go from $\eta>0$ (below the top) to $\eta<0$ (above) without crossing these poles one must pass through $\Im\eta<0$;
\item $\eta$ decreases with $b$, since $K'(b)<0$; hence $\Im\eta<0$ corresponds to $\Im b>0$.
\end{itemize}
The continuation must therefore pass \emph{above} $b_c$. This gives $S_+(b)$ with
\begin{equation}
\Im S_+(b)>0\qquad(b>b_c),
\end{equation}
i.e.\ an exponentially small reflected wave, $|e^{ikS_+}|=e^{-k\Im S_+}$, as expected physically for a wave that is mostly absorbed by the black hole. Passing below $b_c$ would cross the poles and give $S_-(b)=\bar S_+(b)$, an exponentially growing reflected wave. This is the ``physically consistent choice'' of Section~\ref{sec.four}, now derived from the wave equation. The poles $\eta=i\pi(n+\tfrac12)$ are the photon-sphere (barrier-top) resonances \cite{SchutzWill}. The logarithmic branch point of $S(b)$ at $b=b_c$ is the large-$k$ remnant of the term $\frac{i\eta}{\pi}\ln\big(\frac{i\eta}{\pi}\big)$ in Stirling's formula: the exact amplitude has no branch point at $b_c$.

\paragraph{Subleading terms (logarithmic sheets).} Going $s$ times around $b_c$ gives, by \eqref{eq:H:monodromy}--\eqref{eq:H:Slog}, the logarithmic sheets
\begin{equation}
S_{(s)}(b)=S(b)+2is\,K(b),\qquad s\in\mathbb Z .
\end{equation}
They appear in the uniform amplitude through the binomial expansion of the factor $(1+e^{-2\eta})^{-1/2}$ in \eqref{eq:H:kemble}:
\begin{itemize}
\item below $b_c$ ($\eta>0$):
\begin{equation}
\big(1+e^{-2\eta}\big)^{-1/2}=\sum_{s\ge0}\binom{-1/2}{s}e^{-2skK},\qquad e^{ikS}e^{-2skK}=e^{ikS_{(s)}};
\end{equation}
\item above $b_c$ ($\eta<0$, with $K\to-\Im S_+$):
\begin{equation}
\big(1+e^{-2\eta}\big)^{-1/2}=e^{\eta}\sum_{s\ge0}\binom{-1/2}{s}e^{2s\eta},\qquad
e^{ik\Re S_+}e^{\eta}e^{2s\eta}=e^{ik(S_++2is\,\Im S_+)} .
\end{equation}
\end{itemize}
The $s$-th term thus lies on sheet $+s$ below $b_c$ and on sheet $-s$ above. Physically, $e^{-2|\eta|}$ is the ratio of transmission to reflection probability below $b_c$ (and its inverse above), so the $s$-th term describes $s$ additional partial transmissions through the photon-sphere barrier. The individual terms are not continuations of one another across $b_c$; only the full sum \eqref{eq:H:kemble} is analytic there. We refer to them as \emph{Kemble terms}. Their contributions to the position-space correlator are discussed in Appendix~\ref{app:GR:sing}.

\subsection{Large impact parameter: the bounce off the singularity}
\label{app:wkb:largeb}

The limit $b=p^t/p^\theta\to\infty$ corresponds to a radial null geodesic, which crosses the horizon and reaches the singularity. Accordingly, as discussed around \eqref{eq:deltalimit}--\eqref{eq:ThetaTinfty},
\begin{equation}\label{eq:H:Tinf}
T(\infty)\equiv\lim_{b\to\infty}\frac{S_+(b)}{b}=2\int_{\mathcal C}\frac{dr}{f(r)},
\end{equation}
where $\mathcal C$ runs from $r=0$ (where the complex turning point goes as $b\to\infty$, see \eqref{eq:approximatrTcblarge}) to the boundary, and passes the horizon pole of $1/f$ on the side selected by $\Im S_+>0$.
\begin{itemize}
\item \emph{Imaginary part (all $d$).} The horizon is a simple pole of $1/f$ with residue $1/f'(r_H)=\beta/4\pi$. Passing it gives half the residue, $i\pi\beta/4\pi$, doubled by the factor $2$:
\begin{equation}
\Im T(\infty)=\frac\beta2 .
\end{equation}
\item \emph{Real part.} The real part,
\begin{equation}
\Re T(\infty)=2\,{\rm PV}\!\int_0^\infty \frac{dr}{f(r)}\equiv t_c\,,
\end{equation}
is the boundary time of the radial null geodesic that bounces off the singularity \cite{Fidkowski:2003nf}. It vanishes for $d=2$ (BTZ)\footnote{In this case $T(\infty)=i\beta/2$ coincides with a thermal image of the light-cone singularity, and no separate bouncing singularity arises; see Appendix~\ref{app:GR:sing} and \cite{Grozdanov:2026cut}.} and is positive for $d\ge3$.
\end{itemize}
The correction $S_+(b)/b-T(\infty)$ comes from the region near $r=0$, where the complex turning point sits. There $f\simeq-\mu r^{2-d}$ and $V_{\rm eik}\simeq-\mu r^{-d}$, so the turning point is at $r\sim(\mu/b^2)^{1/d}$, cf.\ \eqref{eq:approximatrTcblarge}. Expanding $S_+/b=2\int dr\,f^{-1}\sqrt{1-V_{\rm eik}/b^2}$ about $2\int dr/f$, the correction scales as $r^{d-1}\sim b^{-2(d-1)/d}$:
\begin{equation}\label{eq:H:approach}
S_+(b)=b\,T(\infty)+c\,b^{1-q}+\dots,\qquad q\equiv\frac{2(d-1)}{d}.
\end{equation}
By \eqref{eq:H:legendre} this is equivalent to \eqref{eq:seriesTThetaClargeb}, with
\begin{equation}\label{eq:H:c1a1}
\begin{aligned}
T(b)&=T(\infty)+d_1\,b^{-q}+\dots,\qquad &d_1&=-\frac{d-2}{d}\,c\,,\\
\Theta(b)&=a_1\,b^{1-q}+\dots,\qquad &a_1&=-\frac{2(d-1)}{d}\,c\,.
\end{aligned}
\end{equation}
Note that the combination entering the position-space integrand at $\theta=0$ is
\begin{equation}\label{eq:H:phase0}
b\,t-S_+(b)=b\big(t-T(b)\big)+\Theta(b)=b\Big(t-T(\infty)-c\,b^{-q}+\dots\Big).
\end{equation}

Let us provide explicit formulas in $d=4$. The poles of $1/f$ are at $r^2=r_\pm^2$, with
\begin{equation}
r_\pm^2=\frac{-1\pm\sqrt{1+4\mu}}{2}\,.
\end{equation}
Here $r_+=r_H$ is the horizon, while $r_-^2=-\tilde r^2<0$, with $\tilde r^2=r_H^2+1$. We find
\begin{equation}
\frac1f=\frac{1}{r_H^2+\tilde r^2}\Big[\frac{r_H^2}{r^2-r_H^2}+\frac{\tilde r^2}{r^2+\tilde r^2}\Big].
\end{equation}
The first term has vanishing principal value on $(0,\infty)$ and contributes only the half-residue; the second integrates to $\pi\tilde r/\big(2(r_H^2+\tilde r^2)\big)$. Therefore
\begin{equation}\label{eq:H:Tinf4}
T(\infty)=\pi\,\frac{\sqrt{1+r_H^2}+i\,r_H}{1+2r_H^2}=t_c+\frac{i\beta}{2},\qquad
t_c=\frac{\pi\sqrt{1+r_H^2}}{1+2r_H^2},\qquad \frac\beta2=\frac{\pi r_H}{1+2r_H^2},
\end{equation}
in agreement with \eqref{eq:trdef}.

To compute $c$ in $d=4$, we set $q=\tfrac32$, so that $a_1=-3c/2$ and $d_1=-c/2$. Rescaling the radial coordinate, so that
\begin{equation}\label{eq:H:sigma4}
r=\Big(\frac{\mu}{b^2}\Big)^{1/4}\zeta\,,\qquad f\simeq-\frac{\mu}{r^2}\,,\qquad 1-\frac{V_{\rm eik}}{b^2}\simeq1+\zeta^{-4}\,,
\end{equation}
leads to
\begin{equation}\label{eq:H:c}
c=-2\mu^{-1/4}\Big[\int_{\zeta_c}^\infty\Big(\sqrt{\zeta^4+1}-\zeta^2\Big)d\zeta-\frac{\zeta_c^3}{3}\Big],\qquad \zeta_c=e^{-i\pi/4}.
\end{equation}
Here $\zeta_c$ is the root of $\zeta^4=-1$ selected by $\Im S_+>0$, and we split the integral into two terms, one above and one below $\zeta_c$, the latter giving the term $-\zeta_c^3/3$. Numerically (to 15 digits) the term in square brackets equals $e^{i\pi/4}\Gamma(\tfrac14)^2/(6\sqrt{2\pi})$, so that
\begin{equation}\label{eq:H:c4}
c=-e^{i\pi/4}\,\frac{\Gamma(\frac14)^2}{3\sqrt{2\pi}}\,\mu^{-1/4},\qquad \arg c=-\frac{3\pi}{4}.
\end{equation}
In particular $\Theta(b)=-\tfrac32c\,b^{-1/2}\to0$: the radial geodesic sweeps no angle, as noted below \eqref{eq:ThetaTinfty}.

\begin{figure}[t]\centering
\includegraphics[width=\textwidth]{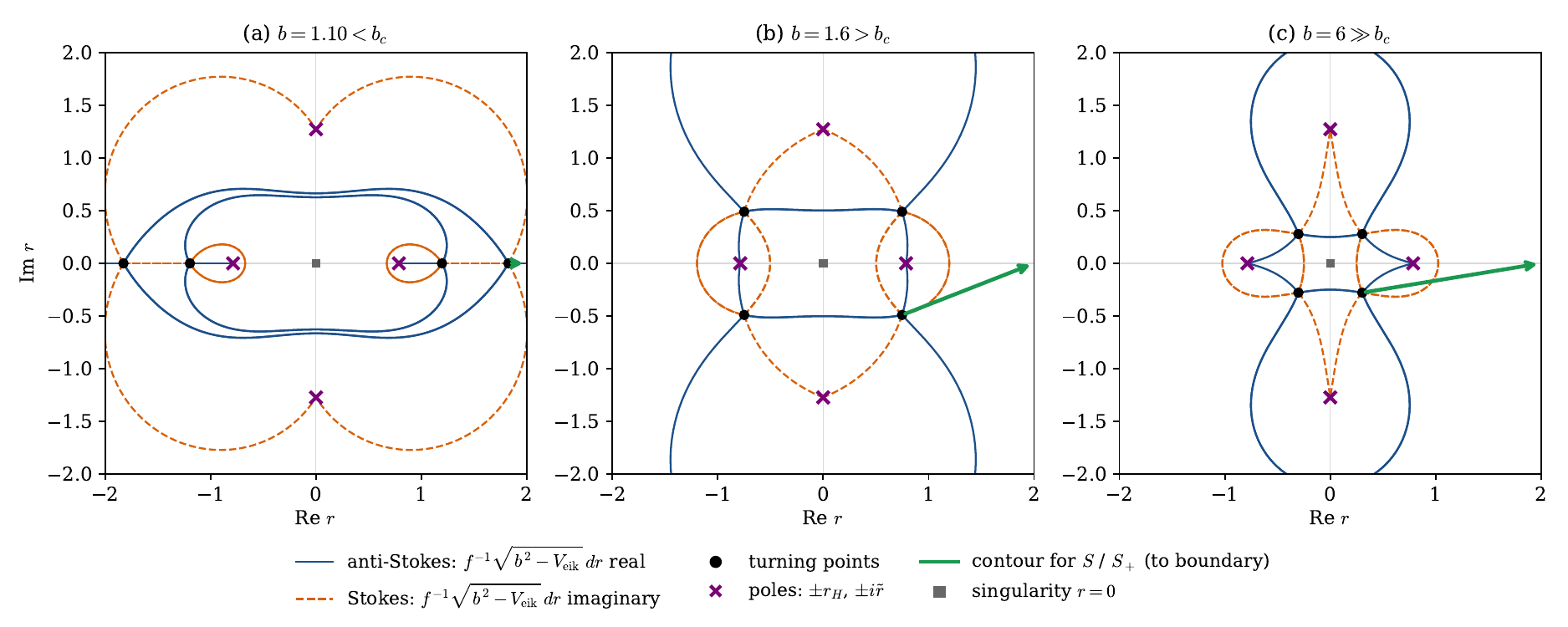}
\caption{Stokes geometry of the phase integral \eqref{eq:H:S} in the complex $r$-plane, for $d=4$ and $\mu=1$ ($b_c\simeq1.118$). Solid (dashed) lines are anti-Stokes (Stokes) lines: curves emanating from the turning points along which $f^{-1}\sqrt{b^2-V_{\rm eik}}\,dr$ is real (imaginary). Dots are the turning points, crosses the poles of $1/f$ at $r=\pm r_H$ and $r=\pm i\tilde r$, and the square the singularity $r=0$. The arrow is the integration contour from the turning point to the boundary, which gives $S(b)$ in (a) and $S_+(b)$ in (b) and (c). (a) $b<b_c$: real turning points, with the barrier between $r_T^<$ and $r_T^>$. (b) $b>b_c$: the turning points are complex. (c) $b\gg b_c$: four turning points approach $r=0$.}
\label{fig:H:stokes}
\end{figure}

The geometry of the phase integral \eqref{eq:H:S} in the complex $r$-plane is shown in Fig.~\ref{fig:H:stokes} for $d=4$. For $1<b<b_c$ the turning points are real, and the contour runs along the real axis from $r_T^>$ to the boundary. For $b>b_c$ they move into the complex plane, and the contour starting at the turning point in the lower half plane, passing below the horizon pole, gives $\Im S_+>0$. As $b\to\infty$ four turning points approach $r=0$ together, at $r\simeq(\mu/b^2)^{1/4}e^{i\pi(2j+1)/4}$, $j=0,\dots,3$, while in $d=4$ the integrand itself is regular at $r=0$. The one in the lower half plane, $j=3$, corresponds to the root $\zeta_c=e^{-i\pi/4}$ of \eqref{eq:H:c}.

\subsection{$\delta_\infty$: correction to the phase near the singularity}
\label{app:wkb:delta}

We just saw that, as $b\to\infty$, all complex turning points merge near the singularity, where the null geodesic bounces off before reaching the other boundary. In the Regge limit, i.e.\ keeping at each order in $\mu$ only the leading term at small $\sigma$, with $\sigma$ defined in \eqref{eq:sigmarhodef}, equivalently the leading power of $k$ at fixed $b$, the reflection there is that of an isolated turning point, and
\begin{equation}\label{eq:H:XRegge}
X\ \xrightarrow{\ b\to\infty\ }\ e^{ikS_+(b)}\,.
\end{equation}
This is what we use in the rest of the paper. Beyond the Regge limit, the connection at $r=0$ is not that of an ordinary turning point \cite{Motl:2003cd,Festuccia:2005pi,Festuccia:2008zx,Afkhami-Jeddi:2025wra}, and $X$ acquires an extra phase,
\begin{equation}\label{eq:H:Xdeltainf}
X\ \xrightarrow{\ b\to\infty\ }\ e^{i\delta_\infty}\,e^{ikS_+(b)}\,.
\end{equation}
In this subsection we describe the origin of $\delta_\infty$, and explain why it is absent in the Regge limit. Its effect on the coefficient of the singularity will be discussed elsewhere.

The phase $\delta_\infty$ arises in the region near $r=0$, whose contribution to the overall phase is given by \eqref{eq:H:approach}. It is therefore a function of the variable
\begin{equation}\label{eq:I:lambda}
\lambda\equiv k\,b^{1-q}=k^{q}\,\omega^{1-q}.
\end{equation}
This region is described by a local equation, in any $d$. Generalizing \eqref{eq:H:sigma4}, we set
\begin{equation}\label{eq:H:sigmad}
r=\Big(\frac{\mu}{b^2}\Big)^{1/d}\zeta\,,\qquad f\simeq-\mu\,r^{2-d}\,,\qquad \psi=\zeta^{\frac{d-2}{2}}\,\phi\,.
\end{equation}
The radial equation \eqref{eq:eqSch} then reduces to
\begin{equation}\label{eq:H:local}
\phi''(\zeta)+\Big[\lambda^2\mu^{-2/d}\,\zeta^{d-4}\big(1+\zeta^d\big)+\frac{1}{4\zeta^2}\Big]\phi=0\,,
\end{equation}
up to corrections suppressed by powers of $r^2$ and $1/k^2$. The three terms in the bracket have the following origin:
\begin{itemize}
\item $\lambda^2\mu^{-2/d}\zeta^{2d-4}$ comes from $\omega^2$;
\item $\lambda^2\mu^{-2/d}\zeta^{d-4}$ comes from the eikonal potential $k^2V_{\rm eik}$, with $V_{\rm eik}\simeq-\mu\,r^{-d}$;
\item $1/(4\zeta^2)$ comes from the term $\mu(d-1)^2/(4r^d)$ in \eqref{Vldef}, together with the change of variables \eqref{eq:H:sigmad}, and has the same coefficient in every $d$.
\end{itemize}
The mass term, $f\,(\nu^2-\tfrac14)$, is subleading near $r=0$. Hence $\delta_\infty=\delta_\infty\big(\lambda\,\mu^{-1/d}\big)$, and it does not depend on $\Delta$. The eikonal turning points, i.e.\ the zeros of the terms proportional to $\lambda^2$ in \eqref{eq:H:local}, are at $\zeta^d=-1$, and the one selected by $\Im S_+>0$ is $\zeta_c=e^{-i\pi/d}$. They are the large-$\lambda$ limits of the zeros of the full bracket in \eqref{eq:H:local}.

The phase $\delta_\infty$ follows from a connection problem for \eqref{eq:H:local} at $r=0$. It vanishes for $\lambda\to\infty$, where $\zeta_c$ is an isolated turning point, and tends to $-i\log2$ for $\lambda\to0$, i.e.\ $b\to\infty$ at fixed $k$, in any $d$, where \eqref{eq:H:local} reduces to the Bessel equation of order zero. This limit agrees with the reflection coefficient at the singularity found in \cite{Afkhami-Jeddi:2025wra} for the planar AdS$_5$ black hole at fixed spatial momentum and large frequency ($R=-2$ in their conventions), which is the analogue of $\lambda\to0$ there. Corrections to the isolated-turning-point connection are suppressed by powers of the large parameter $\lambda\mu^{-1/d}$, so that at large $\lambda$, $\delta_\infty$ is expected to be of the order of $\mu^{1/d}/\lambda$. A thorough examination of this regime will be presented elsewhere.

\paragraph{$\delta_\infty$ and the Regge limit.} The phase $\delta_\infty$ is absent in the Regge limit:
\begin{itemize}
\item it is subleading in $k$ at fixed $b$. At fixed $b$ and large $k$, $\lambda\to\infty$ and $\delta_\infty$ is expected to be of the order of $\mu^{1/d}b^{q-1}/k$, a $1/k$ correction relative to the eikonal phase $kS_+(b)$;
\item it depends on $\mu$ only through $\lambda\mu^{-1/d}$, and its expected leading large-$\lambda$ behavior above is proportional to $\mu^{1/d}$, which is not analytic in $\mu$. This is consistent with its absence from the expansion in $\mu$, but an all-orders statement would require the full asymptotic expansion of the connection; for the Regge limit, the previous point suffices;
\item it arises from a term that the Regge limit drops. In \eqref{eq:H:local}, the terms proportional to $\lambda^2$ come from $\omega^2$ and $k^2V_{\rm eik}$, whereas $1/(4\zeta^2)$ comes from the term $\mu(d-1)^2/(4r^d)$ in \eqref{Vldef}, which carries no factor of $k^2$ and is discarded in the eikonal limit. This term makes $\zeta=0$ a regular singular point with a double exponent, which leads to the Bessel equation of order zero and to the limit $\delta_\infty\to-i\log2$.
\end{itemize}
The resummed Regge phase is therefore $kS_+(b)$, \eqref{eq:H:XRegge}, which is the limit $\lambda\to\infty$ of \eqref{eq:H:Xdeltainf}. However, the approximation of keeping the leading power of $k$ holds at fixed $b$, but not uniformly in $b$: at any large $k$, for $b\sim k^{1/(q-1)}$, i.e.\ $\lambda=O(1)$, the discarded term is as large as those kept. This is precisely the region that generates the singularity at $t=T(\infty)$: near it the integral is dominated by $b\sim b_*\to\infty$ and $k\sim b^{q-1}$, both large, but at fixed $\lambda$, and $\lambda$ is invariant under the rescaling that generates the singular term (Appendix~\ref{app:GR:coef}). As a consequence:
\begin{itemize}
\item the coefficient $C_\infty$ of \eqref{eq:I:Cinf}, obtained in the Regge limit, is multiplied in the full correlator by a factor determined by $\delta_\infty(\lambda)$ at $\lambda=O(1)$: the two limits do not commute for the coefficient. This factor is a weighted integral of $e^{i\delta_\infty(\lambda)}$, whose evaluation requires the connection at finite $\lambda$ and its analytic continuation, and is left for future work; a preliminary evaluation suggests that it does not vanish. In the related regime of fixed spatial momentum and large frequency, where the connection reduces to its $\lambda\to0$ limit, the corresponding singularity of the retarded function of the planar black hole, at $t=\beta(1+i)/2$, has a non-vanishing coefficient for generic $\Delta$ \cite{Afkhami-Jeddi:2025wra}. Moreover, for the planar black hole in $d=4$, the stress-tensor sector of the position-space correlator has a singularity with the same exponent, $2\Delta-\frac d2$, whose coefficient was found numerically to be non-vanishing for $\tfrac54\lesssim\Delta\lesssim15$ \cite{Ceplak:2024bja};
\item provided this factor does not vanish, the location $t=T(\infty)$ and the exponent $\frac d2-2\Delta$ of the singularity, obtained in the Regge limit, are not changed by $\delta_\infty$;
\item the same holds for the singularities at $N\,T(\infty)+in\beta$.
\end{itemize}
The quasinormal modes computed at fixed $\ell$ in Appendix~\ref{app:num:qnm}, i.e.\ outside the Regge limit, exhibit this phase.

\section{WKB and the bulk phase shift}
\label{appendixa}

In this appendix we formulate an alternative way to obtain the eikonal phase, without resorting to the boundary condition at the horizon. The motivation lies in the following. Whilst the bulk cone singularities obtained from the bulk phase shift are in principle dependent upon both the stress-tensor sector OPE data and the double-trace OPE data, the bouncing singularity can be constructed solely from near boundary data. By near boundary data we mean data that can be obtained without imposing the boundary condition at the horizon, such as the OPE data of the stress-tensor sector, which follow from bulk computations that do not reach the horizon. There is a strong indication that the momentum-space phase shift in the Regge limit is of this type: it is independent of the double traces, which contribute only upon transforming back to position space (Section~\ref{sec.six}). The singularity found in Section~\ref{sec.five} further supports this expectation, as it is naturally attributed to the stress-tensor sector \cite{Ceplak:2024bja}; the double traces have been shown to cancel completely the singularity of the stress-tensor sector at $t=t_c-i\beta/2$, inside the strip of analyticity, and to restore the KMS condition \cite{Buric:2025fye, Buric:2025anb}. We therefore replace the boundary condition at the horizon by two conditions, at zeroth and first order in $\mu$, which fix the coefficients of the WKB solution and lead to (\ref{eq:GTansatz}).

We use the radial equation (\ref{eq:eqSch}) and the notation of Appendix~\ref{app:wkb:waves}. In the eikonal limit considered here, it reduces to
\begin{equation}\label{eq:eqEikonal}
\left[\partial_z^2+k^2 (b^2-V_{eik})\right] \psi=0
\end{equation}
where the eikonal potential is given in (\ref{eq:Veikdef}) and $k\equiv \ell+{d-2\over 2}$.
We are interested in the case $1<b<b_c$ where there are two real turning points, $r_T^>>r_T^<$, as can be seen in Fig.~\ref{fig:H:schematic}.

The greater of the two turning points is closer to the boundary and is the turning point for the null geodesic. The WKB solution in the region $r>r_T^>$ is, as in \eqref{eq:H:WKBwaves}:
\begin{equation}
\label{eq:WKBsolgen}
\psi(z)={c_1\over \left(b^2-{f(r)\over r^2}\right)^{1\over 4}} e^{i k \int_{r_T}^r {dr\over f(r)} \sqrt{b^2-{f(r)\over r^2}}} +{c_2\over \left(b^2-{f(r)\over r^2}\right)^{1\over 4}} e^{-i k \int_{r_T}^r {dr\over f(r)} \sqrt{b^2-{f(r)\over r^2}}}\,.
\end{equation}
The approximation breaks down  in the near boundary region $z\sim {1\over k}$. The near boundary behavior of (\ref{eq:WKBsolgen}),
\begin{equation}
\label{eq:WKBsolboundary}
\psi(z)={c_1\over (b^2-1)^{1\over 4}} e^{i {\delta\over 2}}e^{-i k z\sqrt{b^2-1}} +{c_2\over (b^2-1)^{1\over 4}} e^{-i {\delta\over 2}} e^{i k z\sqrt{b^2-1}}
\end{equation}
with $\delta$ as defined in (\ref{eq:deltadef}), should be matched with solutions of the Bessel ODE:
\begin{equation}
\label{eq:eomnearboundary}
\partial_z^2\psi+\left(-{\left(\nu^2-{1\over 4}\right)\over z^2} +k^2 (b^2-1)\right)\psi=0\,,
\end{equation}
which are:
\begin{equation}
\label{eq:Hankelsol}
\psi= \sqrt{z}\left[C_1 H^{(1)}_\nu (k z\sqrt{b^2-1})+C_2 H^{(2)}_\nu (k z\sqrt{b^2-1})\right]\,.
\end{equation}
Here the Hankel functions are more convenient than the Bessel functions $J_{\pm\nu}$ used in Appendix~\ref{app:wkb:boundary}: away from the boundary each of them reduces to a single traveling wave, which is matched directly to one of the two WKB waves in (\ref{eq:WKBsolboundary}).
Far from the boundary (\ref{eq:Hankelsol}) behaves as follows:
\begin{equation}
\label{eq:Hankelasymptotics}
\psi\simeq \sqrt{2\over\pi k \sqrt{b^2-1}}\left[C_1 e^{i\left(k z\sqrt{b^2-1} -\nu {\pi\over 2}-{\pi\over 4}\right)}+C_2 e^{-i\left(k z\sqrt{b^2-1} -\nu {\pi\over 2}-{\pi\over 4}\right)}\right]
\end{equation}
which leads to
\begin{equation}
\label{eq:HankelmatchWKB}
C_1=\sqrt{\tfrac{\pi k}{2}}\;c_2\,e^{-i\delta/2}\,e^{i\left(\nu\frac{\pi}{2}+\frac{\pi}{4}\right)},\qquad
C_2=\sqrt{\tfrac{\pi k}{2}}\;c_1\,e^{+i\delta/2}\,e^{-i\left(\nu\frac{\pi}{2}+\frac{\pi}{4}\right)}
\end{equation}
The Wightman correlator is proportional to the square of the coefficient $C(k,b)$ of the normalisable mode  \cite{Festuccia_2006, Festuccia:2008zx}, which according to (\ref{eq:Hankelsol}) is equal to:
\begin{equation}
\label{eq:coeffcorr}
C(k,b)^2=\frac{i\,e^{i\pi\nu}}{\sin^2\pi\nu}\,\frac{\pi k}{2}
\left[\frac{2^{-\nu}\big(k\sqrt{b^2-1}\big)^{\nu}}{\Gamma(\nu+1)}\right]^2
e^{-i\delta}\Big(c_1e^{i\delta}-i\,c_2\,e^{-i\pi\nu}\Big)^2
\end{equation}
Dividing by the zero temperature result (\ref{eq:GT0norm}) leads to
\begin{equation}
\label{eq:corrc12}
{G^T(k,b)\over G^{T=0}_+(k,b)}=\frac{i\, e^{i\pi\nu}\, k\, \Gamma\left(\frac{d}{2}+\nu\right)}{4\pi^{\frac d2}\sin^2(\pi\nu)\,\Gamma(\nu+1)}\; e^{-i\delta}\Big(c_1e^{i\delta}-i\,c_2\,e^{-i\pi\nu}\Big)^2\,.
\end{equation}
Here we simply expressed the undetermined coefficients of the Hankel solutions by the WKB ones. This is useful because we are searching for a solution in the eikonal regime, and the WKB provides a consistent formula for an expansion in terms of large $k$. To proceed following the standard WKB approach, one would need to consider further regions within the bulk and impose matching conditions as necessary, including boundary conditions at the horizon.
This approach was taken in Appendix~\ref{app:wkb} and in \cite{Dodelson:2023nnr}. The latter determined the bulk phase shift by evaluating the leading Regge-pole behavior.

Here we would like to fix $c_{1,2}$ with different arguments, at least in the region $1<b<b_c$. First, notice that the dependence of (\ref{eq:corrc12}) on $b$ is entirely contained in $\delta(b)$, which is determined by the wave equation in the region $r>r_T^>$. For $1<b<b_c$ this region lies above the photon-sphere barrier and does not reach the horizon. The interior, including the boundary condition at the horizon, enters only through the ratio $c_1/c_2$, the inner reflection amplitude $\mathcal R$ of Appendix~\ref{app:wkb}. For $b<b_c$, away from $b_c$, this ratio is fixed locally, by the connection at the outer turning point. The ingoing condition at the horizon fixes a combination of the two WKB solutions on the inner side of the barrier. Transported through the barrier to $r_T^>$, the solution that grows into the barrier is suppressed by $e^{-2kK}$, with $K$ the barrier action (\ref{eq:H:K}), so that at $r_T^>$ only the solution that decays into the barrier remains. Near the simple turning point $r_T^>$, $b^2-V_{\rm eik}$ is linear in $z-z_T$, and the radial equation reduces to the Airy equation after a rescaling by the slope. The decaying Airy solution connects to the two WKB waves (\ref{eq:WKBsolgen}) with the universal relative phase $\mathcal R=e^{-i\pi/2}$, independent of the slope (Appendix~\ref{app:wkb:waves}). The dependence on $\mu$ of the slope and of the position of $r_T^>$ is contained in the WKB amplitude and in $\delta(b)$, which are factored out of $c_1$ and $c_2$ in (\ref{eq:WKBsolgen}). The same phase arises in pure AdS, where the inner reflection is at the single turning point $r_T=(b^2-1)^{-1/2}$. This holds up to $\mathcal{O}(1/k)$ corrections from the shape of the barrier and corrections of order $e^{-2kK}$ due to tunneling through it; see (\ref{eq:H:kemble}) and the limits below it. Only the latter depend on the horizon. The matching (\ref{eq:HankelmatchWKB}) at $1/k\ll z\ll1$, where the potential differs from its AdS value only by $\mathcal{O}(\mu\,r^{-d})$, relates the Hankel coefficients to $c_{1,2}$. Since it is homogeneous, it does not fix the overall normalization of $c_1$ and $c_2$. We fix it through the flux of the WKB waves in the region $r>r_T^>$. Owing to the factor $\big(b^2-f/r^2\big)^{-1/4}$ in (\ref{eq:WKBsolgen}), each wave carries a constant flux, proportional to $k\,|c_{1,2}|^2$. This normalization does not involve the horizon, and it is independent of $\mu$ at leading order in $k$, since the dependence on $\mu$ is contained in the WKB amplitude and in $\delta(b)$. For $b<b_c$ the incoming and outgoing fluxes are equal, up to the absorption $e^{-2kK}$. In particular, a common $\mu$-dependent rescaling of $c_1$ and $c_2$ is excluded. The value of the normalization constant is fixed by the first of the two conditions below, i.e.\ by the pure-AdS correlator. Near $b_c$, where $kK=\mathcal{O}(1)$, the connection is no longer local and the uniform treatment of Appendix~\ref{app:wkb} is required: $\mathcal R$ acquires the barrier factor $(1+e^{-2\eta})^{-1/2}e^{i\varphi(\eta)}$ of (\ref{eq:H:kemble}), which depends on $\mu$ only through $\eta=kK$. This factor is a universal function of $\eta$, which tends to one for $\eta\gg1$, i.e.\ outside a window $b_c-b=\mathcal{O}(1/k)$. It therefore does not affect the conditions below, which are imposed order by order in $\mu$ at fixed $b$, while $b_c^2-1\propto\mu^{-2/(d-2)}\to\infty$ as $\mu\to0$; nor does it affect the singularity of Section~\ref{sec.five}, which comes from $b\to\infty$. At leading order at large $k$, at fixed $b<b_c$ and apart from exponentially small terms, $c_1$ and $c_2$ therefore do not depend on $\mu$: their ratio by the local connection at $r_T^>$, and their normalization by the flux. Two conditions, at zeroth and first order in $\mu$, then suffice to fix them. Expanding the square in (\ref{eq:corrc12}), the correlator contains three terms: one proportional to $e^{i\delta}$, one without phase, and one proportional to $e^{-i\delta}$. These are, respectively, the eikonal term, the term associated with propagation near the boundary and the complex-conjugate partner of the eikonal term discussed in Section~\ref{sec.six}. As explained there, the Regge limit retains only the eikonal term.

Here, we instead combine (\ref{eq:corrc12}) with two conditions, namely that the correlator:
\begin{itemize}
\item reduces for $\mu=0$ to the eikonal term of the zero temperature correlator in cylinder variables, $G^{T=0}_+\,e^{-i\pi(\nu+1)}e^{i\delta_{AdS}}$, cf.\ (\ref{eq:GTansatz}) and (\ref{eq:Reggecylinder}), and
\item behaves to leading order in $\mu$ as predicted by Regge theory \cite{Kulaxizi:2018dxo}. Equivalently, we could instead impose positivity of the bulk phase shift \cite{Camanho:2014apa, Kulaxizi:2017ixa, Li:2017lmh, Hartman:2015lfa, Costa:2017twz, Caron-Huot:2021enk, Meltzer:2018tnm}.
\end{itemize}
These two conditions result in the standard phase shift expression, (\ref{eq:GTansatz}), with
\begin{equation}
\label{eq:c12sol}
c_1= 2\, e^{\frac{i \pi}{4}}e^{-i\pi\nu}\, \pi ^{d/4} \sqrt{ \frac{\sin^2{(\nu\pi)}\,\Gamma (\nu+1)}{k\, \Gamma \left(\frac{d}{2}+\nu\right)} },\quad c_2=0\,,
\end{equation}
up to an overall sign. The condition $c_2=0$ replaces the boundary condition at the horizon. It should be understood as the statement that only the eikonal term survives in the Regge limit, rather than as a condition on the full correlator, for which $c_1/c_2=\mathcal R$ (see the end of this appendix).

The second condition can alternatively be stated as the condition for positive Shapiro time delay,  since the leading order phase shift is proportional to the time delay produced when a light particle traverses an AdS shock wave geometry \cite{Camanho:2014apa}.  It corresponds to the graviton exchange between a light particle and the black hole (or heavy particle).

This reasoning enables the determination of the bulk phase shift from near boundary data, in the sense explained at the beginning of this appendix.

The argument above determines $\delta(b)$ for $1<b<b_c$. Beyond the critical impact parameter $\delta$ is defined by analytic continuation, and one has to choose between the two continuations $S_\pm$ of Section~\ref{sec.four}. In the complete treatment of Appendix~\ref{app:wkb} the choice $S_+$ follows from the ingoing condition at the horizon or, equivalently, from the analyticity of the retarded function in the upper half $\omega$-plane (Appendix~\ref{app:wkb:continuation}). The latter is a consequence of causality in the CFT, and does not require bulk data. The phase shift beyond $b_c$, and with it the location and the exponent of the bouncing singularity of Section~\ref{sec.five}, would then be determined by near boundary data together with causality. In the Regge limit the coefficient of the singularity is determined in the same way. Beyond it, the coefficient is modified by the phase $\delta_\infty$ of Appendix~\ref{app:wkb:delta}, which is an $\mathcal O(1/k)$ correction at fixed $b$, but becomes of order one for $\lambda=k\,b^{1-q}=\mathcal O(1)$, i.e.\ $b\sim k^{d/(d-2)}$; it arises near the singularity at $r=0$, and is not fixed by near boundary data.

It is curious that the conditions are equivalent to setting $c_2=0$ in the leading order WKB solution. With the identification (\ref{eq:WKBsolboundary}), $c_2$ multiplies the wave traveling from the boundary towards the horizon and $c_1$ the wave traveling back towards the boundary, so $c_2=0$ is not an ingoing condition at the horizon; it retains only the reflected wave. In the complete treatment of Appendix~\ref{app:wkb} neither coefficient vanishes: the ingoing condition at the horizon fixes their ratio $c_1/c_2$ (the inner reflection amplitude), and the retarded function is the ratio (\ref{eq:H:GRY}) rather than the square of a single coefficient. The term linear in $e^{i\delta}$ is then the eikonal term of Appendix~\ref{app:GR}, which reproduces (\ref{eq:GTansatz}) including its constant phase, while the remaining terms are those listed above together with the multiple reflections. Setting $c_2=0$ thus amounts to retaining the eikonal term, as the Regge limit does.

\section{From the retarded function to the position-space integral}
\label{app:GR}

In Section~\ref{sec.five} the position-space correlator was obtained from \eqref{eq:GTansatz}, with the phase continued to $b>b_c$. In this appendix we derive that integral from the exact representation of the thermal Wightman function in terms of the retarded function, using the large-$k$ form \eqref{eq:H:GRY}. This shows which term of the full correlator the integral \eqref{eq:correlatorthetazero} represents, fixes its normalization, and determines what other terms are present and where they are singular. We work at $\theta=0$ throughout, as in \eqref{eq:Gegenbauerthetazero}--\eqref{eq:correlatorthetazero}.

\subsection{Exact representation}
\label{app:GR:exact}

The thermal Wightman function is determined by the spectral function
\begin{equation}
\tilde\rho(\omega,k)=-i\big[G_R(\omega,k)-G_A(\omega,k)\big]=2\Im G_R(\omega,k)
\end{equation}
through the fluctuation--dissipation relation,
\begin{equation}
G_+(\omega,k)=\tilde\rho(\omega,k)\,\big[1+n_B(\omega)\big],\qquad n_B(\omega)=\frac{1}{e^{\beta\omega}-1}\,.
\end{equation}
At $\theta=0$, by the first line of \eqref{eq:correlatorthetazero},
\begin{equation}\label{eq:I:exact}
G_+(t,0)=\sum_{\ell}\frac{k\,\Gamma(\frac d2)}{(d-2)\pi^{d/2}}\,C_\ell^{\frac{d-2}{2}}(1)\int_{-\infty}^{\infty}\frac{d\omega}{2\pi}\,e^{-i\omega t}\,\tilde\rho(\omega,k)\,\big[1+n_B(\omega)\big],
\end{equation}
with $C_\ell^{\frac{d-2}{2}}(1)$ given by \eqref{eq:Gegenbauerthetazero}. At $\theta=0$ no asymptotic form of the angular functions is needed.

\subsection{Bounce expansion in pure AdS}
\label{app:GR:AdS}

As a check of the boundary analysis of Appendix~\ref{app:wkb:boundary}, consider global AdS$_{d+1}$. The mode regular at $r=0$ is
\begin{equation}
\phi\propto r^\ell(1+r^2)^{\omega/2}\,{}_2F_1\Big(\tfrac{\ell+\omega+\Delta}{2},\tfrac{\ell+\omega+d-\Delta}{2};\ell+\tfrac d2;-r^2\Big),
\end{equation}
and the connection formula of the hypergeometric function at $r\to\infty$ gives \cite{Son:2002sd}
\begin{equation}
G_R^{\rm AdS}\propto\frac{\Gamma(-\nu)}{\Gamma(\nu)}\;
\frac{\Gamma\big(\tfrac{\ell+\omega+\Delta}{2}\big)}{\Gamma\big(\tfrac{\ell+\omega+d-\Delta}{2}\big)}\;
\frac{\Gamma\big(\tfrac{\ell-\omega+\Delta}{2}\big)}{\Gamma\big(\tfrac{\ell-\omega+d-\Delta}{2}\big)},
\end{equation}
with $\omega\to\omega+i\epsilon$, and poles at the normal modes $\omega=\Delta+\ell+2m$. The first of the three ratios of $\Gamma$-functions is independent of $\omega$ and $\ell$. In the limit where $(\omega,k)$ are large and their ratio $b>1$ is fixed,
\begin{itemize}
\item the second ratio has large positive arguments and, by Stirling's formula, tends to $\big(k(b+1)/2\big)^{\nu}$;
\item the third ratio has large negative arguments. Using $\Gamma(x)\Gamma(1-x)=\pi/\sin\pi x$ for each factor, the reflected $\Gamma$-functions give $\big(k(b-1)/2\big)^{\nu}$, and what remains is a ratio of sines,
\begin{equation}
\frac{\sin\big(\tfrac\pi2(\omega-\ell-\Delta)+\pi\nu\big)}{\sin\big(\tfrac\pi2(\omega-\ell-\Delta)\big)}=e^{-i\pi\nu}\,\frac{1-e^{2\pi i\nu}Y}{1-Y},\qquad Y=e^{i\pi(\omega-\ell-\Delta)},
\end{equation}
where $|Y|<1$ for $\Im\omega>0$.
\end{itemize}
This is exactly \eqref{eq:H:GRY} with $Y=Y_{\rm AdS}$ of \eqref{eq:H:YAdS}, including the prefactor $\mathcal G_0$. The pure-AdS case thus confirms the boundary reflection phase $e^{-i\pi(\nu+1/2)}$, the Maslov phase $e^{-i\pi/2}$, and the eikonal phase $e^{ik\pi(b-1)}$ separately (the only approximation is Stirling's formula for the second ratio, so that the relative accuracy of \eqref{eq:H:GRY} is $O(1/k^2)$; e.g.\ at $\Delta=2.6$, $b=1.3$, the exact ratio of $\Gamma$-functions divided by \eqref{eq:H:GRY} is $1.00178$, $1.00045$ and $1.00011$ for $k=20$, $40$ and $80$). Expanding in $Y$,
\begin{equation}\label{eq:I:bounce}
G_R\simeq\mathcal G_0\,e^{-i\pi\nu}\Big[1+\big(1-e^{2\pi i\nu}\big)\sum_{N\ge1}Y^N\Big].
\end{equation}
The $N$-th term describes $N$ round trips between the AdS boundary and the inner reflection point, i.e.\ $N$ reflections. In pure AdS the inner reflection point is the turning point $r_T=(b^2-1)^{-1/2}$, where the mode regular at $r=0$ is totally reflected with the Maslov phase $e^{-i\pi/2}$ (Appendix~\ref{app:wkb:boundary}); this is the reflecting-cavity picture of \cite{Amado:2008hw}. For real $\omega$ and $b>1$, $\mathcal G_0$ is real, and $G_A=\overline{G_R}$. The zero-temperature spectral function is therefore
\begin{equation}\label{eq:I:rho0}
\tilde\rho_0=-i\,\mathcal G_0\,e^{-i\pi\nu}\big(1-e^{2\pi i\nu}\big)=-2\sin(\pi\nu)\,\mathcal G_0>0\,.
\end{equation}

\subsection{The black hole}
\label{app:GR:BH}

In the black-hole background the large-$k$ retarded function has the same form \eqref{eq:I:bounce}, with only two ingredients of the round trip changed:
\begin{enumerate}
\item the eikonal phase, $k\pi(b-1)\to kS(b)$;
\item the inner reflection: the regular center (phase $e^{-i\pi/2}$) is replaced by the photon-sphere barrier, with amplitude \eqref{eq:H:kemble}.
\end{enumerate}
The boundary reflection is unchanged at leading order, because the Bessel matching takes place at $1/k\ll z\ll1$, i.e.\ $1\ll r\ll k$, where $V_{\ell}$ differs from its AdS value by $O(\mu r^{-d})$. Hence \eqref{eq:I:bounce} holds with
\begin{equation}\label{eq:I:YBH}
Y=e^{-i\pi(\nu+1)}\,X,\qquad X\ \text{given by \eqref{eq:H:kemble}},\qquad X\ \xrightarrow{b\to\infty}\ e^{ikS_+(b)} .
\end{equation}
The last relation holds in the Regge limit, \eqref{eq:H:XRegge}, in which we work from now on. Beyond it, the connection at the singularity $r=0$ adds the phase $\delta_\infty(\lambda)$ of Appendix~\ref{app:wkb:delta}, which multiplies the $N$-reflection term by $e^{iN\delta_\infty}$. It changes neither the location nor the exponent of the singularities, but it affects their coefficients.

\subsection{Spectral function and normalization}
\label{app:GR:norm}

Since $G_A=\overline{G_R}$, \eqref{eq:I:bounce}, with $Y$ given by \eqref{eq:I:YBH}, yields
\begin{equation}\label{eq:I:rho}
\tilde\rho=\tilde\rho_0\Big[1+\sum_{N\ge1}\big(Y^N+\bar Y^N\big)\Big],
\end{equation}
with $\tilde\rho_0$ given by \eqref{eq:I:rho0}. The $N$-reflection part of $\tilde\rho$ coming from $G_R$ (the ``retarded'' part) is therefore exactly $\tilde\rho_0Y^N$, and the one coming from $G_A$ (the ``advanced'' part) is $\tilde\rho_0\bar Y^N$: the factor $1-e^{2\pi i\nu}$ is absorbed in $\tilde\rho_0$.

\paragraph{The $k$-integral.} We replace the sum over $\ell$ in \eqref{eq:I:exact} by an integral over $k$, change variables from $(\omega,k)$ to $(b,k)$, and use the large-$k$ approximations
\begin{equation}
k\,C_\ell^{\frac{d-2}{2}}(1)\simeq\frac{k^{d-2}}{\Gamma(d-2)}\,,\qquad
\tilde\rho_0\simeq-2\sin(\pi\nu)\,\frac{\Gamma(-\nu)}{\Gamma(\nu)}\Big(\frac k2\Big)^{2\nu}\big(b^2-1\big)^{\nu}.
\end{equation}
Every term of \eqref{eq:I:rho} is a pure exponential in $k$, so the $k$-integral can be done exactly,
\begin{equation}\label{eq:I:kint}
\int_0^\infty dk\,k^{2\Delta-1}e^{-ik[b(t-in\beta)-NS(b)]}=\Gamma(2\Delta)\,\big(i\big[b(t-in\beta)-NS(b)\big]\big)^{-2\Delta},
\end{equation}
for the terms introduced below, and the result is continued in $t$ from the region of convergence, $\Im\big[b(t-in\beta)-NS(b)\big]<0$. Note that each reflection or image term is thereby $O(k^0)$: no term is suppressed by a power of $k$ after the $k$-integral.

The $N=0$ term gives the short-distance singularity, $G_+^{(0)}(t)\propto J_0\,t^{-2\Delta}$, with
\begin{equation}
J_0=\int_1^\infty db\,(b^2-1)^{\nu}\,b^{-2\Delta}=\frac12\,B\Big(\frac{d-1}{2},\nu+1\Big).
\end{equation}
Normalizing the correlator so that $G_+\to t^{-2\Delta}$ as $t\to0$ gives the factor $1/J_0$ below. Using $1+n_B(\omega)=\sum_{n\ge0}e^{-n\beta\omega}$ for $\omega>0$, we write
\begin{equation}
G_+(t)=\sum_{N\ge0,\,n\ge0}G_+^{(N,n)}(t)+\big(\text{advanced and }\omega<0\text{ terms}\big)\,,
\end{equation}
where the retarded $N$-reflection term with thermal image $n$ is
\begin{equation}\label{eq:I:GNj}
\boxed{G_+^{(N,n)}(t)=\frac{e^{-i\pi N(\nu+1)}}{J_0}\int_1^\infty db\,\big(b^2-1\big)^{\Delta-\frac d2}\Big[b\,(t-in\beta)-N\,S(b)\Big]^{-2\Delta}}
\end{equation}
where $S\to S_+$ for $b>b_c$, and where the Kemble terms of Appendix~\ref{app:wkb:continuation} are omitted (they replace $S$ by the sheets $S_{(\pm s)}$, with binomial weights). For $d=4$ the constant phase is $e^{-i\pi N(\Delta-1)}$.

\emph{The integral \eqref{eq:correlatorthetazero}, with the continued phase above $b_c$, is the term $N=1$, $n=0$ of \eqref{eq:I:GNj}.} Two remarks:
\begin{itemize}
\item In \eqref{eq:correlatorthetazero} the $b$-integral starts at $b_c$, to maintain focus on the region beyond the critical impact parameter. The range $1<b<b_c$ is analytic at the singular point $t=T(\infty)$ (it is singular only on the segment $0\le t\le S(b_c)/b_c$ of the real axis; see Fig.~\ref{fig:J:tplane}), and does not affect the analysis of Section~\ref{sec.five}.
\item In momentum space this term is $G_+^{T=0}(\omega,k)\,e^{-i\pi(\nu+1)}e^{ikS}$, with $G_+^{T=0}$ given by \eqref{eq:GT0norm}. This is \eqref{eq:GTansatz}. As explained at the end of Section~\ref{sec.three}, the constant $e^{-i\pi(\nu+1)}$ combines with $e^{ik\pi(b-1)}$ into the Regge-sheet phase $e^{-i\pi\Delta}$ times the shift to the antipodal point.
\end{itemize}

\subsection{Singular points and absence of cancellations}
\label{app:GR:sing}

By the analysis of Section~\ref{sec.five}, a term of the form \eqref{eq:I:GNj} is singular where its phase, divided by $b$, vanishes as $b\to\infty$. Using \eqref{eq:H:Tinf} this gives the singular points listed in Table~\ref{tab:I:sing}. The advanced terms ($\bar Y^N$) contain $e^{-ikN\bar S_+}$, and the terms with $\omega<0$ follow from $\tilde\rho(-\omega)=-\tilde\rho(\omega)$ and $1+n_B(-\omega)=-n_B(\omega)=-\sum_{n\ge1}e^{-n\beta|\omega|}$.

\begin{table}[t]\centering
\begin{tabular}{lll}\toprule
Term & Singular at & Remark\\\midrule
$G_R$ terms, $\omega>0$ & $N\,T(\infty)+in\beta$,\ \ $n\ge0$ & $N=1$, $n=0$: the integral \eqref{eq:correlatorthetazero}\\
$G_A$ terms, $\omega>0$ & $-N\,\overline{T(\infty)}+in\beta$,\ \ $n\ge0$ & $\Re t<0$\\
$G_A$ terms, $\omega<0$ & $-N\,T(\infty)-in\beta$,\ \ $n\ge1$ & $\Im t\le-\tfrac32\beta$\\
$G_R$ terms, $\omega<0$ & $N\,\overline{T(\infty)}-in\beta$,\ \ $n\ge1$ & $\Im t\le-\tfrac32\beta$\\
Kemble terms, $s\ge1$ & retarded points $+\,is\beta$ & formal, see below\\\bottomrule
\end{tabular}
\caption{Singular points of the individual terms of the large-$k$ thermal correlator $G_+$ at $\theta=0$, labeled by the retarded ($G_R$) or advanced ($G_A$) part of the spectral function they come from, with $T(\infty)=t_c+i\beta/2$ and $N\ge1$. $N$ counts round trips, $n$ thermal images and $s$ Kemble terms.}
\label{tab:I:sing}
\end{table}

\begin{itemize}
\item $N$ counts round trips between the AdS boundary and the inner reflection; $N=1$ is the eikonal term. In position space the $N$-reflection term is singular at $N$ times the single round-trip time.
\item $n$ comes from the Bose factor, $1+n_B(\omega)=\sum_{n\ge0}e^{-n\beta\omega}$. Each term is an exact thermal (KMS) image, $t\to t-in\beta$, i.e.\ $n$ windings of the Euclidean time circle. Together with $N$, it generates the lattice of singular points of \cite{Dodelson:2025jff}, in which the number of bounces changes $\Re t$ and the thermal images shift $\Im t$ (see the footnote in Section~\ref{sec.six}).
\item $s$ labels the binomial expansion of the barrier factor in \eqref{eq:H:kemble}, i.e.\ partial transmissions and reflections at the photon-sphere barrier, equivalently the logarithmic sheets $S_{(\pm s)}$. Above $b_c$ they add $2is\,\Im S_+\simeq is\beta b$ to the phase, so their singularities fall on the retarded points shifted by $is\beta$, i.e.\ on the same points as the thermal images.
\end{itemize}
The Kemble terms require a caveat. At $b\to\infty$ they are exponentially small, $e^{-s\beta\omega}$, relative to the leading term, and the uniform formula \eqref{eq:H:kemble} is not guaranteed to capture exponentially small corrections in a region where the turning points approach $r=0$ (which is also where the phase $\delta_\infty$ of Appendix~\ref{app:wkb:delta} arises). Their singularities should therefore be regarded as formal. The numerical status of the three families is summarized in Appendix~\ref{app:num:GR}.

Let us now examine possible cancellations of the singularity. For a second term to be singular at $t=T(\infty)$ one would need one of the following:
\begin{itemize}
\item $(N-1)\,T(\infty)\in i\beta\mathbb Z$ for some $N\ge2$, i.e.\ $(N-1)t_c=0$;
\item $-N\,\overline{T(\infty)}+in\beta=T(\infty)$, i.e.\ $-Nt_c=t_c$;
\item an $\omega<0$ term, which is excluded by the sign of $\Im t$.
\end{itemize}
All are impossible when $t_c\neq0$, which is the case for $d\ge3$. \emph{The singularity at $t=T(\infty)$ therefore comes from the single term \eqref{eq:correlatorthetazero} and cannot be canceled by any other term of the large-$k$ correlator.}

Replacing the sum over $\ell$ by an integral over $k$ neglects the Poisson images $e^{2\pi iwk}$, $w\in\mathbb Z\setminus\{0\}$, which act like $\theta\to2\pi w$. With $\theta\neq0$ the large-$b$ region is dominated by $bt\simeq\theta$, and the singularity at $t=T(\infty)$ is weaker: $(t-T(\infty))^{3-2\Delta}(2\pi w)^{-3}$ in $d=4$. These images cannot cancel the leading term. The situation differs from pure AdS, where $\Theta(b)=\pi$ for all $b$: the sum over $\ell$ then alternates as $(-1)^\ell$, and the continuum eikonal term has a spurious singularity at $t=\pi$, $\theta=0$ which cancels against the $w=1$ image, since the exact AdS correlator is regular there. In the black hole $\Theta(b)\to0$ as $b\to\infty$, so the summand varies slowly with $\ell$ and the continuum term dominates.

\paragraph{$d=2$ and the two-sided correlator.} In BTZ, $t_c=0$ and $T(\infty)=-\overline{T(\infty)}=i\beta/2$ coincides with the advanced term and with the multiple-reflection and image points, so cancellations are possible. This is consistent with the absence of a bouncing singularity in $d=2$: for the planar BTZ black hole the retarded correlator at zero spatial separation is singular only at $t=in\beta$, $n\in\mathbb Z$, and not at $t=i\beta/2$, although the latter point is present at fixed spatial momentum \cite{Grozdanov:2026cut}.

For $d\ge3$ the point $\overline{T(\infty)}=t_c-i\beta/2$, which is the real-time two-sided bounce point and lies inside the strip $-\beta<\Im t<0$, does not appear in Table~\ref{tab:I:sing}. The large-$k$ correlator is therefore regular there, in agreement with \cite{Fidkowski:2003nf,Festuccia:2005pi}. The singularity at $T(\infty)$, with $\Im t=+\beta/2$, lies outside this strip and is reached by analytic continuation in $t$. A complementary analysis of bounce singularities is given in \cite{Dodelson:2023nnr}.

\subsection{Wightman, retarded and advanced functions}
\label{app:GR:orderings}

Table~\ref{tab:I:sing} refers to $G_+$. The other orderings follow from the KMS condition and from hermiticity,
\begin{equation}
\label{eq:I:orderings}
\begin{gathered}
G_-(t,\theta)=G_+(t-i\beta,\theta)=G_+(-t,-\theta)\,,\\
G_R=i\theta(t)\,\big(G_+-G_-\big)\,,\qquad G_A=-i\theta(-t)\,\big(G_+-G_-\big)\,,
\end{gathered}
\end{equation}
or, in momentum space, $G_-(\omega)=e^{-\beta\omega}G_+(\omega)=G_+(-\omega)$ and $G_A(\omega)=\overline{G_R(\omega)}$ for real $\omega$. For the eikonal term ($N=1$, lowest thermal image) the resulting singular points at $\theta=0$ are listed in Table~\ref{tab:I:orderings} of Section~\ref{sec.six}.
Several remarks follow.
\begin{itemize}
\item $G_+$ and $G_-$ are analytic in the strips $-\beta<\Im t<0$ and $0<\Im t<\beta$ respectively, and none of their singular points lies inside the corresponding strip. The four points $\pm(t_c\pm i\beta/2)$ are shared between the two orderings: $t_c+i\beta/2$ and $-t_c+i\beta/2$ belong to $G_+$, and $-t_c-i\beta/2$ and $t_c-i\beta/2$ to $G_-$.
\item For $\omega>0$ the eikonal term of $G_+$ is $\tilde\rho_0Y$, while that of $G_R$ is $i\tilde\rho_0Y$, see (\ref{eq:I:rho}). Near $t=t_c+i\beta/2$ the Wightman and retarded functions therefore differ only by the constant $-i$. They differ elsewhere: $G_R$ contains, through its negative-frequency part $G_R(-\omega)=\overline{G_R(\omega)}$, the point $t_c-i\beta/2$, which belongs to $G_-$, while $G_+$ contains $-t_c+i\beta/2$, which belongs to $G_A$.
\item The integral over $b\in(-\infty,-b_c)$ with weight $G^{T=0}_+(|\omega|,k)$ and phase $e^{ikS(|b|)}$ is the $\omega<0$ part of $G_A$, equivalently of $G_-$, since $G_A(-|\omega|)=G_R(|\omega|)$ and $G_-(-\omega)=G_+(\omega)$. Its singularity at $-t_c-i\beta/2$ is the image of $t_c+i\beta/2$ under $t\to-t$. For $G_+$ itself the $\omega<0$ region carries the Boltzmann factor $e^{-\beta|\omega|}$, which shifts the corresponding points to $-t_c-3i\beta/2$ and $t_c-3i\beta/2$.
\item The two-sided correlator in the thermofield double state is $G_{12}(t)=\langle{\rm TFD}|\mathcal{O}_1(t)\,\mathcal{O}_2(0)|{\rm TFD}\rangle=Z^{-1}{\rm Tr}\big(e^{-\beta H/2}\mathcal{O}(t)\,e^{-\beta H/2}\mathcal{O}(0)\big)=G_+(t-i\beta/2)$, where $\mathcal{O}_{1,2}$ denote the operator $\mathcal{O}$ in the two copies of the CFT, so its singular points are those of $G_+$ shifted by $+i\beta/2$.
\end{itemize}

\subsection{The coefficient of the singularity}
\label{app:GR:coef}

Here we evaluate the coefficient of the singular term in the Regge limit, where $\delta_\infty$ is absent (Appendix~\ref{app:wkb:delta}). We consider the leading singular term of
\begin{equation}\label{eq:I:Idef}
I(t)\equiv\int_1^\infty db\,\big(b^2-1\big)^{\Delta-\frac d2}\big[b\,t-S(b)\big]^{-2\Delta}
\end{equation}
at $t=T(\infty)$, in general $d$. Let $\tilde{t}=t-T(\infty)$. We split the integral into two parts, $\int_1^{b_0}+\int_{b_0}^\infty$, for some finite $b_0>b_c$.
\begin{enumerate}
\item \emph{Only large $b$ contributes.} For any finite endpoint $b_0$, the integral $\int_1^{b_0}$ in \eqref{eq:I:Idef} is analytic at $\tilde{t}=0$, since $S(b)/b\neq T(\infty)$ at finite $b$.
\item \emph{Large-$b$ integrand.} By \eqref{eq:H:phase0},
\begin{equation}
\big(b^2-1\big)^{\Delta-\frac d2}\big[b\,t-S_+(b)\big]^{-2\Delta}\sim b^{-d}\big[\tilde{t}-c\,b^{-q}\big]^{-2\Delta},\qquad b\to\infty.
\end{equation}
The zero of the bracket, $b_*=(c/\tilde{t})^{1/q}$, collides with the endpoint $b=\infty$ as $\tilde{t}\to0$.
\item \emph{Scaling variable.} Set $w=c\,b^{-q}/\tilde{t}$, so that $\tilde{t}-cb^{-q}=\tilde{t}(1-w)$. Hence
\begin{equation}
\int_{b_0}^\infty\simeq\frac1q\,c^{-d/2}\,\tilde{t}^{\frac d2-2\Delta}\int_0^{w_0}w^{\frac d2-1}(1-w)^{-2\Delta}dw,\qquad|w_0|\to\infty .
\end{equation}
\item \emph{Singular part.} The tail $\int_{w_0}^\infty\propto w_0^{d/2-2\Delta}\propto\tilde{t}^{2\Delta-d/2}$ cancels the prefactor and is analytic in $\tilde{t}$; it combines with $\int_1^{b_0}$ and the $b_0$-dependence cancels. The singular part is obtained from $\int_0^\infty$ along the ray $\arg w=\arg(c/\tilde{t})$.
\item \emph{Evaluation.} The ray passes through $w=1$ only if $t$ lies on the curve $\gamma=\{S_+(b)/b:b>b_c\}$, which ends at $T(\infty)$ along the direction $\arg c$ (Fig.~\ref{fig:J:tplane}). Off $\gamma$, rotate the ray onto the negative real axis, by $+\pi$ if $\arg(c/\tilde{t})\in(0,\pi)$ and by $-\pi$ if $\arg(c/\tilde{t})\in(-\pi,0)$. With $w=ue^{\pm i\pi}$,
\begin{equation}
\int_0^\infty w^{\frac d2-1}(1-w)^{-2\Delta}dw=e^{\pm i\pi d/2}\,B\Big(\frac d2,2\Delta-\frac d2\Big).
\end{equation}
\end{enumerate}
Therefore
\begin{equation}\label{eq:I:Cinf}
I(t)\supset C_\infty\,\big(t-T(\infty)\big)^{\frac d2-2\Delta},\qquad
C_\infty=\frac1q\,c^{-d/2}\,e^{\pm i\pi d/2}\,B\Big(\frac d2,2\Delta-\frac d2\Big).
\end{equation}
The exponent agrees with \eqref{eq:deltaIfinal} and with the stress-tensor resummation of \cite{Ceplak:2024bja}. For even $d$ the two sides of $\gamma$ give the same coefficient, so $\gamma$ carries no discontinuity at this order.

\paragraph{$d=4$.} With $q=\tfrac32$, $B(2,2\Delta-2)=1/[(2\Delta-1)(2\Delta-2)]$ and $c$ from \eqref{eq:H:c4},
\begin{equation}\label{eq:I:C4}
\boxed{C_\infty=\frac{2}{3c^2(2\Delta-1)(2\Delta-2)}=-\frac{12\pi i\,\sqrt\mu}{\Gamma(\frac14)^4\,(2\Delta-1)(2\Delta-2)}\neq0 .}
\end{equation}
In the normalized correlator, by \eqref{eq:I:GNj}, the singular term is
\begin{equation}
G_+(t)\supset\frac{e^{-i\pi(\Delta-1)}}{J_0}\,C_\infty\,\big(t-T(\infty)\big)^{2-2\Delta},\qquad J_0=\frac12B\Big(\frac32,\Delta-1\Big).
\end{equation}
It is non-zero for all $\Delta>1$. At $\Delta=1$ the unnormalized integral $I(t)$ has a logarithmic singular term. In the normalized correlator, however, $J_0$ also diverges as $\Delta\to1$: with $\epsilon=\Delta-1$, $J_0\simeq1/(2\epsilon)$ and $C_\infty\simeq1/(3c^2\epsilon)$, so that $C_\infty/J_0\to2/(3c^2)$ while $\tilde t^{\,2-2\Delta}\to1$, and the coefficient of the logarithm vanishes. The endpoint $\Delta=1$ of the normalized correlator therefore requires a separate treatment. In general $d$, $J_0=\frac12B\big(\frac{d-1}{2},\Delta-\frac d2+1\big)$ is finite and non-zero above the unitarity bound, $\Delta>\frac d2-1$, and $C_\infty$ in (\ref{eq:I:Cinf}) is non-zero for $2\Delta>\frac d2$, since the Beta function has no zeros; at $2\Delta=\frac d2$ the singular term becomes logarithmic.

\paragraph{Beyond the Regge limit.} Beyond the Regge limit, the phase $\delta_\infty(\lambda)$ of Appendix~\ref{app:wkb:delta} makes the $k$-integral \eqref{eq:I:kint} not a pure power. The singular part comes from the scaling region $b\sim b_*$, where $b\,t-S_+\simeq b\,(\tilde{t}-c\,b^{-q})\sim c\,b^{1-q}$, so that the $k$-integral is dominated by $k\sim b^{q-1}$, i.e.\ by $\lambda=O(1)$. The variable $\lambda$ is invariant under the rescaling $\tilde{t}\to s\,\tilde{t}$, $b\to s^{-1/q}b$, $k\to s^{1/q-1}k$, which generates the singular term. Hence the coefficient $C_\infty$ is multiplied by a factor determined by $\delta_\infty(\lambda)$ at $\lambda=O(1)$, which we do not evaluate here; a preliminary evaluation suggests that it does not vanish (see also \cite{Afkhami-Jeddi:2025wra} for the analogous singularity at fixed spatial momentum, where its coefficient is non-vanishing for generic $\Delta$, and \cite{Ceplak:2024bja}, where the coefficient of the singularity of the stress-tensor sector, with the same exponent, was found numerically to be non-vanishing in $d=4$). Provided it does not vanish, the location and the exponent of the singularity are unchanged. The term $N=0$, which contains no reflection, is not affected. Away from the singular points the position-space integral is dominated by finite $b$, where $\lambda\propto k\to\infty$ and $\delta_\infty\to0$.

\paragraph{Comparison with the integral restricted to $b<b_c$.} It is natural to ask whether the singularity could be obtained instead by continuing in $t$ the integral over $1<b<b_c$ alone, onto another sheet. Every such sheet comes from a contour with both endpoints at finite $b$, deformed around the zero $b_*(t)$. In the scaling variable $1/w$ these contours only produce
\begin{equation}
\int_0^1 x^{A-1}(1-x)^{-2\Delta}dx=\frac{\Gamma(A)\,\Gamma(1-2\Delta)}{\Gamma(1-d/2)},\qquad A=2\Delta-\frac d2,
\end{equation}
times integer combinations of phases, whereas the integral up to $b=\infty$ also picks up the range from $b_*$ to infinity. In even $d$, $1/\Gamma(1-d/2)=0$, so no sheet of the restricted integral has the leading singularity $(t-T(\infty))^{d/2-2\Delta}$, while the full integral does. The continued integral is therefore not a sheet of the restricted one. Numerical evidence is in Appendix~\ref{app:num:trunc}.

\subsection{Quasinormal modes}
\label{app:GR:qnm}

By \eqref{eq:H:GRY}, the poles of $G_R$ are at
\begin{equation}\label{eq:I:BS}
Y(\omega,k)=1,
\end{equation}
a Bohr--Sommerfeld condition: the round-trip phase is a multiple of $2\pi$. In pure AdS it gives the normal modes $\omega=\Delta+\ell+2m$ exactly. For AdS black holes this quantization condition, with complex turning points, was derived in \cite{Festuccia:2008zx,Festuccia:2005pi}; see also \cite{Dodelson:2022eiz}. The asymptotics of AdS$_5$ black-hole quasinormal modes were computed in \cite{Starinets:2002br,Nunez:2003eq,Natario:2004jd}; for a review see \cite{Berti:2009kk}.

It is instructive to consider the asymptotic spacing at large $b$. For $b\gg1$, by \eqref{eq:H:XRegge},
\begin{equation}\label{eq:I:Yinf}
Y=e^{-i\pi(\nu+1)}X\ \to\ e^{-i\pi(\nu+1)}\,e^{ikS_+(b)},
\end{equation}
where, by \eqref{eq:H:approach} and \eqref{eq:H:c1a1},
\begin{equation}\label{eq:I:Srecall}
\begin{aligned}
S_+(b)&=b\,T(\infty)+c\,b^{1-q}+\dots,\\
T(b)&=S_+'(b)=T(\infty)+d_1\,b^{-q}+\dots,\qquad d_1=(1-q)\,c=-\frac{d-2}{d}\,c\,,
\end{aligned}
\end{equation}
with $q=2(d-1)/d$. Then \eqref{eq:I:BS} gives
\begin{equation}\label{eq:I:qnm}
\omega_m\,T(\infty)=2\pi m+\pi(\nu+1)-c\,k^q\,\omega_m^{1-q}+\dots,\qquad
\omega_{m+1}-\omega_m\ \to\ \frac{2\pi}{T(\infty)} .
\end{equation}
The asymptotic spacing tests $T(\infty)$, both its real and imaginary parts.\footnote{For a large black hole in $d=4$, $r_H\gg1$, one has $t_c\to\beta/2$, so $T(\infty)\to\tfrac\beta2(1+i)$ and $2\pi/T(\infty)\to2\pi T_H(1-i)$, with $T_H=1/\beta$. This is the known asymptotic spacing of AdS$_5$ black-brane quasinormal modes \cite{Starinets:2002br,Nunez:2003eq}.} At fixed $\ell$ and $|\omega|\to\infty$, as in the numerics of Appendix~\ref{app:num:qnm}, one has $\lambda=k^q\omega^{1-q}\to0$, which lies outside the Regge limit. There $Y$ acquires the factor $e^{i\delta_\infty(\lambda)}$ of \eqref{eq:H:Xdeltainf}, which adds $-\delta_\infty(\lambda_m)$ to the right-hand side of \eqref{eq:I:qnm}, but does not change the asymptotic spacing.

Near a pole, $1-Y\simeq-i\,T(b_m)\,(\omega-\omega_m)$, since $\partial_\omega\big(kS_+(b)\big)=S_+'(b)=T(b)$. Hence, from \eqref{eq:H:GRY},
\begin{equation}\label{eq:I:residue}
{\rm Res}_{\omega=\omega_m}G_R=\frac{i\,\mathcal G_0\,e^{-i\pi\nu}\big(1-e^{2\pi i\nu}\big)}{T(b_m)},
\end{equation}
with $T(b)$ as in \eqref{eq:I:Srecall}. Beyond the Regge limit, the $\lambda$-dependence of $\delta_\infty$ adds $(1-q)\,\delta_\infty'(\lambda_m)\,b_m^{-q}$ to the denominator, a correction of the same order as $d_1b_m^{-q}$; it is visible in the numerics of Appendix~\ref{app:num:qnm}. The poles with these residues, spaced by $2\pi/T(\infty)$, are the resummation of the multiple-reflection series \eqref{eq:I:bounce}; in position space they generate the singularities at $N\,T(\infty)$ for all $N$.

\section{Bulk phase shift -- CFT basics}
\label{appendixh}

In this appendix, we spell out the CFT setup we consider relevant for the thermal correlator in the Regge limit. Our notations/definitions stem from the HHLL correlator introduced and employed in \cite{Kulaxizi:2018dxo}, which we also review here.

The starting point is the thermal correlator on the cylinder\footnote{We set the radius of the sphere $R=1$ for convenience.}:
\begin{equation}
\label{eq:corrdeftry}
G_{\rm Regge}^T(t',\theta')\equiv\langle \mathcal{O}^{cyl}(z\, e^{2\pi i},\bar{z})\mathcal{O}^{cyl}(1)\rangle_T
\end{equation}
where $(z,\bar{z})$ are defined in terms of the coordinates on the cylinder as:
\begin{equation}
\label{eq:zzbarexact}
 z= e^{ i x^+},\qquad \bar z = e^{ i x^-}
\end{equation}
with $x^{\pm} = t\pm \theta$ the cylinder light-cone coordinates and $(t',\theta')$ the Regge variables (\ref{eq:Reggevariables}).
Note that the correlator in (\ref{eq:corrdeftry}) is defined with $z$ analytically continued around zero, which in the standard $x^\pm$ coordinates corresponds to\footnote{We do not expect the physics to change by switching the roles of $z$ and $\bar{z}$, {\it i.e.}, analytically continuing $\bar{z}$ and thus shifting $x^-$ instead.}
\begin{equation}
\label{eq:xplusminusac}
x^+\rightarrow x^++2\pi, \quad x^-\rightarrow x^-
\end{equation}
The Regge limit corresponds to
\begin{equation}
\label{eq:Reggelimit}
|1-z|,|1-\bar{z}|\ll 1, \quad \frac{1-z}{1-\bar{z}}=\text{fixed}\,.
\end{equation}
In this regime, in terms of the Regge variables $x'^\pm=t'\pm\theta'$,
\begin{equation}
z=e^{ix'^+}\simeq 1+ix'^{+}\,,\qquad \bar z=e^{ix'^-}\simeq 1+ix'^{-}\,,
\end{equation}
so that (\ref{eq:corrdeftry}) probes the correlator close to the point shifted by $\pi$ in time and angle with respect to the second operator, where $x'^\pm$ act as light-cone coordinates of the Poincar\'e patch centered there.
It is customary in the Regge limit for the correlator to be multiplied by appropriate wavefunctions. Their precise determination we leave for future work.

The definitions above originate from those in \cite{Kulaxizi:2018dxo} where the same correlator, expressed in terms of a heavy-heavy-light-light four-point function was analyzed to leading order in $\mu$ using Regge theory. For completeness, we review the related discussion here.

Assuming thermalization, (\ref{eq:corrdeftry}) can be equivalently expressed as
\begin{equation}
\label{eq:cylinderhhll}
G_{\rm Regge}^T(t',\theta')= \langle  \mathcal{O}_{H}^{cyl.}(\tau_4=+\infty) |\mathcal{O}_{L}^{cyl.}(t_3,\hat{n}_3) \mathcal{O}_{L}^{cyl.}(t_2,\hat{n}_2) |\mathcal{O}_{H}^{cyl.}(\tau_1=-\infty) \rangle
\end{equation}
where $\tau=-i t$ relates the Euclidean and Lorentzian times and $\hat{n}$ a point on the $(d-1)$-dimensional unit sphere. The heavy operators, with dimensions $\Delta_H\sim C_T$, are placed at $\tau_1=-\infty$ and $\tau_4=+\infty$.
The two light operators, of dimensions $\Delta\sim C_T^0=1$, are inserted close to two reference points, $(P_2,P_3)$ which differ by $\pi$ in Lorentzian time and angular distance on the cylinder. Note that $P_{2,3}$ define Poincar\'e patches centered on them.

We can use the translational symmetry in $t$ together with the rotational symmetry to fix the position of $\mathcal{O}_2$ to be that of $P_2$, and parametrize the position of $\mathcal{O}_3(t_3,\hat{n}_3)$ by
\begin{equation}
\label{eq:diftaus}
-i\delta t=(\tau_3-\tau_2)=-i(t'+\pi), \quad t'\geq 0
\end{equation}
and $\hat{n}_3\equiv \hat{n}$ or equivalently, the angular variable $\theta'$ which measures the distance between the point $P_3$ (which has $\hat{n}(P_3)=-\hat{n}(P_2)$) and $\hat{n}$.
The Regge variables $x'^\pm=t'\pm\theta'$ of (\ref{eq:Reggevariables}) thus parametrize the location of $\mathcal{O}_3$ in terms of its distance from the reference point $P_3$; the separation on the cylinder is $(t'+\pi,\theta'+\pi)$.

The correlator on the cylinder can be transformed to the plane $\mathbb{R}\times\mathbb{R}^{d-1}$ via the usual map from the Euclidean time on the
cylinder to the radial polar coordinate $r = e^\tau$.
Using this, we may write
\begin{equation}
\label{eq:stateheavy}
   \langle  \mathcal{O}_{H}^{cyl.} |\mathcal{O}_{L}^{cyl.} \mathcal{O}_{L}^{cyl.} |\mathcal{O}_{H}^{cyl.} \rangle
 =(r_2 r_3)^{\Delta}
 \lim_{x_4\rightarrow \infty}  (x_4^2)^{\Delta_H}  \langle  \mathcal{O}^{pl}_H(x_4) \mathcal{O}^{pl}_L(x_3) \mathcal{O}^{pl}_L(x_2)\mathcal{O}^{pl}_H(0) \rangle \,,
 \end{equation}
where the factor $(r_2 r_3)^{\Delta}$ appears due to the conformal transformation from the cylinder
to  $\mathbb{R}\times\mathbb{R}^{d-1}$.
This can be further written as
\begin{equation}
\label{eq:fptplane}
 \langle  \mathcal{O}^{cyl.}_H   |  \mathcal{O}^{cyl}_L(x_3) \mathcal{O}^{cyl.}_L(x_2)    |  \mathcal{O}^{cyl}_H \rangle    =
 (r_2 r_3)^{\Delta}\times {\mathcal{A}(u,\upsilon) \over x_{32}^{2\Delta}}
 \end{equation}
 where we have defined the partial amplitude $\mathcal{A}(u,\upsilon)$ which only depends on cross-ratios and can be expanded in conformal blocks. In our conventions and setup, the cross-ratios are
 \begin{equation}
 \label{eq:crossu}
 z \bar z  = u ={x_2^2\over x_3^2}  = e^{2(\tau_2-\tau_3)}    = e^{2 i \delta t}= e^{2 i t'}
 \end{equation}
 and
 \begin{equation}
 \label{eq:crossvprel}
 (1-z ) (1-\bar z ) = \upsilon = {x_{32}^2 \over x_3^2} = 1+ e^{2(\tau_2-\tau_3)} - 2 e^{\tau_2-\tau_3} \hat n_3 \cdot \hat n_2  =
  1 + e^{2 i \delta t}+2e^{i \delta t} \cos \theta'\,.
  \end{equation}
Substituting (\ref{eq:diftaus}) this becomes
\begin{equation}
\label{eq:crossv}
(1-z ) (1-\bar z ) =  1 + e^{2 i t'} - 2 e^{i t'} \cos \theta' \, .
\end{equation}
We can solve (\ref{eq:crossu}) and (\ref{eq:crossv}) exactly to obtain:
 \begin{equation}
 \label{eq:zzbarsolved}
 z= e^{ i x'^+},\qquad \bar z = e^{ i x'^-}
 \end{equation}
 with $x'^\pm = t'\pm \theta'$, in agreement with (\ref{eq:zzbarexact}), since $e^{ix'^\pm}=e^{ix^\pm}$.

The Lorentzian correlator in the Regge limit as discussed in \cite{Cornalba:2006xk, Cornalba:2006xm, Cornalba:2007fs, Cornalba:2007zb}, corresponds to the case where the deviation from $P_3$ is small. The configuration can be obtained by starting from the correlator where $\mathcal{O}_3$ and $\mathcal{O}_2$ are inserted close to each other on a spatial circle and at the same time. To reach the configuration where $\mathcal{O}_3$ is inserted close to $P_3$, we need to shift $x^+ \rightarrow x^+ + 2\pi$ which corresponds to $z \rightarrow  e^{2 \pi i } \ z$, as anticipated.
Note that in (\ref{eq:zzbarsolved}) we could have had the opposite assignments of $z ,\bar z$ --  there is a complete symmetry of the correlator, before the analytic continuation.
The bulk phase shift $\delta(p)$ is then defined via the Fourier transform of the correlator in (\ref{eq:fptplane}) in
the Regge limit where $x'^\pm$ are small.\footnote{Note that our $x$
here differs by a minus sign from  that in \cite{Kulaxizi:2017ixa}.}

The $i\epsilon$-prescription in the Fourier transform will be inherited from the ordering of the operators, which translates into sending $\delta t= t_3-t_2\rightarrow \delta t -i\epsilon$ with $\epsilon >0$, as in Section~\ref{sec.three}. This then becomes $x^2 \rightarrow x^2 +i\epsilon x^0$, after a rescaling of $\epsilon$.

\section{Relations between different Regge parametrizations in the CFT correlator}
\label{appendixf}

Here we make contact with parametrizations used in previous work considering four-point functions in the Regge limit. Following Appendix B of  \cite{Kulaxizi:2017ixa}, we describe the Lorentzian cylinder using embedding-space coordinates $Z^A$. A point is represented by a null ray in $\mathbb{R}^{2,d}$, so $Z$ and $\lambda Z$ describe the same point.

\begin{equation}
-(Z^{-1})^2 - (Z^0)^2 + \sum_{i=1}^{d} (Z^i)^2 = 0, \quad Z \sim \lambda Z.
\end{equation}
We use the rescaling freedom to choose the section $(Z^0)^2+(Z^{-1})^2=1$. The null condition then implies $\sum_{i=1}^{d}(Z^i)^2=1$, so the spatial coordinates define a unit vector $n\in S^{d-1}$. This gives the parametrization:
\begin{equation}
Z^{0} + iZ^{-1} = e^{it}, \quad Z^i = n^i, \quad (Z^0)^2 + (Z^{-1})^2 = 1,
\end{equation}
For simplicity we restrict to $d=4$ in this appendix; the generalization to arbitrary $d$ is straightforward. For $d=4$, the unit vector $n\in S^3$ can be parametrized by the three angles $(\theta,\phi,\psi)$. Thus, $t$ is the cylinder time, while $(\theta,\phi,\psi)$ describe the spatial sphere.
\begin{equation}
\begin{aligned}
Z^{-1} &= \sin t, \\
Z^0 &= \cos t, \\
Z^1 &= \sin\theta \sin\phi \cos\psi=n^1, \\
Z^2 &= \sin\theta \sin\phi \sin\psi=n^2, \\
Z^3 &= \sin\theta \cos\phi=n^3, \\
Z^4 &= \cos\theta=n^4
\label{embedding}
\end{aligned}
\end{equation}

\subsection{Parametrization by cylinder insertion points\label{C1}}

In this subsection, we adopt the choice of insertion points from \cite{Kulaxizi:2018dxo}. Following eq.~(3.3) of \cite{Kulaxizi:2018dxo}, setting $\tau_3 = \tau_2 - i\delta t$ with $\delta t = \pi + x^0$ ($x^0 \ge 0$), and choosing $\tau_2 = 0$, we have:



\begin{equation}
\begin{aligned}
O_H^{(1)} &: \quad (\tau_1, \hat{n}_1) = (-\infty, \hat{n}_1), \\
O_L^{(2)} &: \quad (\tau_2, \hat{n}_2) = (0, \hat{n}_2), \\
O_L^{(3)} &: \quad (\tau_3, \hat{n}_3) = \left( -i(\pi + x^0),\hat{n}_3 \right), \\
O_H^{(4)} &: \quad (\tau_4, \hat{n}_4) = (+\infty, \hat{n}_4).
\end{aligned}
\end{equation}
$\tau$ is the Euclidean time, related to Lorentzian time by $t = i\tau$.  And here \(\hat n_i\) specifies the position of the \(i\)-th operator insertion on the spatial sphere \(S^{d-1}\) of the cylinder.

\begin{equation}
\hat{n}^A = (\sin\theta \sin\phi \cos\psi, \, \sin\theta \sin\phi \sin\psi, \, \sin\theta \cos\phi, \, \cos\theta),\quad A=1,2,3,4
\end{equation}
where
\begin{equation}
0 \le \theta \le \pi, \quad 0 \le \phi \le \pi, \quad 0 \le \psi < 2\pi.
\end{equation}
It is important to distinguish the reference point $P_3$
from the actual position of the third insertion.  With
$\tau_2=0$, the two reference points may be written as
\[
 P_2=(0,\hat n_2),
 \qquad
 P_3=(-i\pi,\hat n(P_3)),
 \qquad
 \hat n(P_3)=-\hat n_2.
\]
More explicitly, if $\hat{n}_2 = \hat{n}(\theta_2, \phi_2, \psi_2)$, then, in general, $\hat{n}(P_3)$ can be written as:
\begin{equation}
\hat{n}(P_3) = \hat{n}(\pi - \theta_2, \pi - \phi_2, \psi_2 + \pi) = -\hat{n}_2,
\end{equation}
To make the angular separation explicit, we use the \(SO(4)\) rotational symmetry to choose $\theta=\phi=\frac{\pi}{2}$ for all three directions \(\hat n_2\), \(\hat n(P_3)\), and \(\hat n_3\). This choice places the three directions on the same great circle and does not impose any physical restriction. On this great circle,
\begin{equation}
\hat{n}\left( \frac{\pi}{2}, \frac{\pi}{2}, \psi \right) = (\cos \psi, \sin \psi, 0, 0),
\end{equation}
so that the position along the circle is described entirely by the \(2\pi\)-periodic coordinate \(\psi\). The actual third insertion is separated from \(P_3\) by the angle \(\alpha\). Therefore, on this great circle, $\hat{n}_2$, $\hat{n}(P_3)$, and $\hat{n}_3$ can be written as:

\begin{equation}
\hat{n}_2 = \hat{n}\left( \frac{\pi}{2}, \frac{\pi}{2}, \psi_2 \right) = (\cos\psi_2, \sin\psi_2, 0, 0).
\end{equation}

\begin{equation}
\hat{n}(P_3) = \hat{n}\left( \frac{\pi}{2}, \frac{\pi}{2}, \psi_2+\pi \right) = (-\cos\psi_2, -\sin\psi_2, 0, 0).
\end{equation}

\begin{equation}
\hat{n}_3 = \hat{n}\left( \frac{\pi}{2}, \frac{\pi}{2}, \psi_2+\pi-\alpha \right) = (-\cos(\psi_2-\alpha), -\sin(\psi_2-\alpha), 0, 0).
\end{equation}
It follows immediately that
\begin{equation}
\hat{n}(P_3) \cdot \hat{n}_3 = \cos\alpha,
\end{equation}

\begin{equation}
\hat{n}_2 \cdot \hat{n}_3 =  -\cos\alpha.
\end{equation}
Equivalently, the angular separation between the actual directions \(\hat n_2\) and \(\hat n_3\) is \(\pi-\alpha\). The adapted frame is introduced only to make the angular geometry explicit. In the calculation below, we keep \(\hat n_2\) and \(\hat n_3\) general and use only the coordinate-independent relation $\hat{n}_2 \cdot \hat{n}_3 =  -\cos\alpha$

For the coordinate transformation below, it is convenient to separate the first three components from the fourth embedding component,
\begin{equation}
\hat{n} = (\vec{n}, n^4), \quad \vec{n} = (n^1, n^2, n^3), \quad n^4 = \cos\theta,
\end{equation}
The Minkowski coordinates $y^\mu$ are obtained from the embedding coordinates (\ref{embedding}) by
\begin{equation}
y^\mu=\frac{(Z^{-1},Z^1,Z^2,Z^3)}{Z^0+Z^4}\,,\qquad\text{i.e.}\qquad y^0=\frac{\sin t}{\cos t+\cos\theta}\,,\quad \vec y=\frac{\vec n}{\cos t+\cos\theta}\,,
\end{equation}
see also Table~\ref{tab:coordinate-comparison}. Substituting the four cylinder insertion points, with $t=i\tau$, yields:
\begin{equation}
\begin{aligned}
y_1 &= (-i, \vec{0}), \\
y_2 &= \left( 0, \frac{\vec{n}_2}{1 + n_2^4} \right), \\
y_3 &= \left( \frac{\sin x^0}{\cos x^0 - n_3^4}, \, -\frac{\vec{n}_3}{\cos x^0 - n_3^4} \right), \\
y_4 &= (i, \vec{0}).
\end{aligned}
\end{equation}

We now compute the conformal cross-ratios directly in the \(y\) coordinate.
\begin{equation}
u=\frac{y^2_{12}y^2_{34}}{y^2_{13}y^2_{24}}
\end{equation}
\begin{equation}
\upsilon=\frac{y^2_{14}y^2_{23}}{y^2_{13}y^2_{24}}
\end{equation}
where
\begin{equation}
y_{ij}^2 = -(y_i^0 - y_j^0)^2 + |\vec{y}_i - \vec{y}_j|^2.
\end{equation}
We first evaluate the four distances entering the first cross-ratio.
\begin{equation}
y_{12}^2=y_{24}^2 = \frac{2}{1 + n_2^4}.
\end{equation}

\begin{equation}
y_{13}^2 = \frac{2e^{-ix^0}}{\cos x^0 - n_3^4}.
\end{equation}

\begin{equation}
y_{34}^2 = \frac{2e^{ix^0}}{\cos x^0 - n_3^4}.
\end{equation}
Substituting these expressions into $u$ yields:
\begin{equation}
u = e^{2ix^0}
\end{equation}

We next turn to the second cross-ratio.
\begin{equation}
y_{14}^2 = 4.
\end{equation}

\begin{equation}
y_{23}^2 = \frac{2(\cos x^0 + \hat{n}_2 \cdot \hat{n}_3)}{(1 + n_2^4)(\cos x^0 - n_3^4)}
\end{equation}
Substituting these expressions into $\upsilon$ yields:
\begin{equation}
\upsilon = 2e^{ix^0} (\cos x^0 + \hat{n}_2 \cdot \hat{n}_3)=1 + e^{2ix^0} + 2e^{ix^0} \hat{n}_2 \cdot \hat{n}_3=1 + e^{2ix^0} - 2e^{ix^0} \cos\alpha
\end{equation}

where $\alpha$ is the angle between $\hat{n}(P_3) = -\hat{n}_2$ and $\hat{n}_3$. In the notation of Section~\ref{sec.three}, $x^0$ and $\alpha$ correspond to the Regge variables $t'$ and $\theta'$ of (\ref{eq:Reggevariables}).

Thus, although the individual \(y_i\) depend on the particular stereographic-type parametrization and on the fourth components \(n_2^4\) and \(n_3^4\), these quantities cancel from the conformal cross-ratios. The final result depends only on the Lorentzian time displacement \(x^0\) and on the relative angular position of the two light insertions, in agreement with \cite{Kulaxizi:2018dxo}.

Using the definition of $u=z \bar{z}$ and $\upsilon=(1-z)(1-\bar{z})$ we could solve for:

\begin{equation}
z = e^{ix^+}, \quad \bar{z} = e^{ix^-}
\label{zzbar}
\end{equation}
where $x^+ = x^0 + \alpha$, $x^- = x^0 - \alpha$.

\subsection{Parametrization in light-patch coordinates\label{C2}}
\begin{figure}[htbp]
    \centering
    \includegraphics[width=0.38\textwidth]{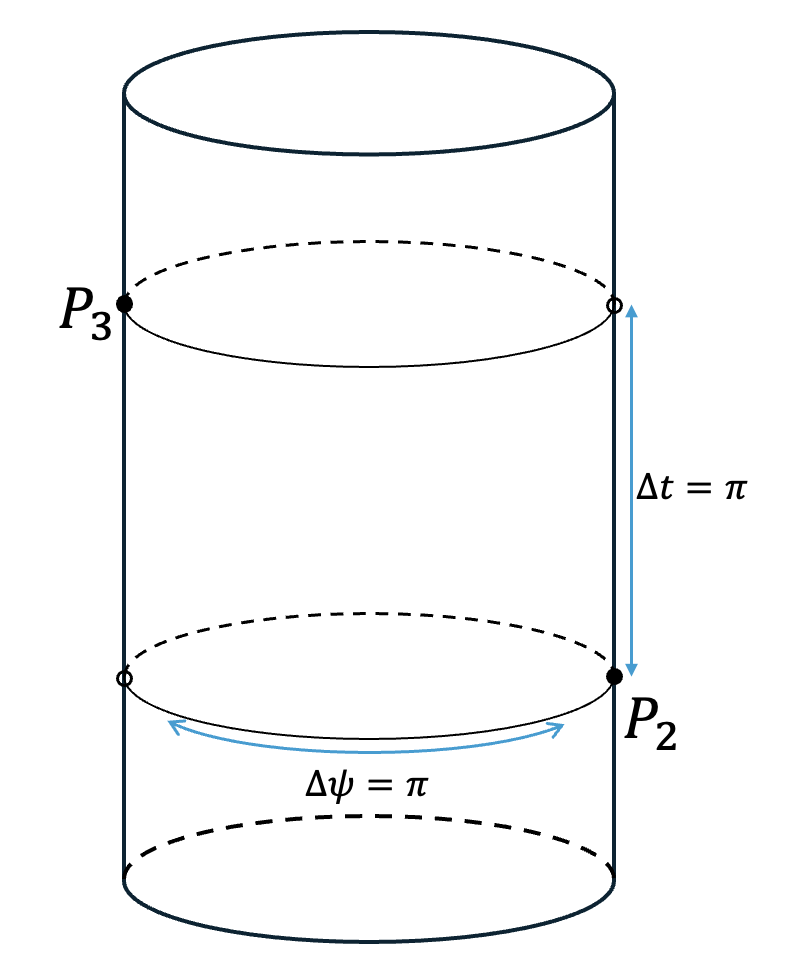}
    \caption{Reference configuration of $P_2$ and $P_3$ on the
    Lorentzian cylinder.}
    \label{cylinder}
\end{figure}

Here, we keep the heavy insertions unchanged,
\begin{equation}
\begin{aligned}
y_1 = (-i, \vec{0}),\qquad y_4 = (i, \vec{0})\,,
\end{aligned}
\end{equation}
but consider an alternative parametrization for the two light insertions inspired by Section 2 of \cite{Kulaxizi:2017ixa}. For the two reference points, we continue to adopt those given in eq.~(2.3) of \cite{Kulaxizi:2017ixa}:
\begin{equation}
\begin{aligned}
P_2 : \quad Z^{-1} &= -1, \quad Z^0 = 0, \quad Z^1 = 1, \\[1.5ex]
P_3 : \quad Z^{-1} &= 1, \quad Z^0 = 0, \quad Z^1 = -1,
\end{aligned}
\end{equation}
According to the mapping between embedding coordinates and cylindrical coordinates in Equation (\ref{embedding}), we obtain:
\begin{equation}
t_2 = -\frac{\pi}{2}, \qquad \hat{n}(P_2) = (1, 0, \ldots, 0).
\end{equation}

\begin{equation}
t_3 = \frac{\pi}{2},\qquad \hat{n}(P_3) = (-1, 0, \ldots, 0).
\end{equation}
Therefore, the two reference points have the following relationship:

\begin{equation}
t_3 - t_2 = \pi, \quad \hat{n}(P_3) = -\hat{n}(P_2).
\end{equation}
Thus $\hat{n}(P_2)\cdot\hat{n}(P_3)=-1$: the two spatial directions are antipodal, with angular separation $\pi$ on $S^3$. The corresponding reference configuration on the Lorentzian cylinder is shown in
Figure~\ref{cylinder}.

The actual insertion points, which deviate from the reference points \(P_2\) and \(P_3\), are parametrized as in Eq.~(2.17) of \cite{Kulaxizi:2017ixa}.
\begin{equation}
y_2^-=-y_3^- = -\frac{2}{\bar{x}^+}, \quad y_2^+ =-y_3^+= -\frac{\bar{x}^2}{2\bar{x}^+}, \quad y_{2\perp} =y_{3\perp}= \frac{\bar{x}_{\perp}}{\bar{x}^+}
\end{equation}
Here
\begin{equation}
\bar{x}^\mu = (\bar{x}^+, \bar{x}^-, \bar{x}_{\perp}), \quad \bar{x}^2 = -\bar{x}^+ \bar{x}^- + \bar{x}_{\perp}^2,\quad \bar{x}^+ + \bar{x}^-=2\bar{x}^0
\end{equation}
with $y^\pm=y^0\pm y^1$ and $\bar x^\pm=\bar x^0\pm\bar x^1$; $y_{j\perp}$ and $\bar x_\perp$ denote the two transverse components.

We now compute the conformal cross-ratios. First, calculate $y_{ij}$ in $u$ respectively:
\begin{equation}
y_{34}^2 = y_{12}^2 = \frac{4\bar{x}^0 + i(4 + \bar{x}^2)}{2\bar{x}^+}
\end{equation}

\begin{equation}
y_{24}^2 = y_{13}^2 = \frac{4\bar{x}^0 - i(4 + \bar{x}^2)}{2\bar{x}^+}
\end{equation}
The common factors of \(\bar x^+\) cancel in the definition of \(u\). Therefore
\begin{equation}
u = \left[ \frac{1 + \frac{\bar{x}^2}{4} - i\bar{x}^0}{1 + \frac{\bar{x}^2}{4} + i\bar{x}^0} \right]^2
\end{equation}

Next, calculate the components of $\upsilon$:
\begin{equation}
y_{14}^2 = 4.
\end{equation}

\begin{equation}
y_{23}^2 = -\frac{4\bar{x}^2}{(\bar{x}^+)^2}
\end{equation}

Thus
\begin{equation}
\upsilon = \frac{4\bar{x}^2}{\left( 1 + \frac{\bar{x}^2}{4} + i\bar{x}^0 \right)^2}
\end{equation}

To relate this result directly to the two-vector parametrization of \cite{Kulaxizi:2017ixa}, we note that under the light-patch transformation given below Table~\ref{tab:coordinate-comparison} the fixed positions \(y_1=(-i,\vec0)\) and \(y_4=(i,\vec0)\) are mapped to themselves, $\hat x_1=(-i,\vec 0)$ and $\hat x_4=(i,\vec 0)$. Their difference defines the fixed auxiliary vector
\begin{equation}
\begin{aligned}
x \equiv \hat x_1-\hat x_4= (-2i, \vec{0})
\end{aligned}
\end{equation}
For this vector, we have:
\begin{equation}
\begin{aligned}
x^2 &= 4 \\
x \cdot \bar{x} &= -(-2i)\bar{x}^0 = 2i\bar{x}^0.
\end{aligned}
\end{equation}
Substituting the above expressions into Eq.~(2.19) of \cite{Kulaxizi:2017ixa} yields the same cross-ratios \(u\) and \(\upsilon\) as those obtained above.

Using the definition of $u=z \bar{z}$ and $\upsilon=(1-z)(1-\bar{z})$ we could solve for:

\begin{equation}
z = \left[ \frac{1 - \frac{i}{2} \left(\bar{x}^0 - \sqrt{(\bar{x}^0)^2 + \bar{x}^2}\right)}{1 + \frac{i}{2} \left(\bar{x}^0 - \sqrt{(\bar{x}^0)^2 + \bar{x}^2}\right)} \right]^2\approx 1 - 2i \left( \bar{x}^0 - \sqrt{(\bar{x}^0)^2 + \bar{x}^2} \right)
\end{equation}

\begin{equation}
\bar{z} = \left[ \frac{1 - \frac{i}{2} \left(\bar{x}^0 + \sqrt{(\bar{x}^0)^2 + \bar{x}^2}\right)}{1 + \frac{i}{2} \left(\bar{x}^0 + \sqrt{(\bar{x}^0)^2 + \bar{x}^2}\right)} \right]^2\approx  1 - 2i \left( \bar{x}^0 + \sqrt{(\bar{x}^0)^2 + \bar{x}^2} \right)
\end{equation}
For real $\bar x^\mu$ the moduli of the above two expressions are both 1, so they can be written as:

\begin{equation}
z = e^{i\tilde{x}^+}, \quad \bar{z} = e^{i\tilde{x}^-}
\end{equation}
where
\begin{equation}
\tilde{x}^+ = -4 \arctan \left[ \frac{\bar{x}^0 - \sqrt{(\bar{x}^0)^2 + \bar{x}^2}}{2} \right]
\end{equation}

\begin{equation}
\tilde{x}^- = -4 \arctan \left[ \frac{\bar{x}^0 + \sqrt{(\bar{x}^0)^2 + \bar{x}^2}}{2} \right]
\end{equation}
Comparing the above expressions with equation (\ref{zzbar}), we obtain:

\begin{equation}
x^0 + \alpha = -4 \arctan \left[ \frac{\bar{x}^0 - \sqrt{(\bar{x}^0)^2 + \bar{x}^2}}{2} \right]
\end{equation}

\begin{equation}
x^0 - \alpha = -4 \arctan \left[ \frac{\bar{x}^0 + \sqrt{(\bar{x}^0)^2 + \bar{x}^2}}{2} \right]
\end{equation}
Rearranging these, we obtain the following relationship:
\begin{equation}
\bar{x}^0 = -\tan \left( \frac{x^0 + \alpha}{4} \right) - \tan \left( \frac{x^0 - \alpha}{4} \right)
\end{equation}

\begin{equation}
\bar{x}^2 = -4 \tan \left( \frac{x^0 + \alpha}{4} \right) \tan \left( \frac{x^0 - \alpha}{4} \right).
\end{equation}
The above equations give the explicit relation between the parametrizations used in Sections~\ref{C1} and~\ref{C2}



We have $z = 1 - \sigma e^{\rho}$ and $ \bar{z} = 1 - \sigma e^{-\rho}.$ It is easy to get $\upsilon = (1 - z)(1 - \bar{z}) = \sigma^2$ and $e^{2\rho} = \frac{1 - z}{1 - \bar{z}}.$ Therefore:

\begin{equation}
\sigma^2 = \frac{4\bar{x}^2}{\left( 1 + \frac{\bar{x}^2}{4} + i\bar{x}^0 \right)^2}\approx 4\bar{x}^2
\label{sigma}
\end{equation}
\begin{equation}
e^{2\rho} = \frac{\bar{x}^0 - \sqrt{(\bar{x}^0)^2 + \bar{x}^2}}{\bar{x}^0 + \sqrt{(\bar{x}^0)^2 + \bar{x}^2}} \left[ \frac{1 + \frac{i}{2} \left( \bar{x}^0 + \sqrt{(\bar{x}^0)^2 + \bar{x}^2} \right)}{1 + \frac{i}{2} \left( \bar{x}^0 - \sqrt{(\bar{x}^0)^2 + \bar{x}^2} \right)} \right]^2\approx \frac{\bar{x}^0 - \sqrt{(\bar{x}^0)^2 + \bar{x}^2}}{\bar{x}^0 + \sqrt{(\bar{x}^0)^2 + \bar{x}^2}}
\label{rho}
\end{equation}

In the limit $\bar{x}^{\mu}\to0$ with its direction held fixed, the two light insertions approach the reference points $P_2$ and $P_3$. Equation~(\ref{sigma}) implies that $\sigma\to0$, while Equation~(\ref{rho}) is invariant under a common rescaling of all components of $\bar{x}^{\mu}$. Therefore, $\rho$ remains fixed as $\bar{x}^{\mu}\to0$, provided that the ratios between its components are held fixed. Consequently, $z,\bar{z}\to1$ with $\rho$ fixed. Thus, the parametrization of \cite{Kulaxizi:2017ixa} reproduces the same Regge kinematics as that used in Section~\ref{C1}. In both cases the Regge limit also requires the analytic continuation $z\to ze^{2\pi i}$ before $z,\bar z\to1$, see (\ref{eq:xplusminusac}).


\subsection{Comparison table}

A point on the Euclidean cylinder $\mathbb{R}_{\tau}\times S^3$ is labeled by $(\tau,\hat n)$, where $\tau$ is Euclidean time and $\hat n$ is a unit vector on the spatial sphere:
\[
\hat n
=
\bigl(
\sin\theta\sin\phi\cos\psi,\,
\sin\theta\sin\phi\sin\psi,\,
\sin\theta\cos\phi,\,
\cos\theta
\bigr)
\]
The Lorentzian cylinder time $t$ is related to $\tau$ by $t=i\tau$. The corresponding cylinder metrics are
\[
ds_{\mathrm{cyl},E}^2=d\tau^2+d\Omega_3^2,
\qquad
ds_{\mathrm{cyl},L}^2=-dt^2+d\Omega_3^2.
\]

We denote by $X^A$ the Euclidean Cartesian coordinates obtained from the radial map $X^A=e^\tau n^A$. These coordinates become complex after the continuation $\tau=-it$. By contrast, $y^\mu$ are Minkowski Cartesian coordinates, with
\[
ds_y^2=-(dy^0)^2+\sum_{a=1}^{3}(dy^a)^2.
\]
The angle $\alpha$ is measured from the antipodal direction $-\hat n_2$ to $\hat n_3$, so that $\hat n_2\cdot\hat n_3=-\cos\alpha$.

\begingroup
\small
\setlength{\tabcolsep}{4pt}
\renewcommand{\arraystretch}{1.35}

\newlength{\coordavailablewidth}
\setlength{\coordavailablewidth}{%
    \dimexpr\textwidth-8\tabcolsep-5\arrayrulewidth\relax
}

\newcommand{\coordmath}[1]{%
    \par
    \begingroup
    \centering
    \vspace*{4pt}%
    \noindent
    \(\displaystyle #1\)%
    \par
    \vspace*{4pt}%
    \endgroup
}

\newcommand{\coordtablehead}{%
    \hline
    &
    \textbf{Radial map~\cite{Kulaxizi:2018dxo}}

    &
    \textbf{Section~\ref{C1} \(y\) coordinates}

    &
    \textbf{Section~\ref{C2}  \(\bar x\) parametrization}

    \\
    \hline
}

\begin{longtable}{
    |>{\raggedright\arraybackslash}m{0.13\coordavailablewidth}
    |>{\centering\arraybackslash}m{0.29\coordavailablewidth}
    |>{\centering\arraybackslash}m{0.29\coordavailablewidth}
    |>{\centering\arraybackslash}m{0.29\coordavailablewidth}|
}

\caption{Comparison of the coordinate maps and insertion points.}
\label{tab:coordinate-comparison}\\

\coordtablehead
\endfirsthead

\coordtablehead
\endhead

\hline
\multicolumn{4}{|r|}{Continued on the next page}\\
\hline
\endfoot

\hline
\endlastfoot

Heavy operators on the cylinder
&
\multicolumn{3}{
    >{\centering\arraybackslash}
    m{\dimexpr
        0.87\coordavailablewidth
        +4\tabcolsep+2\arrayrulewidth
      \relax}|
}{
    \coordmath{
        \begin{aligned}
        O_H^{(1)}:\quad
        &(\tau_1,\hat n_1)=(-\infty,\hat n_1),\\[4pt]
        O_H^{(4)}:\quad
        &(\tau_4,\hat n_4)=(+\infty,\hat n_4).
        \end{aligned}
    }
}
\\
\hline

Light operators on the cylinder
&
\multicolumn{2}{
    >{\centering\arraybackslash}
    m{\dimexpr
        0.58\coordavailablewidth
        +2\tabcolsep+\arrayrulewidth
      \relax}|
}{
    \coordmath{
        \begin{aligned}
        O_L^{(2)}:\quad
        &(\tau_2,\hat n_2)=(0,\hat n_2),\\[4pt]
        O_L^{(3)}:\quad
        &(\tau_3,\hat n_3)
          =\left(-i(\pi+x^0),\hat n_3\right),\\[4pt]
        &\hat n_2\cdot\hat n_3=-\cos\alpha.
        \end{aligned}
    }
    \par
    The reference points correspond to
    \(x^0=0\) and \(\alpha=0\).
}
&
Reference points:
\par
\coordmath{
    \begin{aligned}
    P_2:
    &(t_2,\hat n(P_2))=(-\pi/2,\mathbf e_1),\\[4pt]
    P_3:
    &(t_3,\hat n(P_3))=(\pi/2,-\mathbf e_1).
    \end{aligned}
}
\par
The actual points are parametrized by \(\bar x^\mu\).
\\
\hline

Cylinder-to-plane map
&
\coordmath{
    \begin{aligned}
    X^1&=e^\tau\sin\theta
          \sin\phi\cos\psi,\\[3pt]
    X^2&=e^\tau\sin\theta
          \sin\phi\sin\psi,\\[3pt]
    X^3&=e^\tau\sin\theta\cos\phi,\\[3pt]
    X^4&=e^\tau\cos\theta.
    \end{aligned}
}
&
\multicolumn{2}{
    >{\centering\arraybackslash}
    m{\dimexpr
        0.58\coordavailablewidth
        +2\tabcolsep+\arrayrulewidth
      \relax}|
}{
    \coordmath{
        \begin{aligned}
        y^0&=\frac{\sin t}
                  {\cos t+\cos\theta},\\[4pt]
        y^1&=\frac{\sin\theta\sin\phi\cos\psi}
                  {\cos t+\cos\theta},\\[4pt]
        y^2&=\frac{\sin\theta\sin\phi\sin\psi}
                  {\cos t+\cos\theta},\\[4pt]
        y^3&=\frac{\sin\theta\cos\phi}
                  {\cos t+\cos\theta}.
        \end{aligned}
    }
}
\\
\hline

Metric
&
\coordmath{
    ds_X^2=e^{2\tau}ds_{\mathrm{cyl},E}^2.
}
&
\multicolumn{2}{
    >{\centering\arraybackslash}
    m{\dimexpr
        0.58\coordavailablewidth
        +2\tabcolsep+\arrayrulewidth
      \relax}|
}{
    \coordmath{
        ds_y^2=
        \frac{ds_{\mathrm{cyl},L}^2}
             {(\cos t+\cos\theta)^2}.
    }
}
\\
\hline

Heavy operators on the plane
&
\coordmath{
    \begin{aligned}
    X_1&=0,\\[3pt]
    X_4&=\infty.
    \end{aligned}
}
&
\multicolumn{2}{
    >{\centering\arraybackslash}
    m{\dimexpr
        0.58\coordavailablewidth
        +2\tabcolsep+\arrayrulewidth
      \relax}|
}{
    \coordmath{
        \begin{aligned}
        y_1&=(-i,\vec 0),\\[3pt]
        y_4&=(i,\vec 0).
        \end{aligned}
    }
}
\\
\hline

Light operators on the plane
&
\coordmath{
    \begin{aligned}
    X_2&=\hat n_2,\\[4pt]
    X_3&=-e^{-ix^0}\hat n_3.
    \end{aligned}
}
&
\coordmath{
    \begin{aligned}
    y_2^0&=0,\\[4pt]
    \vec y_2&=\frac{\vec n_2}{1+n_2^4},\\[4pt]
    y_3^0&=\frac{\sin x^0}
                   {\cos x^0-n_3^4},\\[4pt]
    \vec y_3&=\frac{-\vec n_3}
                    {\cos x^0-n_3^4}.
    \end{aligned}
}
&
\coordmath{
    \begin{aligned}
    y_2^-=-y_3^-&=-\frac{2}{\bar x^+},\\[4pt]
    y_2^+=-y_3^+&=-\frac{\bar x^2}{2\bar x^+},\\[4pt]
    y_{2\perp}=y_{3\perp}
        &=\frac{\bar x_\perp}{\bar x^+}.
    \end{aligned}
}
\\
\hline

Cross ratios
&
\multicolumn{2}{
    >{\centering\arraybackslash}
    m{\dimexpr
        0.58\coordavailablewidth
        +2\tabcolsep+\arrayrulewidth
      \relax}|
}{
    \coordmath{
        \begin{aligned}
        u&=e^{2ix^0},\\[4pt]
        \upsilon&=1+e^{2ix^0}
           -2e^{ix^0}\cos\alpha.
        \end{aligned}
    }
}
&
{\footnotesize
\coordmath{
    \begin{aligned}
    u&=
    \left(
    \frac{1+\bar x^2/4-i\bar x^0}
         {1+\bar x^2/4+i\bar x^0}
    \right)^2,\\[6pt]
    \upsilon&=
    \frac{4\bar x^2}
         {\left(1+\bar x^2/4+i\bar x^0\right)^2}.
    \end{aligned}
}
}
\\

\end{longtable}
\endgroup

Here $y^\pm$, $\bar x^\pm$, $y_{j\perp}$ and $\bar x_\perp$ are as in Section~\ref{C2}; in the cylinder-to-$y$ map, $\vec y$ denotes the full three-dimensional spatial vector.

The parametrization in the last column follows from the light-patch conformal transformation
\[
(\hat x_j^+,\hat x_j^-,\hat x_{j\perp})
=
-\frac{(1,y_j^2,y_{j\perp})}{y_j^-},
\qquad j=2,3,
\]
followed by the choice
$\hat x_2=\bar x/2$ and $\hat x_3=-\bar x/2$.

Matching the cross ratios near the reference configuration gives
\[
\begin{aligned}
\bar x^0
&=
-\tan\frac{x^0+\alpha}{4}
-\tan\frac{x^0-\alpha}{4},\\
\bar x^2
&=
-4\tan\frac{x^0+\alpha}{4}
  \tan\frac{x^0-\alpha}{4}.
\end{aligned}
\]

\section{Integrals in general spacetime dimensions}
\label{appendixb}

Here we express the relevant integrals (time delay and angular deflection) in terms of Lauricella functions, which generalize the hypergeometric function, for $1<b<b_c$, with $r_T^>$ the outer turning point and $\mathbf b=(b^2-1)^{-1/2}$ of \eqref{eq:bdef}. To do so, we first change variables to $y={r_T^>/ r}$ and express the relevant polynomials as follows:
\begin{equation}
\label{eq:polynomial1}
\begin{aligned}
b^2-{f(r)\over r^2} &= {1\over (r_T^>)^d } \left(\mathbf{b}^{-2} (r_T^>)^d -(r_T^>)^{d-2}y^{2} +\mu\, y^d\right)=\\
&={\mu\over (r_T^>)^d } \left({\mathbf{b}^{-2} (r_T^>)^d\over\mu} - {(r_T^>)^{d-2}\over \mu}\, y^2+y^d  \right)=\\
&=(-1)^{d}\, {\mu\over (r_T^>)^d }\, (1-y)(y_T^>-y) \prod_{i=1}^{d-2} (y_i-y)\,,
\end{aligned}
\end{equation}
where $(1,  y_T^> , y_i)$ are the solutions of
\begin{equation}
{\mathbf{b}^{-2} (r_T^>)^d\over\mu} - {(r_T^>)^{d-2}\over \mu}\, y^2+y^d =0\,,
\end{equation}
with $y_T^>=r_T^>/r_T^<>1$ and $y_i\in \mathbb{C}$. Similarly,
\begin{equation}
\label{eq:polynomial2}
\begin{aligned}
f(r)&=r^2+1-{\mu\over r^{d-2}}=y^{-2}( r_T^>)^2 \left(1+ y^2 (r_T^>)^{-2}-y^d \mu\, (r_T^>)^{-d}\right)=\\
&=y^{-2}( r_T^>)^{-d+2} \mu  \left(\mu^{-1}(r_T^>)^d + \mu^{-1}(r_T^>)^{d-2}\, y^2-y^d\right)=\\
&=(-1)^{d+1} \,y^{-2}( r_T^>)^{-d+2} \mu \, \prod_{j=1}^d (\tilde y_j-y)\,,
\end{aligned}
\end{equation}
where $\tilde y_j$, $j=1,\dots,d$, are the solutions of
\begin{equation}
\mu^{-1}(r_T^>)^d + \mu^{-1}(r_T^>)^{d-2}\, y^2-y^d=0\,,
\end{equation}
and $\tilde y_1=r_T^>/r_H$ corresponds to the horizon. Using $(-1)^d\,y_T^>\prod_iy_i=\mathbf b^{-2}(r_T^>)^d/\mu$ and $(-1)^{d+1}\prod_j\tilde y_j=(r_T^>)^d/\mu$, we can now write the integrals (\ref{eq:TThetadefR}) as follows:
\begin{equation}
 \begin{aligned}
T(b)&=2\, b\,\mu^{-{3\over 2}}   ( r_T^>)^{{3d\over2}-1}    \,\int_0^1 dy\, (1-y)^{-{1\over 2}}(y_T^>-y) ^{-{1\over 2}} \Big[(-1)^d\prod_{i=1}^{d-2} (y_i-y)\Big]^{-{1\over 2}} \, \Big[(-1)^{d+1}\prod_{j=1}^d (\tilde y_j-y)\Big]^{-1}=\\
&= {2\, b\,\mathbf{b}\over r_T^>}   \,\int_0^1 dy\, (1-y)^{-{1\over 2}}\left(1-{y\over y_T^>}\right) ^{-{1\over 2}} \prod_{i=1}^{d-2} \left(1-{y\over y_i}\right)^{-{1\over 2}} \, \prod_{j=1}^d \left(1-{y\over \tilde y_j}\right)^{-1} =\\
&={4\, b\,\mathbf{b}\over r_T^>}\, F_D^{(2d-1)}\Big[1;\,\tfrac12,\dots,\tfrac12,1,\dots,1;\,\tfrac32;\,(y_T^>)^{-1},y_1^{-1},\dots,y_{d-2}^{-1}, \tilde y_1^{-1},\dots,\tilde y_d^{-1}\Big]\,,
 \end{aligned}
\label{octp91}
\end{equation}
with $d-1$ parameters equal to $\tfrac12$ and $d$ equal to $1$, and similarly
\begin{equation}
 \begin{aligned}
\Theta(b)&=2\,  \mu^{-{1\over 2}} (r_T^>)^{{d\over2}-1}  \,\int_0^1 dy\, (1-y)^{-{1\over 2}}(y_T^>-y) ^{-{1\over 2}} \Big[(-1)^d\prod_{i=1}^{d-2} (y_i-y)\Big]^{-{1\over 2}}= \\
&={2\,\mathbf{b}\over r_T^>}\, \int_0^1 dy\, (1-y)^{-{1\over 2}}\left(1-{y\over y_T^>}\right) ^{-{1\over 2}} \,\prod_{i=1}^{d-2} \left(1-{y\over y_i}\right)^{-{1\over 2}}= \\
&= {4\,\mathbf{b}\over r_T^>}\, F_D^{(d-1)}\Big[1;\,\tfrac12,\dots,\tfrac12;\,\tfrac32;\,(y_T^>)^{-1},y_1^{-1},\dots,y_{d-2}^{-1}\Big]\,.
 \end{aligned}
\label{octp92}
\end{equation}
Here we used the integral representation of the Lauricella function,
\begin{equation}
F_D^{(n)}\big[a;\,b_1,\dots,b_n;\,c;\,x_1,\dots,x_n\big]=\frac{\Gamma(c)}{\Gamma(a)\Gamma(c-a)}\int_0^1dt\,t^{a-1}(1-t)^{c-a-1}\prod_{i=1}^n(1-x_it)^{-b_i}\,,
\end{equation}
with $a=1$ and $c=\tfrac32$, for which the prefactor is $\Gamma(\tfrac32)/\Gamma(\tfrac12)=\tfrac12$.

\section{Gegenbauer polynomials for large spin}
\label{appendixc}

The starting point is the representation of the Gegenbauer polynomials in terms of hypergeometric functions \cite{Dodelson:2023nnr}:
\begin{equation}
  \begin{aligned}
C_{\ell}^{({d-2\over 2})}(\cos{\theta})&=e^{i\pi\ell} C_{\ell}^{({d-2\over 2})}(-\cos\theta)=e^{i\pi\ell}\frac{i(2\sin \theta)^{1-2\alpha}\Gamma(\ell+d-2)}{\Gamma(\frac{d-2}{2})\Gamma(\ell+\frac {d}{2})}\times\\
&\times\left(   e^{-i(1+\ell)(\pi-\theta)}f(\ell,\theta)-e^{i(1+\ell)(\pi-\theta)}f(\ell,\pi-\theta)\right)
  \end{aligned}
\label{eq:gp1}
\end{equation}
where
\begin{equation}
  \begin{aligned}
f(\ell,\theta)={}_2F_1(2-\frac{d}{2},1+\ell,\ell+{d\over 2},e^{2i\theta})
  \end{aligned}
\label{eq:gp2}
\end{equation}
and $\alpha\equiv\frac{d-2}{2}$.
In the limit $\ell\gg 1$, the hypergeometric behaves as follows
\begin{equation}
  \begin{aligned}
{}_2F_1(1-\alpha,1+k-\alpha,1+k,e^{2i\theta})&=\sum_{n=0}^{\infty}\frac{(1-\alpha)_n(1+k-\alpha)_n}{(1+k)_n}\frac{(e^{2i\theta})^n}{n!}\approx\\
&\approx\sum_{n=0}^{\infty}(1-\alpha)_n\frac{(e^{2i\theta})^n}{n!}=(-2ie^{i\theta}\sin \theta)^{\alpha-1}
  \end{aligned}
\label{eq:gp3}
\end{equation}
where we made the variable change $\ell=k-{d-2\over 2}$, to express the result in the variable $k$ used in the main text. For odd $d$, $\alpha-1$ is a half-integer, and the power on the right-hand side is understood on the principal branch, $(-2ie^{i\theta}\sin\theta)^{\alpha-1}=(2\sin\theta)^{\alpha-1}e^{i(\alpha-1)(\theta-\pi/2)}$, which is unambiguous for $0<\theta<\pi$.

Substituting the large $k$ behavior of the hypergeometrics, (\ref{eq:gp3}), into (\ref{eq:gp1}) and simplifying the overall $\Gamma$-functions accordingly in this limit, leads, for $0<\theta<\pi$ and in any $d$, to:
\begin{equation}
  \begin{aligned}
C_{\ell}^{({d-2\over 2})}(\cos\theta)\approx  \frac{1}{2}\left(\frac{2}{k}  \right)^{{4-d\over 2}}\frac{e^{-i\pi(d-2)/4}}{\Gamma({d-2\over 2})(\sin \theta)^{{d-2\over 2}}}\left(   e^{ik\theta}+e^{i\pi\alpha}e^{-ik\theta}\right)
  \end{aligned}
\label{eq:gp4}
\end{equation}

\section{Leading order phase shift and absence of singularity}
\label{appendixLO}

In this appendix we consider the correlator to leading order in $\mu$, and its behavior at the ``bulk point'' \cite{Maldacena:2015iua,Simmons-Duffin:2016gjk} (see also section 3 of \cite{Kulaxizi:2017ixa} for the bulk point in relation to the Regge limit). Since we expand around pure AdS, it is natural to use the Regge variables $(t',\theta')$ of (\ref{eq:Reggevariables}), as in \cite{Kulaxizi:2018dxo,Kulaxizi:2017ixa} and Appendix~\ref{appendixh}. In these variables the zero temperature correlator is $G^{T=0}_{\rm Regge}$, and the phase is $\delta_{CFT}=\delta-\delta_{AdS}$, (\ref{eq:deltaCFTvsbulk}), which starts at $\mathcal{O}(\mu)$. Following \cite{Kulaxizi:2018dxo}, we parametrize
\begin{equation}
\label{eq:sigmarhodef}
i(t'+\theta')\sim\sigma \, e^\rho,\quad i(t'-\theta')\sim\sigma \,e^{-\rho}\,.
\end{equation}
The Regge limit corresponds to $\sigma\ll 1$ at fixed $\rho$, and $\rho$ plays the role of an impact parameter in position space.

\paragraph{The bulk point and the bouncing singularity.} The bulk point is $\rho\to0$, i.e.\ $\theta'/t'\to0$. In cylinder variables it lies near $\theta=\pi$, $t=\pi$, the point where all null geodesics of pure AdS emanating from the first operator refocus. This is not where the singularity of Section~\ref{sec.five} appears. That singularity is at $\theta=0$, $t=t_c+i\beta/2$, i.e.\ at $\theta'=-\pi$, $t'=t_c-\pi+i\beta/2$: it is neither at small $\sigma$ nor at $\rho\to0$. In the geodesic picture, the bulk point is reached by geodesics that return to the antipodal point $\theta=\pi$, as all null geodesics do in pure AdS, whereas the singularity of Section~\ref{sec.five} is due to the radial geodesic, $b\to\infty$, which crosses the horizon and, after bouncing off the singularity, emerges with $\Theta(\infty)=0$, i.e.\ at $\theta=0$. The expansion in $\mu$ cannot interpolate between the two, because it is not uniform in $b$. At $\mathcal{O}(\mu)$ the phase grows at large $b$ (see below), and $T-\pi\propto\mu\, b^{d-3}$, $\Theta-\pi\propto\mu\, b^{d-2}$ for $d>3$; the corrections to $\Theta$, and to the phase per unit $k$, become of order one at $b\sim\mu^{-1/(d-2)}$, which is the scale of $b_c$, since $b_c^2-1\propto\mu^{-2/(d-2)}$ by (\ref{eq:photonsphere}). The region $b\gtrsim b_c$ thus lies beyond the reach of the expansion at any fixed order. Relatedly, for $d>3$, $\Im T(\infty)=\beta/2\propto r_H\sim\mu^{1/(d-2)}$ is not analytic at $\mu=0$, so the singularity of Section~\ref{sec.five} cannot be seen at any finite order in $\mu$. In $d=3$ this argument does not apply, since $\beta/2=2\pi(\mu-4\mu^3+\dots)$ is analytic; the non-uniformity in $b$ described above still holds.

\paragraph{Why the bulk point is nevertheless relevant.} As discussed extensively in \cite{Kulaxizi:2017ixa}, the correlator obtained by a single analytic continuation around $z=0$, i.e.\ on the Regge sheet, is not expected to be singular at the bulk point, precisely due to the presence of double-trace operators in the OPE. The stress-tensor Regge conformal block alone is proportional to the propagator $\Pi_{d-1}(\rho)$ on $H_{d-1}$, which is singular as $\rho\to0$ (as $\rho^{3-d}$ for $d>3$, and logarithmically for $d=3$). Only following a further analytic continuation, $\rho\rightarrow\rho\pm i\pi$ -- which effectively removes the double-trace operators -- does the correlator become singular. The picture is already evident at leading order in $\mu$ (see the discussion of section 3.3 in \cite{Kulaxizi:2017ixa}), which we now verify directly from the phase shift. This shows that the Fourier transform of the eikonal phase in the regime $b<b_c$, which is responsible for the standard bulk-cone singularities, includes the double-trace contributions.

\paragraph{Leading order computation.} We set $\theta'=0$ and keep the term linear in $\mu$ of $e^{i\delta_{CFT}}$, i.e.\ we replace the exponential in (\ref{eq:correlatorthetazero}) by $i\delta_1$, with $(t,\theta)\to(t',\theta')$ and no shift. The $\mathcal{O}(\mu)$ term in $\delta_{CFT}$ following \cite{Kulaxizi:2018dxo} can be written as:
\begin{equation}
\label{eq:deltaone-leadingmu}
\delta_1\propto\mu \, k\,  (b^2-1)^{d\over 2}   \,_2F_1\Big[1,\frac{d-1}{2},\frac d2+1,-(b^2-1)\Big]\,,
\end{equation}
where $_2F_1$ denotes the hypergeometric function. In writing (\ref{eq:deltaone-leadingmu}) we have omitted an overall numerical constant. Recall that $b$ in eq.~(2.26) of \cite{Kulaxizi:2018dxo} corresponds here to the impact parameter $\mathbf{b}$, related to $b\equiv {\omega\over k}$ here as follows: $\mathbf{b}=(b^2-1)^{-{1\over 2}}$.
Substituting (\ref{eq:deltaone-leadingmu}) into (\ref{eq:correlatorthetazero}) in Regge variables leads to:
\begin{equation}
\label{eq:corr_leading_mu_a}
G^{(1),T}_{\rm Regge}\sim  \mu\int_0^\infty  dk\,\int_1^\infty db\, k^{2\Delta} (b^2-1)^{\Delta} \, e^{-i k(b t'-\theta')}\, _2F_1\Big[1,\frac{d-1}{2},\frac d2+1,-(b^2-1)\Big]\,.
\end{equation}
The integral over $k$ can be easily performed,  leading to:
\begin{equation}
\label{eq:corr_leading_mu_b}
G^{(1),T}_{\rm Regge}\sim \mu \int_1^\infty  db\,\frac{(b^2-1)^{\Delta}}{(b\, t'-\theta')^{2\Delta+1}}\,  {}_2F_1\Big[1,\frac{d-1}{2},\frac d2+1,-(b^2-1)\Big]\,.
\end{equation}
Besides the expected lightcone singularity from $b\sim 1$, the integral could diverge for large $b\gg 1$, which is the head-on, $\mathbf{b}\to0$, region. For $b\gg1$ the hypergeometric function behaves as $\frac{d}{d-3}(b^2-1)^{-1}$ for $d>3$ (and as $(b^2-1)^{-1}\ln(b^2-1)$, up to a constant factor, for $d=3$), so that $\delta_1\propto\mu\,\omega\, b^{d-3}$. This is the momentum-space image of the singularity of $\Pi_{d-1}$ at zero impact parameter. Nevertheless, the integrand of (\ref{eq:corr_leading_mu_b}) behaves as $b^{-3}$ (up to a logarithm in $d=3$), and the integral converges. At $\theta'=0$, i.e.\ at $\rho=0$, (\ref{eq:corr_leading_mu_b}) is therefore simply $\mu\, t'^{-2\Delta-1}$ times a finite constant: the $\mathcal{O}(\mu)$ correlator shows the Regge growth $\sigma^{-1}$ relative to the vacuum, but no enhanced singularity at the bulk point, in contrast with the stress-tensor Regge block. Regularity therefore implies that double-trace operators contribute in the OPE together with the stress tensor at this order \cite{Kulaxizi:2017ixa}.

\paragraph{Large $b$.} To linear order in $\mu$ the correlator does not exhibit a singularity associated with $b\to\infty$ either. This is to be expected: at this order there is neither a photon sphere nor a horizon, and as explained above the singularity of Section~\ref{sec.five} arises from the resummed, analytically continued phase $S_+(b)\simeq b\,T(\infty)$ for $b\gg b_c$, whose imaginary part is non-analytic in $\mu$.

\section{Numerical tests in $d=4$}
\label{app:num}

In this appendix we test the analytic statements of Appendices~\ref{app:wkb} and~\ref{app:GR} numerically for AdS$_5$--Schwarzschild. The tests fall into three groups:
\begin{itemize}
\item analytic properties of the phase $S(b)$: its logarithmic branch point at $b_c$ and its large-$b$ behavior (Appendix~\ref{app:num:phase});
\item comparisons with the exact radial equation: the round-trip amplitude, including the uniform barrier amplitude and the multiple-reflection expansion, extracted from the exact retarded function, and the quasinormal modes, including the phase $\delta_\infty$ (Appendices~\ref{app:num:GR}, \ref{app:num:qnm});
\item the position-space integral near $t=T(\infty)$ (Appendices~\ref{app:num:coef}, \ref{app:num:trunc}).
\end{itemize}

\subsection{Conventions}
\label{app:num:conv}

All results are for $d=4$, $\mu=1$, for which
\begin{equation}
\begin{gathered}
r_H=0.786151,\qquad \beta=\frac{4\pi}{f'(r_H)}=2.209027,\qquad b_c=\frac{\sqrt5}{2}=1.118034,\\
T(\infty)=t_c+\frac{i\beta}{2}=1.787140+1.104514\,i ,
\end{gathered}
\end{equation}
the last from \eqref{eq:H:Tinf4}. The value of $\Delta$ is given separately for each test. Arbitrary-precision computations use \texttt{mpmath} \cite{mpmath}; ordinary differential equations are integrated with an explicit Runge--Kutta method of order 8 (DOP853) \cite{Hairer,SciPy}. The closed form \eqref{eq:d4delta}--\eqref{eq:d4tpreal} of the phase is evaluated at $b+i\epsilon$, with $\epsilon>0$ infinitesimal, for $b>b_c$, which gives $S_+$. Where we quote extrapolated values we use Richardson extrapolation \cite{Richardson,BenderOrszag}: if $A(h)=A_0+a\,h^\kappa+o(h^\kappa)$ with known $\kappa$, two evaluations give $A_0\simeq\big(h_1^\kappa A(h_2)-h_2^\kappa A(h_1)\big)/(h_1^\kappa-h_2^\kappa)$.

\subsection{The phase $S(b)$}
\label{app:num:phase}

\paragraph{The logarithmic branch point.} The coefficient of $\ln(b_c-b)$ in \eqref{eq:H:Slog} should be $K(b)/\pi$. We extracted it independently from the discontinuity of $S$ across the cut on $b>b_c$, i.e.\ from $-\Im S_+/\pi$, by fitting it and extrapolating the fit to $b<b_c$:
\begin{center}\begin{tabular}{lcccc}\toprule
$b$ & 1.02 & 1.05 & 1.10 & 1.115\\\midrule
$K/\pi$ from \eqref{eq:H:K} & 0.070331 & 0.046286 & 0.0115892 & 0.00192379\\
$K/\pi$ from the cut & 0.069951 & 0.046258 & 0.0115891 & 0.00192379\\\bottomrule
\end{tabular}\end{center}
The small differences at $b=1.02,1.05$ are due to the extrapolation of the fit. We also checked that $P=S-(K/\pi)\ln(b_c-b)$ is smooth across $b=b_c$: a single polynomial fits $P$ on both sides to $10^{-9}$ in its imaginary part.

\paragraph{Large $b$.} To extract $c$ we computed $\big(S_+(b)/b-T(\infty)\big)\,b^{3/2}$ at $b=10^5$ and $10^6$, and removed the $O(1/b)$ error by Richardson extrapolation with $h=1/b$, $\kappa=1$. The result agrees with \eqref{eq:H:c4} to 10 digits for $\mu=\tfrac12,1,2$; for $\mu=1$, $c=-1.236049785\,(1+i)$. This tests \eqref{eq:H:approach} and \eqref{eq:H:c4} directly from the continued phase $S_+$.

\subsection{The exact retarded function: multiple reflections and Kemble terms}
\label{app:num:GR}

\paragraph{What is tested.}
\begin{itemize}
\item The multiple-reflection expansion \eqref{eq:I:bounce}, i.e.\ all the $N$-reflection terms with their weights;
\item the barrier factor of \eqref{eq:H:kemble}, i.e.\ all the Kemble terms $s$.
\end{itemize}

\paragraph{Method.} We compute the exact $G_R(\omega,k)$ by the power-series method of \cite{Horowitz:1999jd}. With $u=1/r^2$, $g(u)=1+u-\mu u^2$ and $\phi=e^{-i\omega t}Y_\ell U(u)$, the radial equation reads
\begin{equation}\label{eq:J:ueq}
4u^2g^2U''+4ug\,(g'u-g)\,U'+\big[\omega^2u-\big(\ell(\ell+2)u+m^2\big)g\big]U=0 .
\end{equation}
\begin{itemize}
\item At the horizon, $u_H=1/r_H^2=1.618$, we use the ingoing Frobenius series $U_{\rm in}=(u-u_H)^{-i\omega/f'(r_H)}\sum_nc_n(u-u_H)^n$, with $f'(r_H)=5.6886$.
\item At the boundary we use the two series $U_N=u^{\Delta/2}(1+\dots)$ and $U_{NN}=u^{(4-\Delta)/2}(1+\dots)$, i.e.\ $r^{-\Delta}$ and $r^{\Delta-4}$ with unit leading coefficients. A non-integer $\Delta-2$ avoids logarithms.
\item The series are matched at $u_\ast=0.3$, within the radii of convergence ($1.618$ and $0.618$ respectively, limited by $u=0$ and by the root $u=-0.618$ of $g$).
\end{itemize}
Writing $U_{\rm in}=A\,U_{NN}+B\,U_N$,
\begin{equation}
A=\frac{\mathcal W[U_{\rm in},U_N]}{\mathcal W[U_{NN},U_N]},\qquad B=\frac{\mathcal W[U_{NN},U_{\rm in}]}{\mathcal W[U_{NN},U_N]},\qquad G_R=\frac BA ,
\end{equation}
where $\mathcal W$ denotes the Wronskian at $u_\ast$. We divide $G_R$ by the exact pure-AdS factor $\mathcal G_0e^{-i\pi\nu}$, computed from $\Gamma$-functions in the same normalization (with no large-$k$ approximation): $\tilde G=G_R/(\mathcal G_0e^{-i\pi\nu})$. Since \eqref{eq:H:GRY} states $\tilde G=(1-e^{2\pi i\nu}Y)/(1-Y)$, we can \emph{extract} the round-trip amplitude by inverting this M\"obius map,
\begin{equation}
Y_{\rm ex}=\frac{\tilde G-1}{\tilde G-e^{2\pi i\nu}},
\end{equation}
and compare it with the prediction $Y_{\rm WKB}=e^{-i\pi(\nu+1)}X$, with $X$ from \eqref{eq:H:kemble}. Note that the inversion itself maps every pole of $\tilde G$ to $Y_{\rm ex}=1$, so that the smoothness of $Y_{\rm ex}$ is automatic and is not a test. The test is the comparison of $Y_{\rm ex}$ with the independently computed $Y_{\rm WKB}$: if the barrier factor or the continued phase were incorrect, $Y_{\rm ex}$ would deviate from $Y_{\rm WKB}$, e.g.\ $|Y_{\rm ex}|$ from $|X|=(1+e^{-2\eta})^{-1/2}$.

\paragraph{Parameters and accuracy.} $\Delta=2.6$; $k=20,40$ with $\ell=k-1$; real $\omega=bk$ with $1.02\le b\le1.26$. We work with 40 digits. The horizon series gives the same $G_R$ to 40 digits with 600 and 900 terms, and the AdS factor reproduces the exact identity $G_R^{\rm AdS}/(\mathcal G_0e^{-i\pi\nu})=(1-e^{2\pi i\nu}Y_{\rm AdS})/(1-Y_{\rm AdS})$ to all digits.

\begin{figure}[t]\centering
\includegraphics[width=\textwidth]{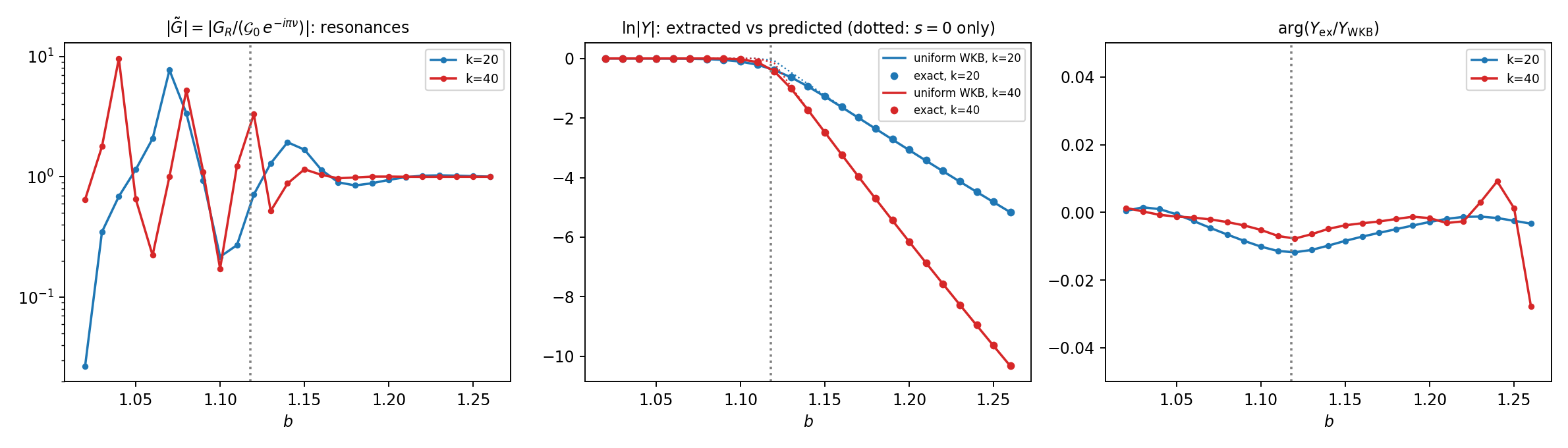}
\caption{Left: $|\tilde G|$, showing the resonances (near-quasinormal modes) below $b_c$. Middle: $\ln|Y|$ extracted from the exact $G_R$ (dots) against the uniform WKB prediction (lines); the dotted lines show the leading term $s=0$ alone. Right: phase difference between extracted and predicted $Y$. The vertical line marks $b_c$.}
\label{fig:J:bounceY}
\end{figure}

\paragraph{Results.} A selection is given in Table~\ref{tab:J:Y}, and the full scan, in steps of $0.01$ in $b$, in Fig.~\ref{fig:J:bounceY}.
\begin{table}[t]\centering\small
\begin{tabular}{l|cccc|cccc}\toprule
& \multicolumn{4}{c|}{$k=20$} & \multicolumn{4}{c}{$k=40$}\\
$b$ & $|\tilde G|$ & $\ln|Y_{\rm ex}|$ & $\ln|Y_{\rm WKB}|$ & $\arg\frac{Y_{\rm ex}}{Y_{\rm WKB}}$ & $|\tilde G|$ & $\ln|Y_{\rm ex}|$ & $\ln|Y_{\rm WKB}|$ & $\arg\frac{Y_{\rm ex}}{Y_{\rm WKB}}$\\\midrule
1.04 & 0.687 & $-0.0006$ & $-0.0006$ & $0.001$ & 9.612 & $-0.0000$ & $-0.0000$ & $-0.001$\\
1.07 & 7.686 & $-0.0088$ & $-0.0091$ & $-0.005$ & 1.005 & $-0.0002$ & $-0.0002$ & $-0.002$\\
1.10 & 0.217 & $-0.1019$ & $-0.1048$ & $-0.010$ & 0.173 & $-0.0261$ & $-0.0265$ & $-0.005$\\
1.12 & 0.712 & $-0.3789$ & $-0.3871$ & $-0.012$ & 3.314 & $-0.4261$ & $-0.4306$ & $-0.008$\\
1.15 & 1.680 & $-1.2650$ & $-1.2796$ & $-0.008$ & 1.153 & $-2.4743$ & $-2.4822$ & $-0.004$\\
1.20 & 0.943 & $-3.0635$ & $-3.0790$ & $-0.003$ & 1.004 & $-6.1468$ & $-6.1558$ & $-0.002$\\
1.25 & 1.012 & $-4.8081$ & $-4.8260$ & $-0.002$ & 1.000 & $-9.6257$ & $-9.6519$ & $0.001$\\\bottomrule
\end{tabular}
\caption{Round-trip amplitude extracted from the exact retarded function, compared with the uniform WKB prediction, for $d=4$, $\mu=1$, $\Delta=2.6$.}
\label{tab:J:Y}
\end{table}
Three features stand out:
\begin{itemize}
\item $\tilde G$ itself varies by a factor of order $300$ over the range, because of the resonances below $b_c$;
\item the extracted $Y_{\rm ex}$ agrees with the independent prediction $Y_{\rm WKB}$ to $|\Delta\ln|Y||\le0.018$ ($k=20$) and $\le0.026$ ($k=40$; $\le0.009$ for $b\le1.22$). The phase agrees to $\le0.012$ ($k=20$) and $\le0.03$ ($k=40$);
\item above $b_c$ the agreement extends down to $|Y|\sim e^{-10}$. The continued phase $S_+$, real and imaginary parts, is thereby confirmed to high accuracy.
\end{itemize}

\paragraph{The multiple-reflection terms $N\ge2$.} Since $Y_{\rm ex}$ agrees with the independently computed single round-trip amplitude $Y_{\rm WKB}$, the full resonant $\tilde G$ is reproduced by the M\"obius form with $Y_{\rm WKB}$, which is equivalent to the geometric series $1+(1-e^{2\pi i\nu})\sum_NY^N$ with all its weights. The multiple-reflection terms are therefore confirmed for $b$ near $b_c$. The regime $b\to\infty$ is probed, at fixed $\ell$, through the quasinormal residues, Appendix~\ref{app:num:qnm}.

\begin{figure}[t]\centering
\includegraphics[width=.6\textwidth]{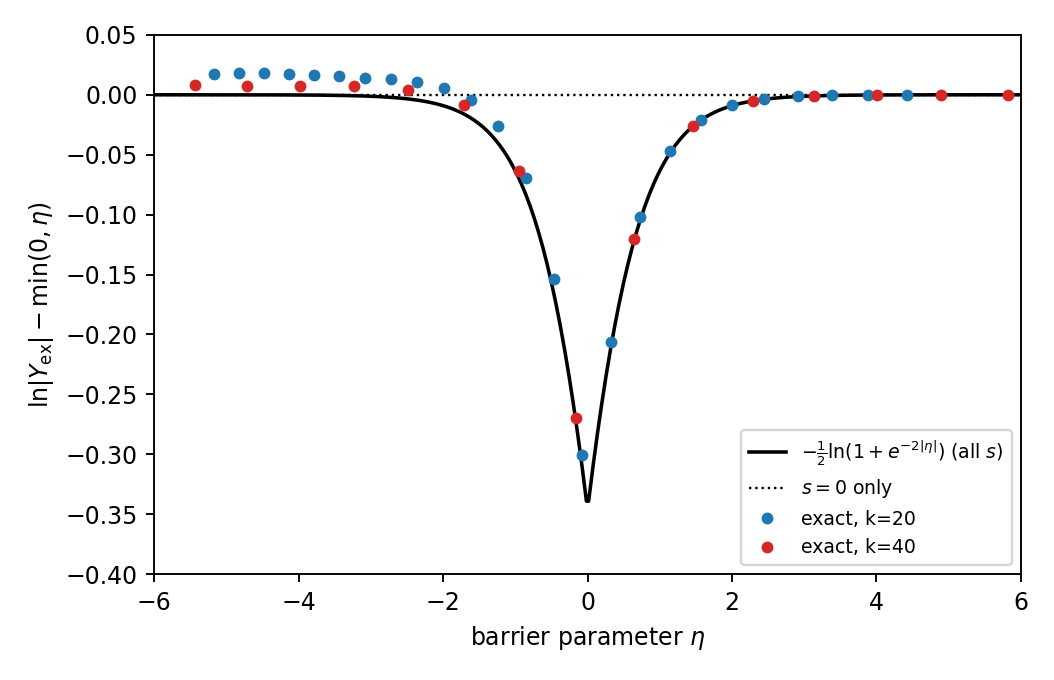}
\caption{Deviation of the exact $\ln|Y|$ from the leading Kemble term $s=0$, as a function of the barrier parameter $\eta$ \eqref{eq:H:eta}. For $|\eta|\lesssim2$ the full barrier factor (all $s$) is required.}
\label{fig:J:kemble}
\end{figure}

\paragraph{The Kemble terms $s\ge1$.} Fig.~\ref{fig:J:kemble} shows $\ln|Y_{\rm ex}|-\min(0,\eta)$, i.e.\ the deviation from the leading term $s=0$ alone, as a function of the barrier parameter $\eta$:
\begin{itemize}
\item for $|\eta|\lesssim2$ the data follow the full barrier factor $-\tfrac12\ln(1+e^{-2|\eta|})$, which differs from the $s=0$ term by up to $0.34$. The $s\ge1$ terms are therefore present, with the binomial weights (the Taylor coefficients of this function);
\item for $|\eta|\gtrsim2$ their contribution, $\simeq-\tfrac12e^{-2|\eta|}$, falls below the $O(1/k)$ residual ($\approx0.015$ at $k=20$ and $0.008$ at $k=40$) and cannot be resolved.
\end{itemize}
The singularities of the Kemble terms come from $b\to\infty$, where these terms are suppressed by $e^{-s\beta\omega}$ relative to the leading one. They are thus neither confirmed nor excluded by the exact retarded function.

\paragraph{The thermal images $n\ge1$.} These follow from the Bose factor and are exact shifts $t\to t-in\beta$ of each reflection term; they require no test.

\subsection{Quasinormal modes}
\label{app:num:qnm}

\paragraph{What is tested.} The asymptotic spacing $2\pi/T(\infty)=2.544039-1.572303\,i$ of \eqref{eq:I:qnm}, which tests $T(\infty)$ and hence the continued phase at $b\to\infty$; and the residues \eqref{eq:I:residue}, which test the multiple-reflection structure in the same regime. Since the modes are computed at fixed $\ell$, i.e.\ outside the Regge limit, they also exhibit the phase $\delta_\infty$ of Appendix~\ref{app:wkb:delta}.

\paragraph{Method.} We use the same series as in Appendix~\ref{app:num:GR}, at $\ell=0$, $\Delta=2.6$. The quasinormal frequencies are the zeros of $U_{\rm in}'/U_{\rm in}-U_N'/U_N$ at $u_\ast$. We work with 40 digits, $350+60m$ terms of the horizon series and $200$ of the boundary series. The roots are stable to 15 digits when increasing the horizon series from $1310$ to $1800$ terms, the boundary series from 200 to 260 terms, and the precision from 40 to 50 digits (checked at $m=16$).

\begin{figure}[t]\centering
\includegraphics[width=\textwidth]{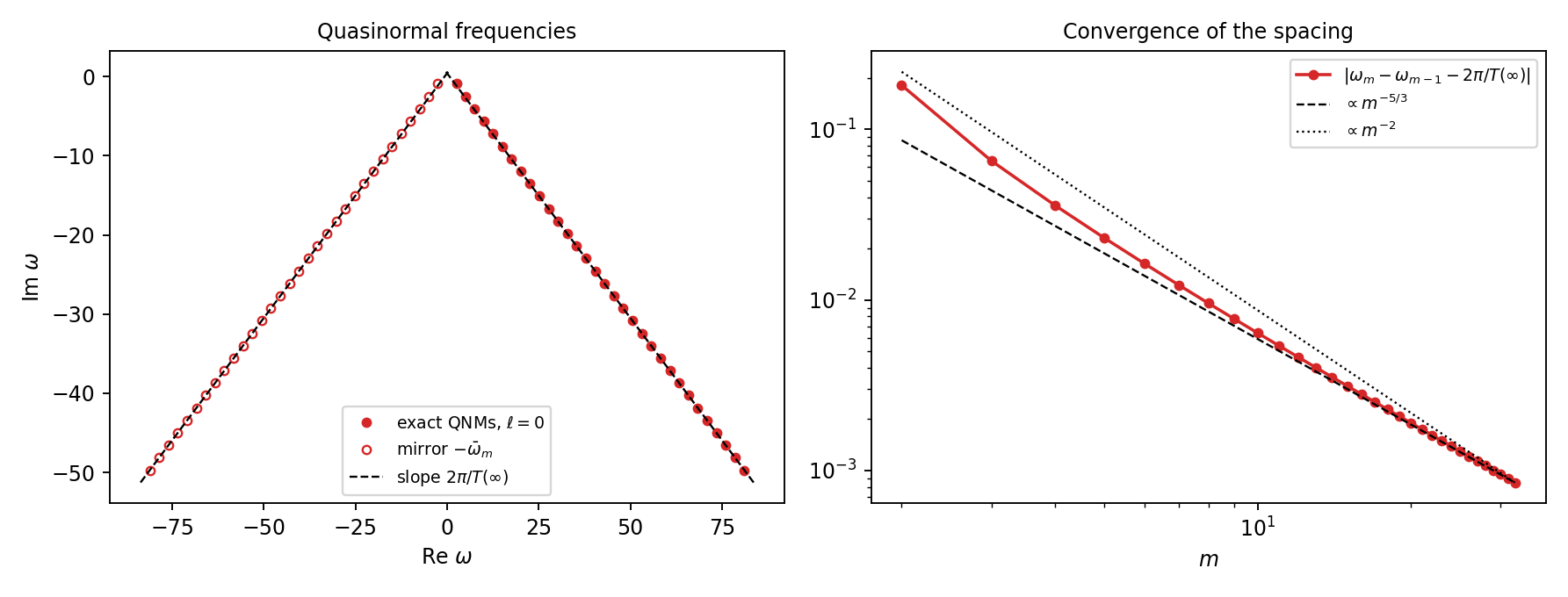}
\caption{Left: exact $\ell=0$ quasinormal frequencies for $d=4$, $\mu=1$, $\Delta=2.6$, $1\le m\le32$, together with their mirror images and lines of slope $2\pi/T(\infty)$. Right: convergence of the spacing to $2\pi/T(\infty)$, with reference lines $\propto m^{-5/3}$ (dashed) and $\propto m^{-2}$ (dotted), normalized at $m=32$.}
\label{fig:J:qnm}
\end{figure}

\paragraph{Spacing.}
\begin{center}\begin{tabular}{rlll}\toprule
$m$ & $\omega_m$ & $\omega_m-\omega_{m-1}$ & $|\omega_m-\omega_{m-1}-2\pi/T(\infty)|$\\\midrule
1 & $2.642021-0.891641\,i$ & & \\
2 & $5.005262-2.477383\,i$ & $2.36324-1.58574\,i$ & 0.181\\
4 & $9.994621-5.644279\,i$ & $2.50934-1.58125\,i$ & 0.036\\
8 & $20.112146-11.950912\,i$ & $2.53491-1.57518\,i$ & 0.0096\\
12 & $30.265327-18.247594\,i$ & $2.53964-1.57376\,i$ & 0.0046\\
16 & $40.428707-24.541083\,i$ & $2.54139-1.57320\,i$ & 0.0028\\\bottomrule
\end{tabular}\end{center}
The deviation decreases with $m$ (Fig.~\ref{fig:J:qnm}). With $\omega_{31}=78.567945-48.132930\,i$ and $\omega_{32}=81.111182-49.705510\,i$, it is $8.49\times10^{-4}$ at $m=32$, so that between $m=16$ and $m=32$ it decreases effectively as $m^{-1.72}$. This is consistent with a correction to the spacing of order $m^{-5/3}$, as expected at fixed $\ell$ from the first perturbation of the connection near $r=0$, which is of relative order $\omega^{-2/3}$ \cite{Musiri:2005ev,Afkhami-Jeddi:2025wra}. Richardson extrapolation from $m=15,16$ assuming $m^{-5/3}$ gives a spacing within $1.4\times10^{-4}$ of $2\pi/T(\infty)$ (assuming $m^{-2}$, within $3.8\times10^{-4}$). This confirms both $\Re T(\infty)=t_c$ and $\Im T(\infty)=\beta/2$.

\paragraph{Residues.} The exact residue is $B/\partial_\omega A$ at $\omega=\omega_m$. Its ratio to \eqref{eq:I:residue} is given below. Here $k=1$, $b_m=\omega_m$, $T(b_m)=S_+'(b_m)$ is computed from the analytic continuation of \eqref{eq:H:S}, and $\mathcal G_0e^{-i\pi\nu}$ is the exact pure-AdS prefactor, computed from the $\Gamma$-functions of Appendix~\ref{app:GR:AdS} at $\omega=\omega_m$, without large-frequency approximation:
\begin{center}\begin{tabular}{lcccccc}\toprule
$m$ & 1 & 2 & 4 & 8 & 12 & 16\\\midrule
$\Re$ (exact$/$\eqref{eq:I:residue}) & 0.9598 & 0.9967 & 0.9996 & 1.00005 & 1.00008 & 1.00008\\
$\Im$ (exact$/$\eqref{eq:I:residue}) & 0.0056 & 0.0043 & 0.0021 & 0.0010 & 0.0006 & 0.0004\\\bottomrule
\end{tabular}\end{center}
The remaining deviation at large $m$, $4\times10^{-4}$ at $m=16$, is of the size of the correction beyond the Regge limit, $(1-q)\,\delta_\infty'(\lambda_m)\,b_m^{-q}$ (Appendix~\ref{app:GR:qnm}).
The quasinormal poles, with these residues and spacing, are the resummed multiple-reflection series. This confirms the multiple-reflection structure at $b\to\infty$ at fixed $\ell$. It does not by itself test the regime of large $k$ at fixed $\lambda$, which determines the coefficients of the singularities at $N\,T(\infty)$.

\paragraph{Beyond the Regge limit: the phase $\delta_\infty$.} Subtracting the known $c\,\omega^{-1/2}$ term of \eqref{eq:I:qnm} (with $k=1$; the overtones in the tables are labeled from $m=1$, so that $m-1$ here plays the role of $m$ in \eqref{eq:I:qnm}):
\begin{center}\begin{tabular}{lccc}\toprule
$m$ & $\omega_mT(\infty)-2\pi(m-1)$ & $-c\,\omega_m^{-1/2}$ & difference\\\midrule
4 & $5.246+0.952\,i$ & $0.260+0.446\,i$ & $4.986+0.506\,i$\\
8 & $5.161+0.856\,i$ & $0.179+0.314\,i$ & $4.982+0.542\,i$\\
16 & $5.110+0.796\,i$ & $0.125+0.222\,i$ & $4.985+0.574\,i$\\\bottomrule
\end{tabular}\end{center}
Compared with $\pi(\nu+1)=\pi(\Delta-1)=5.027$, this gives $\delta_\infty\approx0.04-0.57\,i$. The real part is small, while the imaginary part is clearly non-zero and still drifts slowly with $m$. This is the correction beyond the Regge limit: at fixed $\ell$, $\lambda=k^{3/2}\omega_m^{-1/2}\to0$ (Appendix~\ref{app:wkb:delta}), and the drift reflects the dependence of $\delta_\infty$ on $\lambda$ ($|\lambda|\approx0.14$ at $m=16$), towards the limit $-i\log2=-0.693\,i$. At $\ell=0$, i.e.\ $k=1$, the terms of relative order $1/k^2$ neglected in \eqref{eq:H:local} are not small, so that the drift cannot be attributed entirely to the universal function $\delta_\infty(\lambda)$; the limit $-i\log2$ is not affected by them.

\subsection{The coefficient of the singularity}
\label{app:num:coef}

\paragraph{What is tested.} The coefficient $C_\infty$ of \eqref{eq:I:C4}, for $\Delta=1.3$, where $C_\infty=-0.227267\,i$.

\paragraph{Method.} We evaluate $I(t)$ of \eqref{eq:I:Idef} at $t=T(\infty)+\tilde{t}$ along two rays that avoid the curve $\gamma$, which approaches $T(\infty)$ along $\arg c=-3\pi/4$ (Fig.~\ref{fig:J:tplane}). The $b$-integral is truncated at $b\simeq10^4$, which affects the result at relative order $\lesssim10^{-6}$ for the values of $|\tilde{t}|$ used.

\begin{figure}[t]\centering
\includegraphics[width=.58\textwidth]{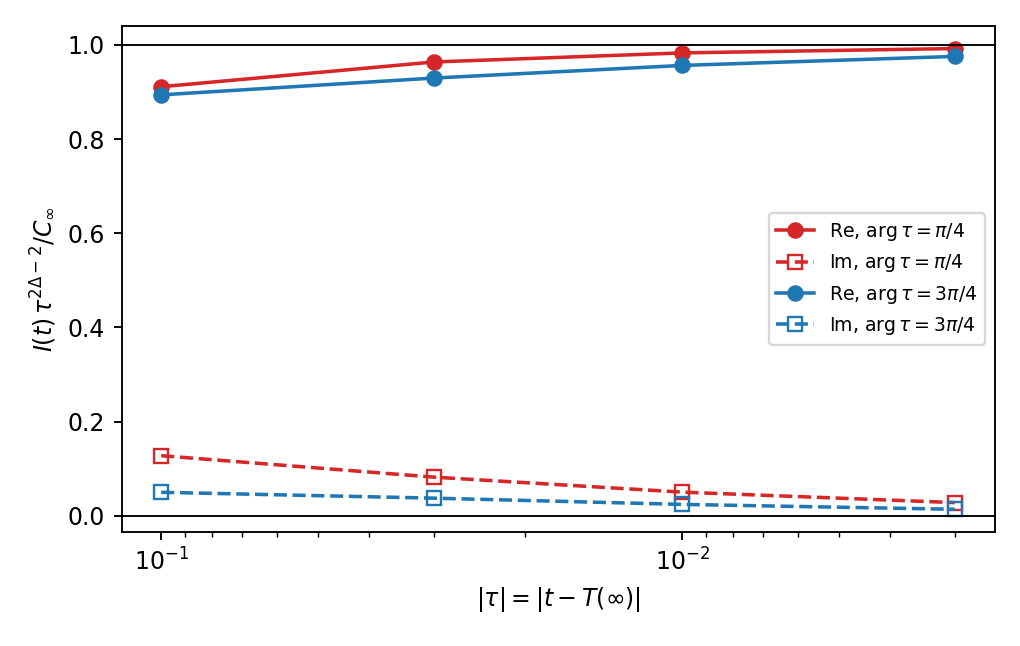}
\caption{$I(t)\,\tilde{t}^{2\Delta-2}/C_\infty\to1$ as $t\to T(\infty)$ along two directions, for $\Delta=1.3$.}
\label{fig:J:coef}
\end{figure}

\begin{center}\begin{tabular}{lcc}\toprule
$|\tilde{t}|$ & $I\,\tilde{t}^{2\Delta-2}/C_\infty$, $\arg\tilde{t}=\pi/4$ & $I\,\tilde{t}^{2\Delta-2}/C_\infty$, $\arg\tilde{t}=3\pi/4$\\\midrule
0.1 & $0.911+0.128\,i$ & $0.894+0.050\,i$\\
0.03 & $0.964+0.082\,i$ & $0.930+0.037\,i$\\
0.01 & $0.983+0.050\,i$ & $0.956+0.024\,i$\\
0.003 & $0.992+0.028\,i$ & $0.976+0.014\,i$\\\bottomrule
\end{tabular}\end{center}
The approach to $1$ goes as $|\tilde{t}|^{2\Delta-2}=|\tilde{t}|^{0.6}$, as expected from the analytic remainder. Extrapolating the last two points gives $1.001+0.007\,i$ and $0.994+0.004\,i$, so $C_\infty$ is confirmed to better than $1\%$. The two directions agree, as expected for even $d$ (Fig.~\ref{fig:J:coef}).

\paragraph{Individual terms.} As a further check of Appendix~\ref{app:GR:sing}, we evaluated the $b$-integrals of two other terms near their predicted singular points, for $\Delta=1.3$:
\begin{itemize}
\item the $N=2$ reflection term ($S\to2S$), singular at $2T(\infty)$ with $c\to2c$;
\item the first Kemble term ($S\to S_{(\pm1)}$), singular at $T(\infty)+i\beta$ with $c\to c+2i\,\Im c$.
\end{itemize}
Each is approached along the ray opposite to the corresponding curve of singular points, $\arg\tilde t=\arg(-c_{\rm eff})$, with $c_{\rm eff}=2c$ and $c+2i\,\Im c$ respectively, i.e.\ $\arg\tilde t=\pi/4$ for $N=2$ and $\arg\tilde t=\arctan3$ for $s=1$. Normalizing by the corresponding coefficient $C$ obtained from \eqref{eq:I:Cinf}:
\begin{center}\begin{tabular}{lcccc}\toprule
term & $|\tilde{t}|=0.03$ & $0.01$ & $0.003$ & extrapolated\\\midrule
$N=2$ & $0.978+0.061\,i$ & $0.989+0.036\,i$ & $0.995+0.020\,i$ & $1.0005+0.004\,i$\\
$s=1$ & $1.058+0.081\,i$ & $1.034+0.045\,i$ & $1.019+0.023\,i$ & $1.0035+0.003\,i$\\\bottomrule
\end{tabular}\end{center}
As integrals, both terms have the predicted singularities with the predicted coefficients. Whether they appear in the correlator with the stated weights is the question addressed in Appendix~\ref{app:num:GR}: yes for the bounces; untested at large $b$ for the Kemble terms.

\subsection{The integral restricted to $b<b_c$}
\label{app:num:trunc}

\begin{figure}[t]\centering
\includegraphics[width=.58\textwidth]{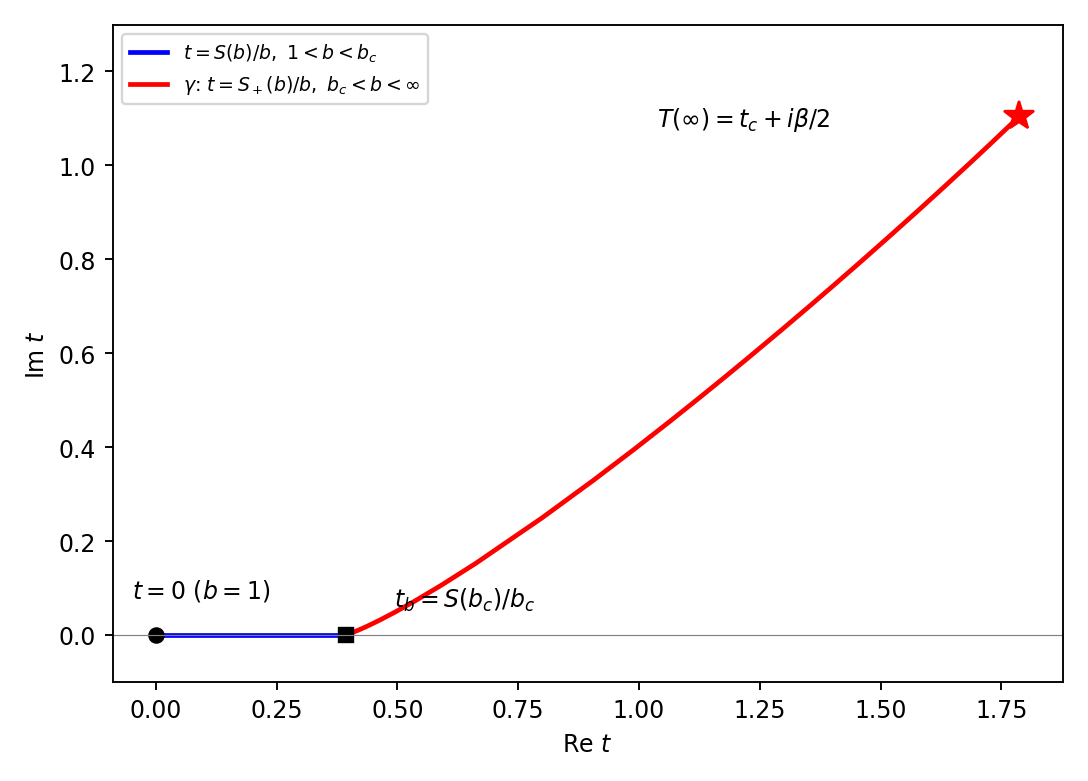}
\caption{Image $t=S(b)/b$ of the $b$-contour in the complex $t$-plane, for $d=4$, $\mu=1$. The range $1<b<b_c$ maps to the segment $[0,t_b]$ (blue), and $b>b_c$ to the curve $\gamma$ (red), which ends at $T(\infty)=t_c+i\beta/2$ (star).}
\label{fig:J:tplane}
\end{figure}

Fig.~\ref{fig:J:tplane} shows the image of the integration contour in the complex $t$-plane, $t=S(b)/b$. The range $1<b<b_c$ maps to the segment $0\le t\le t_b$, with $t_b\equiv S(b_c)/b_c=0.392615$, and the range $b>b_c$ maps to the curve $\gamma$, which ends at $T(\infty)$. We denote by $I_<(t)$ the integral \eqref{eq:I:Idef} restricted to $1<b<b_c$ and by $I_>$ the remainder, so that $I=I_<+I_>$. For $\Delta=1.3$ we found:
\begin{enumerate}
\item \emph{$t_b$ is not a singularity of $I$ on the principal sheet.} At $t=x-0.004\,i$ with $x=t_b$, $I_<\approx-1022+1109\,i$ and $I_>\approx1028-1101\,i$, while their sum varies smoothly, from $8.17+11.80\,i$ to $4.21+5.92\,i$ as $x-t_b$ goes from $-0.06$ to $+0.06$. Above $\gamma$, $|I|$ grows like $|I_<|$ as $t\to t_b$, so $\gamma$ is a branch cut of $I$ starting at $t_b$.
\item \emph{The second sheet of $I_<$}, reached by crossing the segment $(0,t_b)$, differs from $I$:
\begin{center}\begin{tabular}{llll}\toprule
$t$ & $I_<$ (principal sheet) & $I$ & $I_<$ (second sheet)\\\midrule
$0.2+0.1\,i$ & $-16.4-30.0\,i$ & $-16.2-32.9\,i$ & $36.4-5.9\,i$\\
$0.3+0.3\,i$ & $-8.39-4.70\,i$ & $-6.56-5.02\,i$ & $7.38-4.99\,i$\\
$0.8+0.6\,i$ & $-0.34-1.13\,i$ & $-0.29-0.37\,i$ & $0.87+0.18\,i$\\\bottomrule
\end{tabular}\end{center}
The second sheet was computed on a deformed contour, and is path-independent to $\sim10^{-3}$.
\item \emph{Continuation of $I_<$ around $t_b$} (which requires continuing $S$ through its cut) does not reproduce $I$ either:
\begin{center}\begin{tabular}{lllll}\toprule
$t_0$ & $I_<$ & $I$ & clockwise & anticlockwise\\\midrule
$t_b+0.1e^{-1.2i}$ & $2.40+19.99\,i$ & $1.26+7.61\,i$ & $-2.97+0.72\,i$ & $10.38-14.38\,i$\\
$t_b+0.1e^{-1.8i}$ & $-7.84+17.83\,i$ & $0.65+10.43\,i$ & $1.57+7.90\,i$ & $19.88-7.90\,i$\\\bottomrule
\end{tabular}\end{center}
\end{enumerate}
These results are consistent with the analytic statement of Appendix~\ref{app:GR:coef}: the integral including $b>b_c$ is not a sheet of $I_<$.

\subsection{Summary}
\label{app:num:summary}

For $d=4$, $\mu=1$ the numerical results confirm:
\begin{itemize}
\item the logarithmic branch point of $S(b)$ at $b_c$, and its continuation $S_+$ with $\Im S_+>0$, down to reflection amplitudes $\sim e^{-10}$;
\item $T(\infty)=t_c+i\beta/2$, through the asymptotic quasinormal spacing, and the approach \eqref{eq:H:approach} with $c$ of \eqref{eq:H:c4}, directly from $S_+$;
\item the multiple-reflection expansion of the exact retarded function, near $b_c$ and, through the quasinormal residues, at $b\to\infty$;
\item the Kemble terms near $b_c$ ($|\eta|\lesssim2$);
\item the coefficient $C_\infty$ of the singularity at $t=T(\infty)$ in the Regge limit, to better than $1\%$;
\item beyond the Regge limit, the presence of the phase $\delta_\infty$ in the quasinormal modes at fixed $\ell$.
\end{itemize}
Two points remain open: the effect of $\delta_\infty$ on the coefficient of the singularity beyond the Regge limit, and the fate of the Kemble terms at large $b$, which would require control of exponentially small corrections near $r=0$.

\bibliographystyle{JHEP}
\bibliography{refs}
\end{document}